\documentclass[twocolumn,twocolappendix]{aastex63}
\pdfoutput=1 
\usepackage{amsmath,amstext}
\usepackage[utf8]{inputenc}
\usepackage[T1]{fontenc}
\usepackage{apjfonts} 
\usepackage[figure,figure*]{hypcap}
\usepackage{booktabs}
\usepackage[mathscr]{euscript}
\usepackage{verbatim}
\usepackage{rotating}
\usepackage{footmisc}

\newcommand{\kms}{\mbox{km\,s$^{-1}$}}
\newcommand{\cmc}{\mbox{cm$^{-3}$}}
\newcommand{\per}{$^{-1}$}
\newcommand{\pers}{$^{-2}$}
\newcommand{\mJybeam}{\mbox{mJy\,beam$^{-1}$}}
\newcommand{\muJybeam}{\mbox{$\mu$Jy\,beam$^{-1}$}}
\newcommand{\msun}{M$_\odot$}
\newcommand{\HII}{\mbox{\rm H{\small II}}}
\newcommand{\hcn}{HCN $4-3$}
\newcommand{\httcn}{H$^{13}$CN $4-3$}
\newcommand{\cs}{CS $7-6$}
\newcommand{\hcop}{HCO$^+\ 4-3$}
\newcommand{\cott}{CO $3-2$}
\newcommand{\htwo}{H$_2$}
\newcommand{\ncrit}{$n_{\rm crit}$}
\newcommand{\tclean}{{\fontfamily{cmtt}\selectfont tclean}}

\newcommand{\emcee}{{\fontfamily{cmtt}\selectfont emcee}}
\newcommand{\hst}{\textit{HST}}
\newcommand{\tzamsage}{${\rm t_{ZAMS-age}}$}
\newcommand{\tff}{${\rm t_{ff}}$}

\newcommand{\tcross}{${\rm t_{cross}}$}
\newcommand{\tremovegas}{${\rm t_{remove-gas}}$}

\newcommand{\tdep}{${\rm t_{dep}}$}
\newcommand{\change}{}
\newcommand{\changes}{}

\shortauthors{Levy et al.}

\begin{document}

\title{Outflows from Super Star Clusters in the Central Starburst of NGC\,253}

\author[0000-0003-2508-2586]{Rebecca C. Levy}
\affiliation{Department of Astronomy, University of Maryland, College Park, MD 20742, USA}

\author[0000-0002-5480-5686]{Alberto D. Bolatto}
\affiliation{Department of Astronomy, University of Maryland, College Park, MD 20742, USA}

\author[0000-0002-2545-1700]{Adam K. Leroy}
\affiliation{Department of Astronomy, The Ohio State University, Columbus, OH 43210, USA}

\author[0000-0001-6527-6954]{Kimberly L. Emig}
\altaffiliation{\change Jansky Fellow of the National Radio Astronomy Observatory} 
\affiliation{Leiden Observatory, Leiden University, PO Box 9513, 2300-RA Leiden, the Netherlands}
\affiliation{\change National Radio Astronomy Observatory, 520 Edgemont Road, Charlottesville, VA 22903-2475, USA}

\author[0000-0001-9300-354X]{Mark Gorski}
\affiliation{Chalmers University of Technology, Gothenburg, Sweden}

\author[0000-0003-1104-2014]{Nico Krieger}
\affiliation{Max-Planck-Institut f\"ur Astronomie, K\"onigstuhl 17, 69120 Heidelberg, Germany}

\author[0000-0003-4023-8657]{Laura Lenki\'{c}}
\affiliation{Department of Astronomy, University of Maryland, College Park, MD 20742, USA}

\author[0000-0001-9436-9471]{David S. Meier}
\affiliation{Department of Physics, New Mexico Institute of Mining and Technology, 801 Leroy Pl., Socorro, NM, 87801, USA}
\affiliation{National Radio Astronomy Observatory, P. O. Box O, 1003 Lopezville Rd., Socorro, NM, 87801, USA}

\author[0000-0001-8782-1992]{Elisabeth A. C. Mills}
\affiliation{Department of Physics and Astronomy, University of Kansas, 1251 Wescoe Hall Dr., Lawrence, KS 66045, USA}

\author[0000-0001-8224-1956]{J\"urgen Ott}
\affiliation{National Radio Astronomy Observatory, P. O. Box O, 1003 Lopezville Rd., Socorro, NM, 87801, USA}

\author[0000-0002-5204-2259]{Erik Rosolowsky}
\affiliation{Department of Physics, University of Alberta, 4-183 CCIS, Edmonton, Alberta, T6G 2E1, Canada}

\author[0000-0003-1356-1096]{Elizabeth Tarantino}
\affiliation{Department of Astronomy, University of Maryland, College Park, MD 20742, USA}

\author[0000-0002-3158-6820]{Sylvain Veilleux}
\affiliation{Department of Astronomy, University of Maryland, College Park, MD 20742, USA}

\author[0000-0003-4793-7880]{Fabian Walter}
\affiliation{Max-Planck-Institut f\"ur Astronomie, K\"onigstuhl 17, 69120 Heidelberg, Germany}
\affiliation{National Radio Astronomy Observatory, P. O. Box O, 1003 Lopezville Rd., Socorro, NM, 87801, USA}

\author[0000-0003-4678-3939]{Axel Wei{\upshape{\ss}}}
\affiliation{Max-Planck-Institut f\"ur Radioastronomie, Auf dem H\"ugel 69, D-53121 Bonn, Germany}

\author[0000-0003-0101-1804]{Martin A. Zwaan}
\affiliation{European Southern Observatory, Karl-Schwarzschildstrasse 2, D-85748 Garching Bei Muenchen, Germany}

\correspondingauthor{Rebecca C. Levy}
\email{\change rlevy@astro.umd.edu}

\begin{abstract}
Young massive clusters play an important role in the evolution of their host galaxies, and feedback from the high-mass stars in these clusters can have profound effects on the surrounding interstellar medium. The nuclear starburst in the nearby galaxy NGC\,253 at a distance of 3.5 Mpc is a key laboratory in which to study star formation in an extreme environment. Previous high resolution (1.9 pc) dust continuum observations from ALMA discovered 14 compact, massive super star clusters (SSCs) still in formation. We present here ALMA data at 350 GHz with 28 milliarcsecond (0.5 pc) resolution. We detect blueshifted absorption and redshifted emission (P-Cygni profiles) towards three of these SSCs in multiple lines, including CS 7$-$6 and H$^{13}$CN 4$-$3, which represents direct evidence for previously unobserved outflows. The mass contained in these outflows is a significant fraction of the cluster gas masses, which suggests we are witnessing a short but important phase. Further evidence of this is the finding of a molecular shell around the only SSC visible at near-IR wavelengths. We model the P-Cygni line profiles to constrain the outflow geometry, finding that the outflows must be nearly spherical. Through a comparison of the outflow properties with predictions from simulations, we find that none of the available mechanisms completely explains the observations, although dust-reprocessed radiation pressure and O star stellar winds are the most likely candidates. The observed outflows will have a very substantial effect on the clusters' evolution and star formation efficiency.
\end{abstract}


\section{Introduction}
\label{sec:intro}
Many stages of the stellar life cycle inject energy and momentum into the surrounding medium. For young clusters of stars, this feedback can alter the properties of the cluster itself in addition to the host galaxy. Outflows from young clusters, in particular, remove gas that may have otherwise formed more stars, affecting the star formation efficiency (SFE) of the cluster. The gas clearing process can proceed through a number of physical mechanisms that are efficient over different density regimes and timescales, such as photoionization, radiation pressure, supernovae, and stellar winds. Outflows from forming massive clusters have been observed in the Large Magellanic Cloud \citep{nayak19} and in the Antennae \citep{gilbert07,herrera17}.

At high levels of star formation, a larger fraction of stars form in clustered environments \citep{kruijssen12,johnson16,ginsburg18}. The most extreme star forming environments can lead to massive (${\rm M_*>10^5}$ \msun) and compact (${\rm r\sim1}$ pc) so-called "super" star clusters \citep[SSCs; e.g.,][]{portegieszwart10}. Because they are often deeply embedded, observations of young SSCs in the process of forming are rare {\change \citep[e.g.,][]{keto05,herrera17,oey17,turner17,leroy18,emig20}}. Observations and simulations both indicate that SSCs should have high SFEs {\change \citep[e.g.,][]{skinner15,oey17,turner17,krumholz19,lancaster20}}. Given these high SFEs, by what process(es) do these SSCs disperse their natal gas?

NGC\,253 is an ideal target to study massive, clustered star formation in detail. It is one of the nearest \citep[d$\sim$3.5 Mpc;][]{rekola05} starburst galaxies forming stars at a rate of $\sim$2 \msun\ yr\per\ in the central kiloparsec \citep{bendo15,leroy15}. The star formation is concentrated in dense clumps, knots, and clouds of gas \citep{turner85,ulvestad97,paglione04,sakamoto06,sakamoto11,bendo15,leroy15,ando17}. Recent high-resolution data from the Atacama Large Millimeter/submillimeter Array (ALMA) reveal at least 14 dusty, massive proto-SSCs \citep{leroy18} at the heart of these larger dense gas structures. Moreover, the clusters themselves have radio recombination line (RRL) and radio continuum emission (E. A. C. Mills et al. in prep.), further confirmation that these are young clusters. 

However, even at the 1.9 pc (0.11$\arcsec$) resolution of the previous study by \citet{leroy18}, the clusters, which have radii of $0.6-1.5$ pc, are only marginally resolved. Even higher resolution data are needed to spatially resolve these compact clusters to study the cluster-scale kinematics and feedback.

Here we present direct evidence for massive outflows from three forming SSCs in the center of NGC\,253 using new, very high resolution (0.028$\arcsec\ \approx$ 0.48 pc $\approx 10^5$ AU) data from ALMA at 350 GHz. These three clusters show blueshifted absorption and redshifted emission line profiles---P-Cygni profiles---in several lines. Our analysis shows that these profiles are a direct signature of massive outflows from these SSCs. We briefly describe the observations, data processing, and imaging in Section \ref{sec:obs}. We present measured properties of the outflows from three SSCs based on the line profiles and modeling in Section \ref{sec:results}. We discuss the relevant timescales of the outflows and clusters in terms of their evolutionary stage, mechanisms to power the outflows, and comment on the specific case of SSC 5---which is the only cluster visible in the NIR---in Section \ref{sec:discussion}. We summarize our findings in Section \ref{sec:summary}.

The analysis in this paper will be focused on the three clusters with clear P-Cygni profiles: SSCs 4a, 5a, and 14. We will focus on results obtained from the \cs\ and \httcn\ lines for the following reasons. First, these lines show bright emission towards many of the SSCs and are detected at relatively high signal-to-noise ratio (SNR). Because of their high critical densities (\ncrit$\sim3-5\times10^{6}$\,\cmc\ and \ncrit$\sim0.8-2\times10^{7}$\,\cmc\ respectively;  \citealt{shirley15}) they probe gas that is more localized to the clusters themselves, lessening uncertainties introduced by the foreground gas correction (Section \ref{ssec:foregroundgas}). In the case of \httcn, the $^{13}$C/$^{12}$C isotopic ratio means this line is even less likely to be very optically thick. Although strong, the CO $3-2$ line shape is complicated because it has components from clouds along the line of sight that are not associated with the clusters. Though the \hcn\ and \hcop\ lines are bright and also have high critical densities (\ncrit$\sim0.9-3\times10^{7}$\,\cmc\ and \ncrit$\sim2-3.6\times10^{6}$\,\cmc\ respectively; \citealt{shirley15}), they are more abundant and more optically thick than \cs\ and \httcn, and the absorption components in these lines suffer from saturation (i.e. absorption down to zero). This makes determining physical quantities of the outflows---which rely on the absorption depth to continuum ratio (Section \ref{sec:results})---difficult and uncertain. Therefore, we find that the \cs\ and \httcn\ lines provide the best balance between bright lines with sufficient SNR, absorption features which do not suffer from saturation effects, and which probe gas localized to the clusters to minimize uncertainties from the foreground gas correction.

\section{Observations, Data Processing, and Outflow Identification}
\label{sec:obs}

Data for this project were taken with the Atacama Large Millimeter/submillimeter Array (ALMA) as part of project 2017.1.00433.S (P.I. A. Bolatto). We observed the central 16.64$\arcsec$ (280 pc) of NGC\,253 at Band 7 ($\nu\sim350$\,GHz, $\lambda\sim0.85\ \mu$m) using the main 12-m array in the C43-9 configuration. {\change This configuration resulted in baselines spanning from 113~m $-$ 13.9~km and hence a maximum recoverable scale of 0.38\arcsec\ (6.4~pc).} The spectral setup is identical to our previous observations of this region \citep{leroy18,krieger19,krieger20}, spanning frequency ranges of $342.08-345.78$ GHz in the lower sideband and $353.95-357.66$ GHz in the upper sideband. Observations were taken on November $9-11$, 2017 with a total observing time of 5.7 hours of which 2.0 hours were on-source. The visibilities were pipeline calibrated using the Common Astronomy Software Application (CASA; \citealt{casa}) version 5.1.1-5 (L. Davis et al. in prep.). J0038-2459 was the phase calibrator, and J0006-0623 was the bandpass and flux calibrator. 

\subsection{Imaging the Continuum}
\label{ssec:cont}

In this paper, we focus on the line profiles, specifically the \cs\ and \httcn\ lines towards three of the SSCs. We will present the continuum data more fully in a forthcoming paper (R. C. Levy et al., in prep.). Some of the data are presented and used here, and we provide a brief summary.

To extract the 350 GHz continuum data, we flagged {\change channels} that may contain strong lines in the band, assuming a systemic velocity of 243\,\kms\ \citep{koribalski04}. {\change Lines included in the flagging are $^{12}$CO $3-2$, \hcn, \httcn, \cs, \hcop, $^{29}$SiO $8-7$, and $^{12}$CO $3-2$ v=1 (though this line is not detected), and channels within $\pm$200~\kms\ of the rest frequencies of these lines were flagged.} We imaged the line-flagged {\change visibilities} using the CASA version 5.4.1 \tclean\ task with \texttt{specmode=`mfs'}, \texttt{deconvolver=`multiscale'}, \texttt{scales=[0,4,16,64]}, Briggs weighting with \texttt{robust=0.5}, and no $uv$-taper. The baseline was fit with {\change a linear function (\texttt{nterms=2})} to account for any change in slope over the wide band. The continuum was cleaned down to a threshold of 27 \muJybeam, after which the residuals resembled noise. This map has  a beam size of 0.024$\arcsec\times$0.016$\arcsec$ and a cell (pixel) size of 0.0046$\arcsec$ (0.078 pc), so that we place around 4 pixels across the minor axis of the beam. Finally, the map was convolved to a circular beam size of 0.028$\arcsec$ (0.48 pc). The rms noise of the continuum image (away from emission) is 26 \muJybeam (0.3 K).

\begin{figure*}
    \label{fig:cont}
    \centering
    \includegraphics[width=\textwidth]{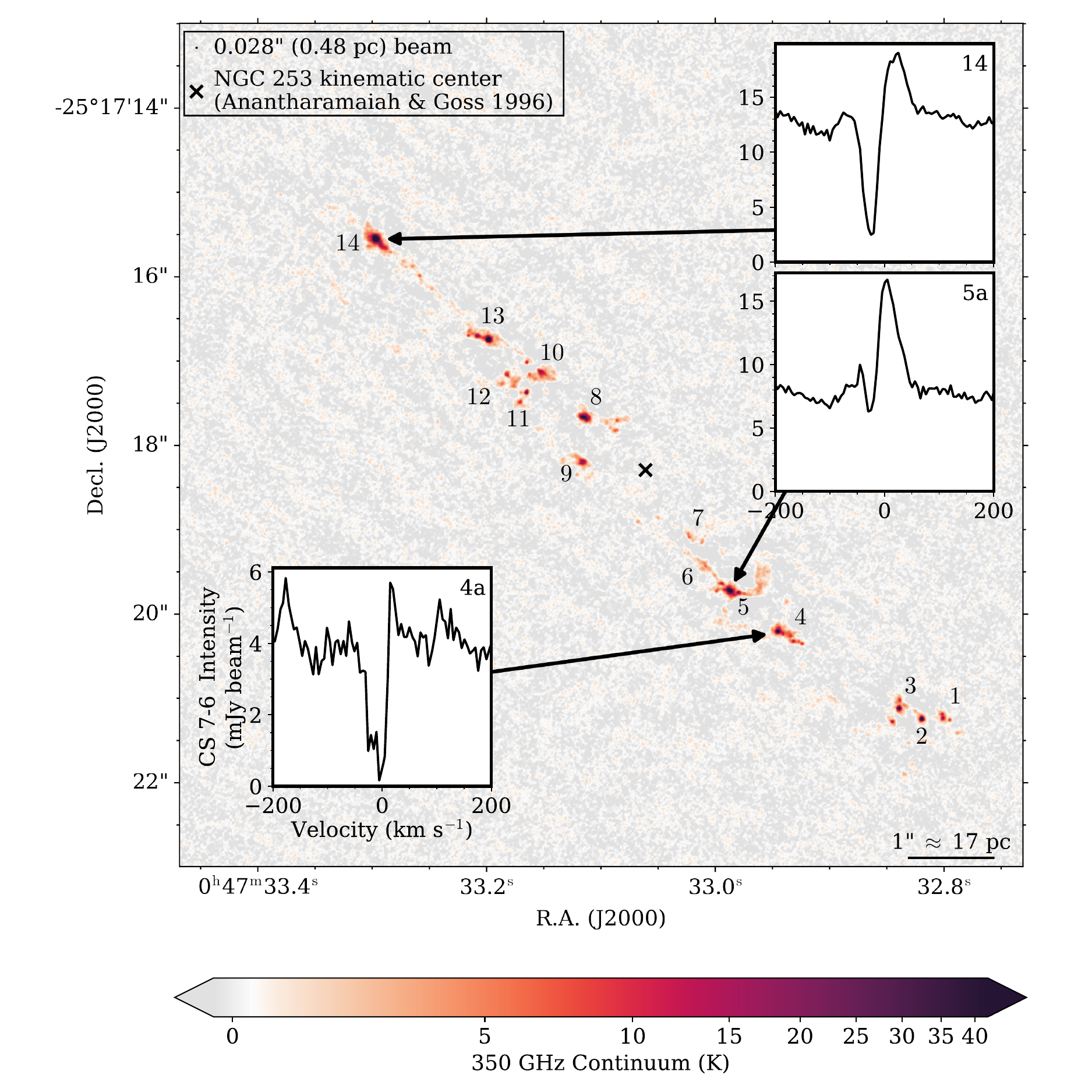}
    \caption{The 350 GHz dust continuum map covering the central 10$\arcsec\times$10$\arcsec$ (170 pc $\times$ 170 pc) of NGC\,253 with an rms noise of 26 \muJybeam (0.3 K). The beam FWHM is shown in the upper left corner. The $\times$ marks the center of NGC\,253 from ionized gas kinematics traced by radio recombination lines \citep{anantharamaiah96}. The 14 clusters identified by \citet{leroy18} are labeled. Despite the increased resolution most of the massive clusters identified by \citet{leroy18} persist, sometimes with a few lower-mass companion objects. The insets show the \cs\ spectra towards the three SSCs analyzed in this paper (SSCs 4a, 5a, and 14), highlighting the characteristic P-Cygni profiles indicative of outflows. These are the only clusters toward which we unambiguously identify blueshifted absorption signaling outflow activity.}
\end{figure*}

From this high resolution 350 GHz continuum image (Figure \ref{fig:cont}), we identify approximately three dozen clumps of dust emission. These are co-spatial with the 14 proto-SSCs identified by \citet{leroy18}, with many sources breaking apart into smaller clumps that likely represent individual clusters. {\change We note that there is appreciable overlap in the spatial scales probed by these observations and those presented by \citet{leroy18}. This means that the larger cluster sizes measured by \citet{leroy18} are due to the lower resolution, with the true cluster sizes better traced by these observations. We further verify this by computing the fluxes in the high resolution continuum image in the same apertures used by \citet[][see their Table 1]{leroy18}. For SSCs 4, 5, and 14 (the focus of this analysis), we recover $>80$\% of the reported flux.}

We follow the SSC naming convention of \citet{leroy18}, adding letters to denote sub-clusters in decreasing order of brightness. For clusters that break apart, there is always a main, brightest cluster. For example, SSC 4 from \citet{leroy18} breaks apart into six dust clumps, where the peak intensity of SSC 4a is $>$4$\times$ that of any of the smaller clumps. For this paper, we will focus only on the primary clusters since those are the most massive, and smaller dust clumps may not be true clusters. We also remove SSC 6 from this analysis, as it is extended and more tenuous in the high resolution continuum map and so may not be a SSC. \citet{leroy18} also note that this clump is weak and may instead be a supernova remnant, which is further supported by a {\change non-negligible} synchrotron component to this cluster's spectral energy distribution (SED; E. A. C. Mills et al., in prep.).

\begin{deluxetable}{ccccc}
\tablecaption{Primary Cluster Positions and Sizes Measured from the 350 GHz Continuum Image\label{tab:contparams}}
\tablehead{\colhead{SSC Number} & \colhead{R.A.} & \colhead{Decl. } & \colhead{$r_\mathrm{half-flux}$ } & \colhead{V$_\mathrm{LSRK}$} \\  & \colhead{(J2000)} & \colhead{(J2000)} & \colhead{(pc)} & \colhead{(km s$^{-1}$)}}
\startdata
1a & $0^\mathrm{h}47^\mathrm{m}32.801^\mathrm{s}$ & $-25^\circ17{}^\prime21.236{}^{\prime\prime}$ & 0.59 & 315\\
2 & $0^\mathrm{h}47^\mathrm{m}32.819^\mathrm{s}$ & $-25^\circ17{}^\prime21.247{}^{\prime\prime}$ & 0.19 & 305\\
3a & $0^\mathrm{h}47^\mathrm{m}32.839^\mathrm{s}$ & $-25^\circ17{}^\prime21.122{}^{\prime\prime}$ & 0.47 & 302\\
4a & $0^\mathrm{h}47^\mathrm{m}32.945^\mathrm{s}$ & $-25^\circ17{}^\prime20.209{}^{\prime\prime}$ & 0.50 & 251\\
5a & $0^\mathrm{h}47^\mathrm{m}32.987^\mathrm{s}$ & $-25^\circ17{}^\prime19.725{}^{\prime\prime}$ & 0.76 & 215\\
7a & $0^\mathrm{h}47^\mathrm{m}33.022^\mathrm{s}$ & $-25^\circ17{}^\prime19.086{}^{\prime\prime}$ & 0.59 & 270\\
8a & $0^\mathrm{h}47^\mathrm{m}33.115^\mathrm{s}$ & $-25^\circ17{}^\prime17.668{}^{\prime\prime}$ & 1.20 & 295\\
9a & $0^\mathrm{h}47^\mathrm{m}33.116^\mathrm{s}$ & $-25^\circ17{}^\prime18.200{}^{\prime\prime}$ & 0.46 & 155\\
10a & $0^\mathrm{h}47^\mathrm{m}33.152^\mathrm{s}$ & $-25^\circ17{}^\prime17.134{}^{\prime\prime}$ & 1.32 & 280\\
11a & $0^\mathrm{h}47^\mathrm{m}33.165^\mathrm{s}$ & $-25^\circ17{}^\prime17.376{}^{\prime\prime}$ & 0.13 & 145\\
12a & $0^\mathrm{h}47^\mathrm{m}33.182^\mathrm{s}$ & $-25^\circ17{}^\prime17.165{}^{\prime\prime}$ & 1.29 & 160\\
13a & $0^\mathrm{h}47^\mathrm{m}33.198^\mathrm{s}$ & $-25^\circ17{}^\prime16.750{}^{\prime\prime}$ & 0.36 & 245\\
14 & $0^\mathrm{h}47^\mathrm{m}33.297^\mathrm{s}$ & $-25^\circ17{}^\prime15.558{}^{\prime\prime}$ & 0.53 & 205\\
\enddata
\tablecomments{The cluster positions and sizes for the primary SSCs identified from the 350 GHz continuum image (Figure \ref{fig:cont}). We follow the cluster naming convention of \cite{leroy18}. Clusters that break apart into multiple sub-clusters have an "a" added to their number. We remove SSC 6 as it is likely not a true cluster due to its morphology. Positions are determined by fitting a 2D Gaussian. $r_\mathrm{half-flux}$ is the half-flux radius determined from the radial profile. The beam (0.028" = 0.48 pc) has been deconvolved from the reported sizes. V$_\mathrm{LSRK}$ is the cluster systemic velocity (see Section \ref{ssec:vsys}). For SSCs 4a, 5{\change a}, and 14, the estimated {\change uncertainty} is $\pm1$ \kms. For the other {\change clusters}, estimated uncertainties are $\pm5$ \kms.}
\end{deluxetable}

The position and orientation of each primary cluster is measured by fitting a 2D Gaussian to each cluster in the continuum image using \emcee\ \citep{emcee}. The reported values are the medians of the marginalized posterior likelihood distributions (Table \ref{tab:contparams}). As a non-parametric measurement of the cluster sizes, we compute the half-flux radius ($r_{\rm half-flux}$) for each cluster. Given the cluster positions and orientations from the 2D Gaussian fit, we construct elliptical annuli in steps of half the beam FWHM. The half-flux radius is the median of the cumulative flux distribution. We use this radius measurement as opposed to those from the 2D Gaussian fit because the cluster light profiles are not necessarily well described by a Gaussian, which will be discussed further in a forthcoming paper (R. C. Levy et al., in prep.). We deconvolve the beam from the radius measurements by removing half the beam FWHM from the half-flux radius in quadrature. The beam deconvolution will be treated more fully in a forthcoming paper focusing on the continuum properties of the SSCs {\change which will also include the lower resolution data} (R. C. Levy et al., in prep.).

\subsection{Imaging the Lines}
\label{ssec:lineimaging}

\begin{figure*}
\label{fig:maps}
\centering
\gridline{\fig{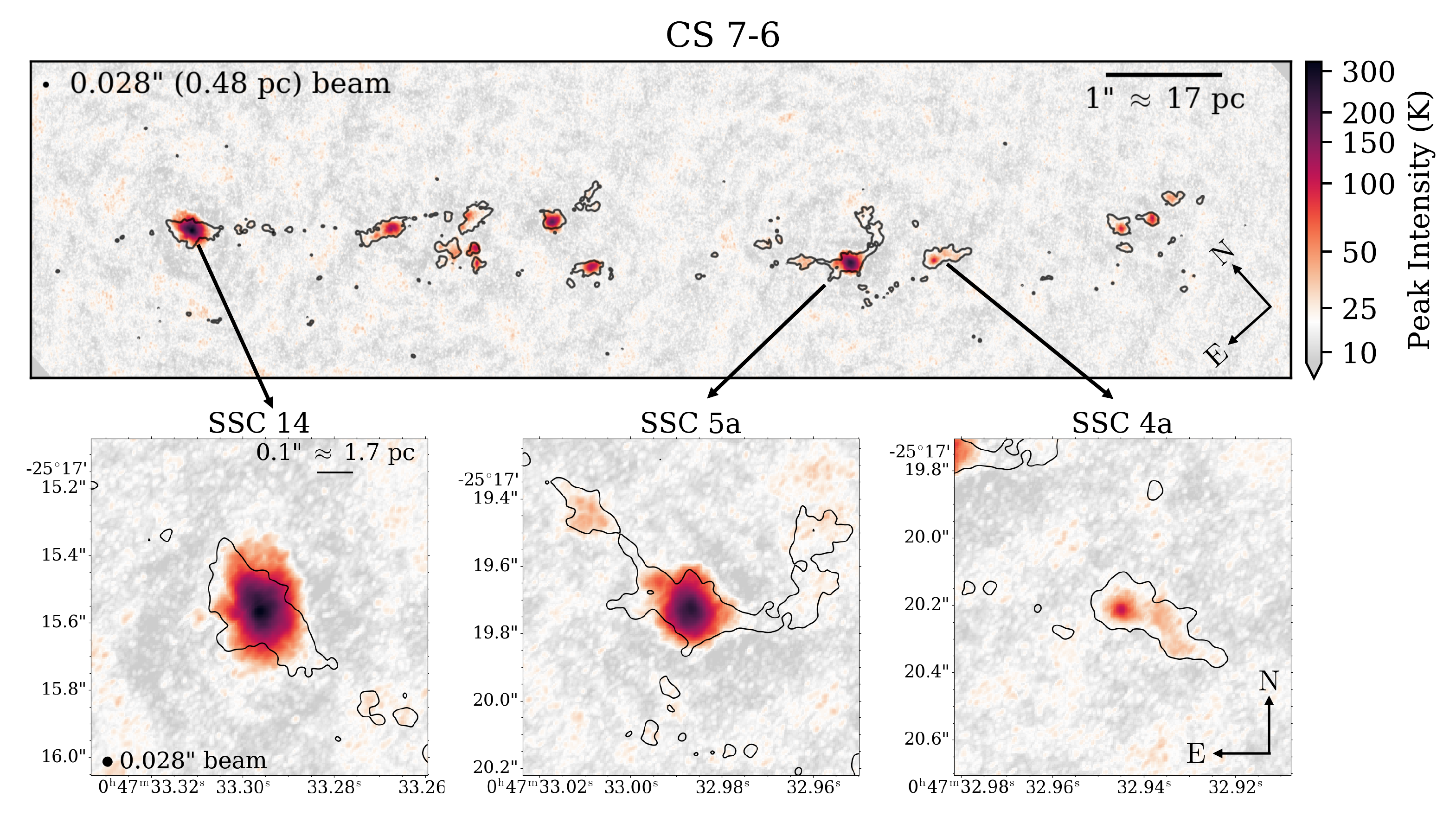}{0.95\textwidth}{}}
\vspace{-5mm}
\gridline{\fig{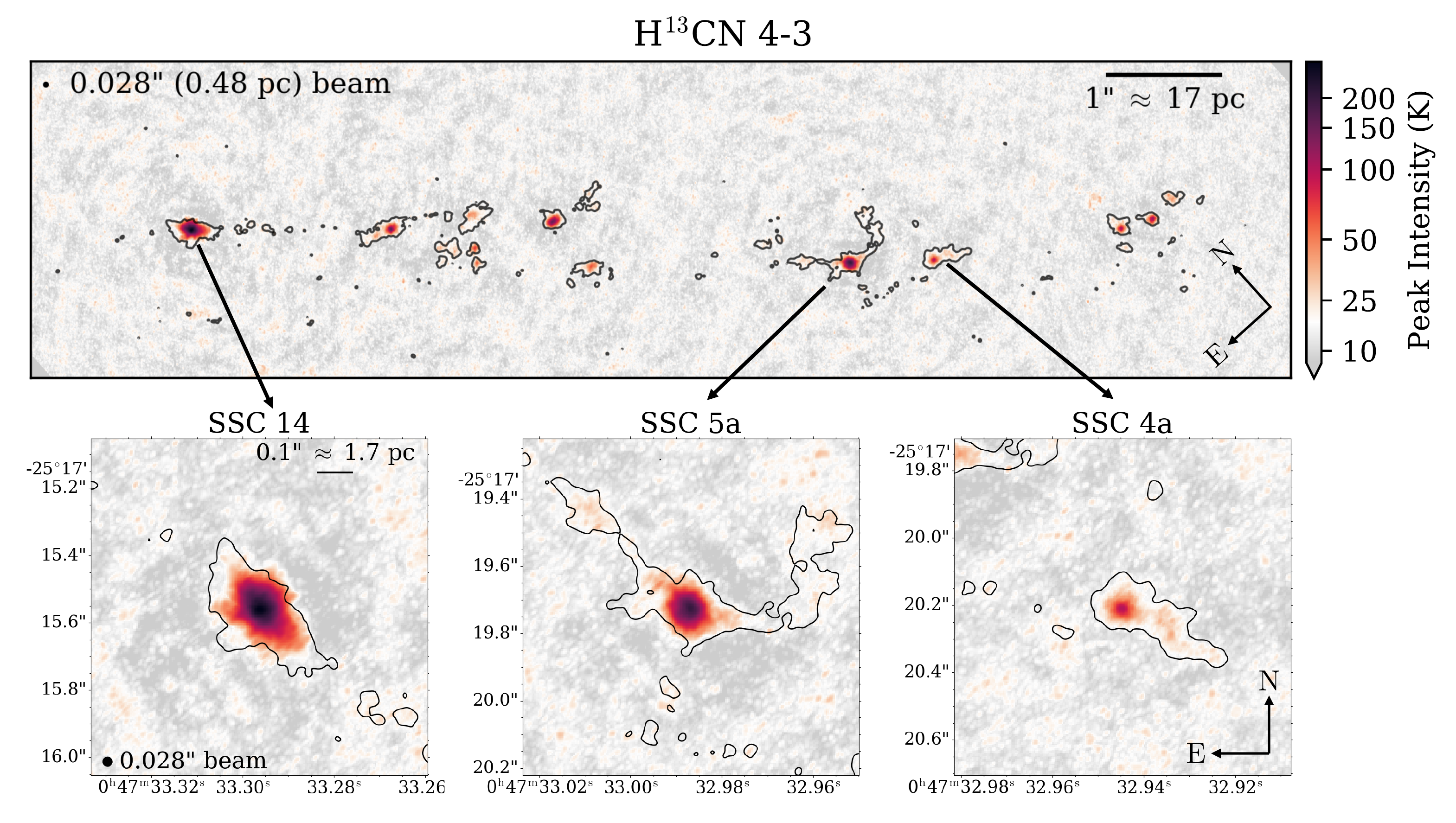}{0.95\textwidth}{}}
\vspace{-24pt}
\caption{Peak intensity maps of  \cs\ (top group) and \httcn\ (bottom group) within $\pm$200 \kms\ of the galaxy systemic velocity to avoid contamination by other strong lines in the band. The top panels in each group have been rotated so the linear structure is horizontal, as indicated by the coordinate axes in the lower right corner. The contours show 5 times the rms noise level in the 350 GHz dust continuum shown in Figure \ref{fig:cont} (5$\times$rms = 1.5 K), and so indicate the extent of the dust structures around the SSCs. The square plots show the peak intensity maps and dust continuum contours zoomed in to 1$\arcsec$ (17 pc) square regions around SSCs  14, 5a, and 4a, highlighting the localized and spatially resolved emission towards these SSCs. These have not been rotated, as indicated by the coordinate axes in the lower right corner.}
\end{figure*}

We image the \cs\ and \httcn\ lines by selecting a 800 \kms\ window around the line center (rest frequencies of 342.883 GHz and 349.339 GHz respectively), assuming a systemic velocity of 243\,\kms\ \citep{koribalski04}. We clean the lines of interest using \tclean\ in CASA version 5.4.1 with \texttt{specmode=`cube'}, \texttt{deconvolver=`multiscale'}, \texttt{scales = [0]}, Briggs weighting with \texttt{robust = 0.5}, and no $uv$-taper. No clean mask was used. The baseline was fit with {\change a linear function (\texttt{nterms=2})} to account for any change in slope over the wide band. The continuum is not removed from {\change the visibilities or the cleaned cubes}. The cell (pixel) size is 0.0046$\arcsec$ (0.078 pc), so that we place around 4 pixels across the minor axis of the beam. The final elliptical beam is 0.027$\arcsec\times$0.019$\arcsec$. Finally, we convolve to a round 0.028$\arcsec$ (0.48 pc) beam. The rms noise is 0.5 \mJybeam\  (6.5 K) per 5 \kms\ channel. Figure \ref{fig:maps} shows the peak intensity maps for \cs\ and \httcn\ within $\pm$200 \kms\ {\change about} the galaxy's systemic velocity to avoid possible contamination by other strong lines (especially by HC$_3$N in the case of \httcn). 

\subsection{Full-Band Spectra and Detected Lines}
\label{ssec:linelist}

\begin{figure*}[p]
    \centering
    \gridline{\fig{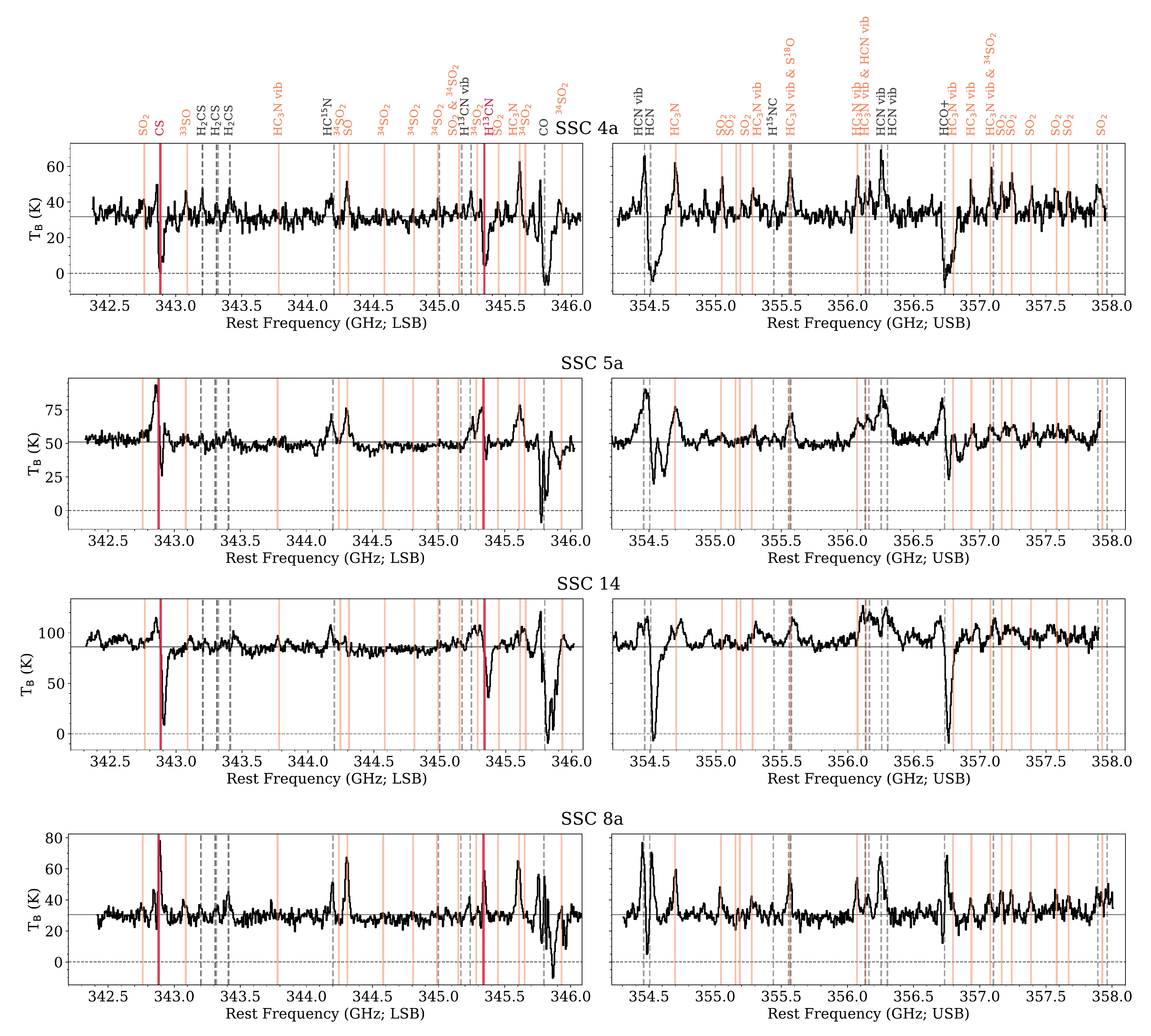}{\textwidth}{}}
    \vspace{-5mm}
    \caption{The full band spectra of SSCs 4a (top), 5a (top-middle), 14 (bottom-middle), and 8a (bottom). These spectra are averaged over the FWHM continuum source size and have been corrected for foreground gas (Section \ref{ssec:foregroundgas}). The continuum has not been removed{\change, and is shown by the solid horizontal gray lines}. There are many {\change spectral} lines detected in these clusters, {\change some of} which are marked by the vertical lines and labels (where vib denotes a ro-vibrational transition; see also \citealt{krieger20}). The main lines used in this study (\cs\ and \httcn) are marked in red (thick solid), lines used to determine the systemic velocity are in orange (thin solid), and other lines are in gray (dashed). The P-Cygni profile is seen in many lines in SSCs 4a, 5a, and 14. In some cases, the absorption is saturated (i.e. absorbed down to zero intensity shown by the horizontal gray {\change dashed} lines) especially in the brightest lines (such as \cott, \hcn, and \hcop). The \cott\ line profile is further complicated due to clouds along the line of sight. SSC 8a is shown as an example of bright source without evidence for outflows, though there may be a hint of inflows in \cs, \httcn, and \hcop. }
    \label{fig:fullbandspec}
\end{figure*}

Though the analysis is focused on the \cs\ and \httcn\ lines, we also make a "dirty" cube covering the entire band and imaged area by running \tclean\ with \texttt{niter=0}. This dirty cube has a cell (pixel) size of 0.02$\arcsec$ and an rms noise away from the emitting regions of 0.57 \mJybeam\ per 5 \kms\ channel. The synthesized beam is 0.05$\arcsec\times$0.025$\arcsec$. The continuum is not removed from {\change the visibilities or the dirty cube}.

In Figure \ref{fig:fullbandspec}, we show the full band spectrum of SSCs 4a, 5a, and 14, extracted from the dirty cube and averaged over the FWHM continuum source size. There are several immediately striking features in these spectra. The brightest lines in SSCs 4a, 5a, and 14 show deep, blueshifted absorption features, extending down to $\sim0$\,\mJybeam{\change, and redshifted emission. This line shape is commonly referred to as a P-Cygni profile --- for the star in which it was first detected --- and is indicative of outflows.} As a comparison, we also show the spectrum of SSC 8a in Figure \ref{fig:fullbandspec}, which does not have these P-Cygni line shapes. We note that the CO $3-2$ line has a complicated structure, likely because it traces lower density gas than other detected lines  \citep[\ncrit$\sim2\times10^4$\,\cmc; e.g.,][]{turner17} and so it arises in many places along the line of sight to the SSCs.

There are many detected lines in addition to the usual dense gas tracers. These include shock tracers such as SO $8-7$ and SO$_2$ lines and vibrationally excited HC$_3$N and \hcn, which are primarily excited through IR pumping \citep[e.g.][]{ziurys86,krieger20}. Many of these have been previously identified by \citet{krieger20}, some of which are marked in Figure \ref{fig:fullbandspec}. 

\subsection{Correcting for Foreground Gas}
\label{ssec:foregroundgas}

\begin{figure*}[p]
\label{fig:chanmaps}
\centering
\gridline{\fig{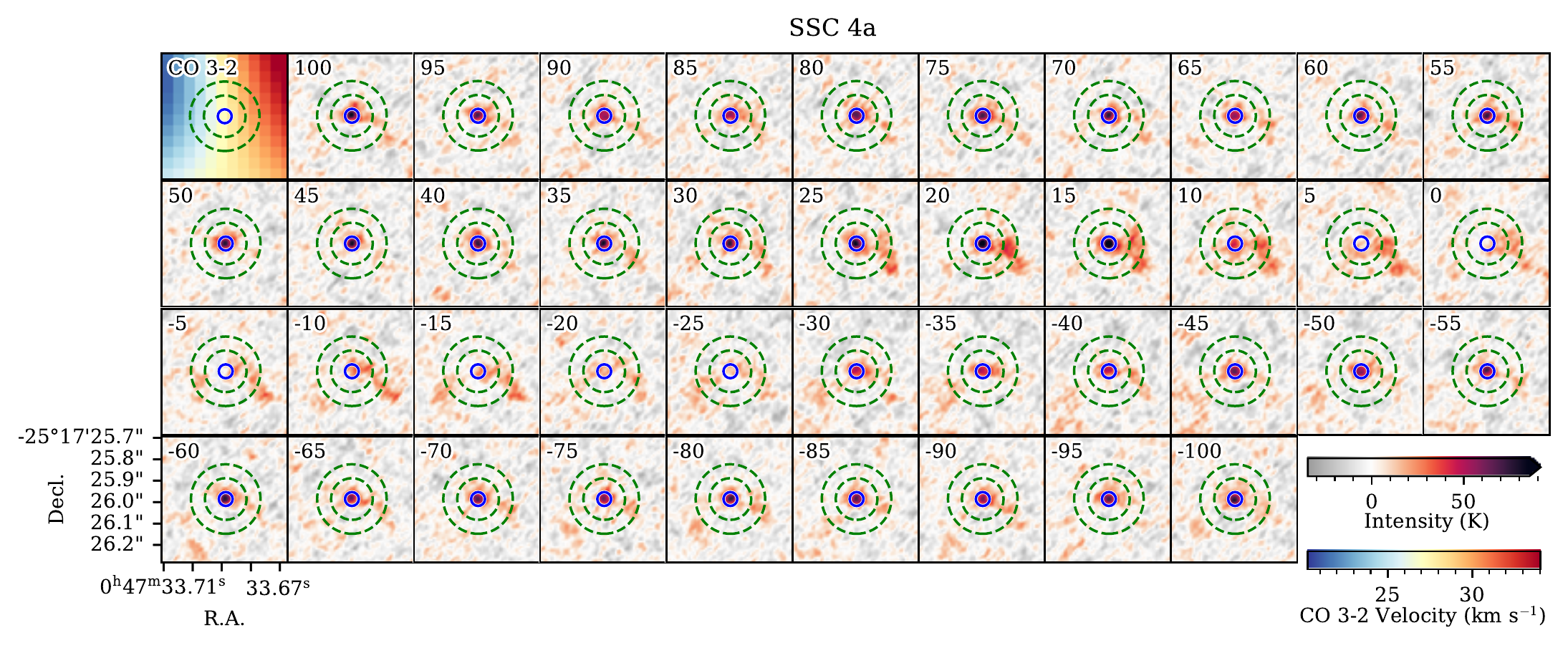}{0.9\textwidth}{}}
\vspace{-10mm}
\gridline{\fig{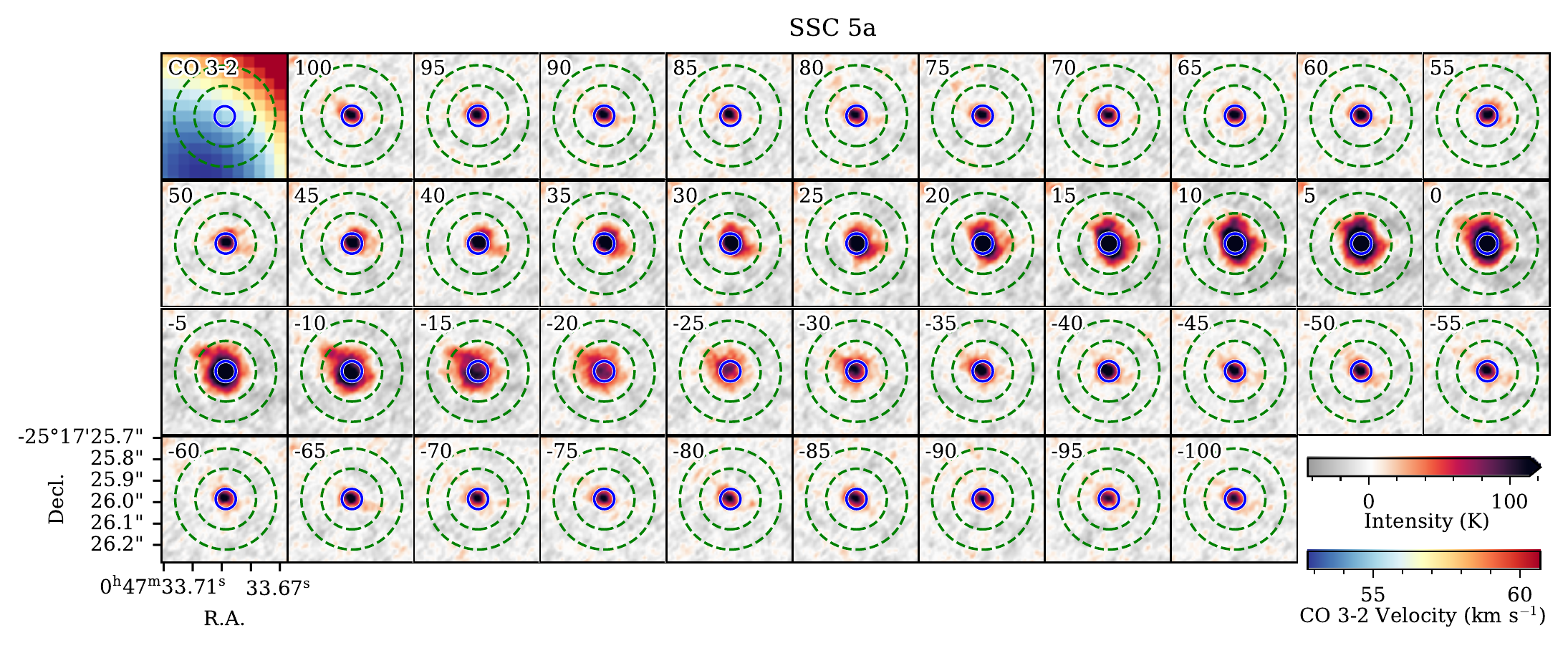}{0.9\textwidth}{}}
\vspace{-10mm}
\gridline{\fig{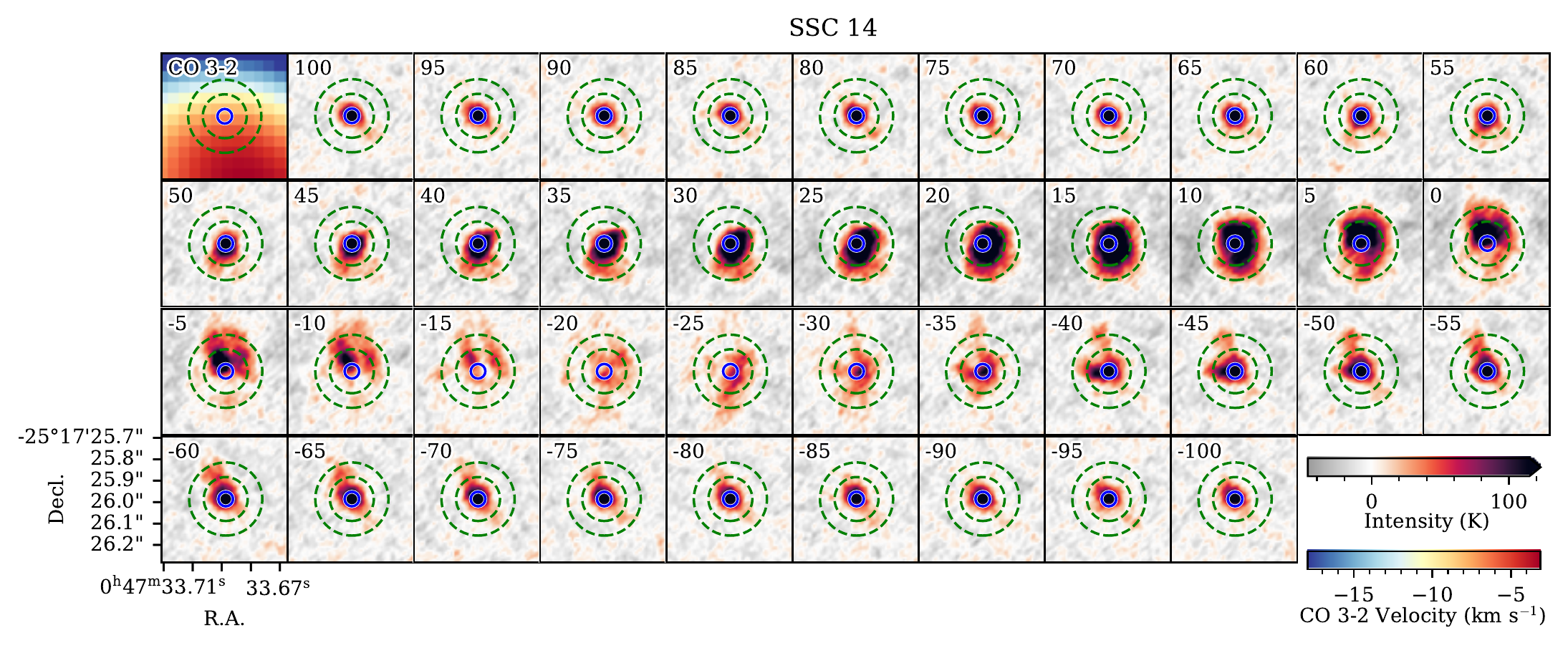}{0.9\textwidth}{}}
\vspace{-10mm}
\caption{Channel maps around \cs\ for SSCs 4a (top), 5a (middle), and 14 (bottom) in 10 pc $\times$ 10 pc regions. The continuum has not been removed. The first panel shows the smoothed CO $3-2$ velocity field in this region \citep{krieger19}, indicative of the large-scale bulk gas motions. The \cott\ colorscale and the labeled \cs\ velocity channels (in \kms) are relative to the cluster systemic velocity (Table \ref{tab:contparams}). The solid blue circles show the continuum source sizes, and the dashed green annuli show the regions used for the foreground correction. The \httcn\ channel maps are similar.}
\end{figure*}

\begin{figure}
\label{fig:foregroundgas}
\centering
\gridline{\fig{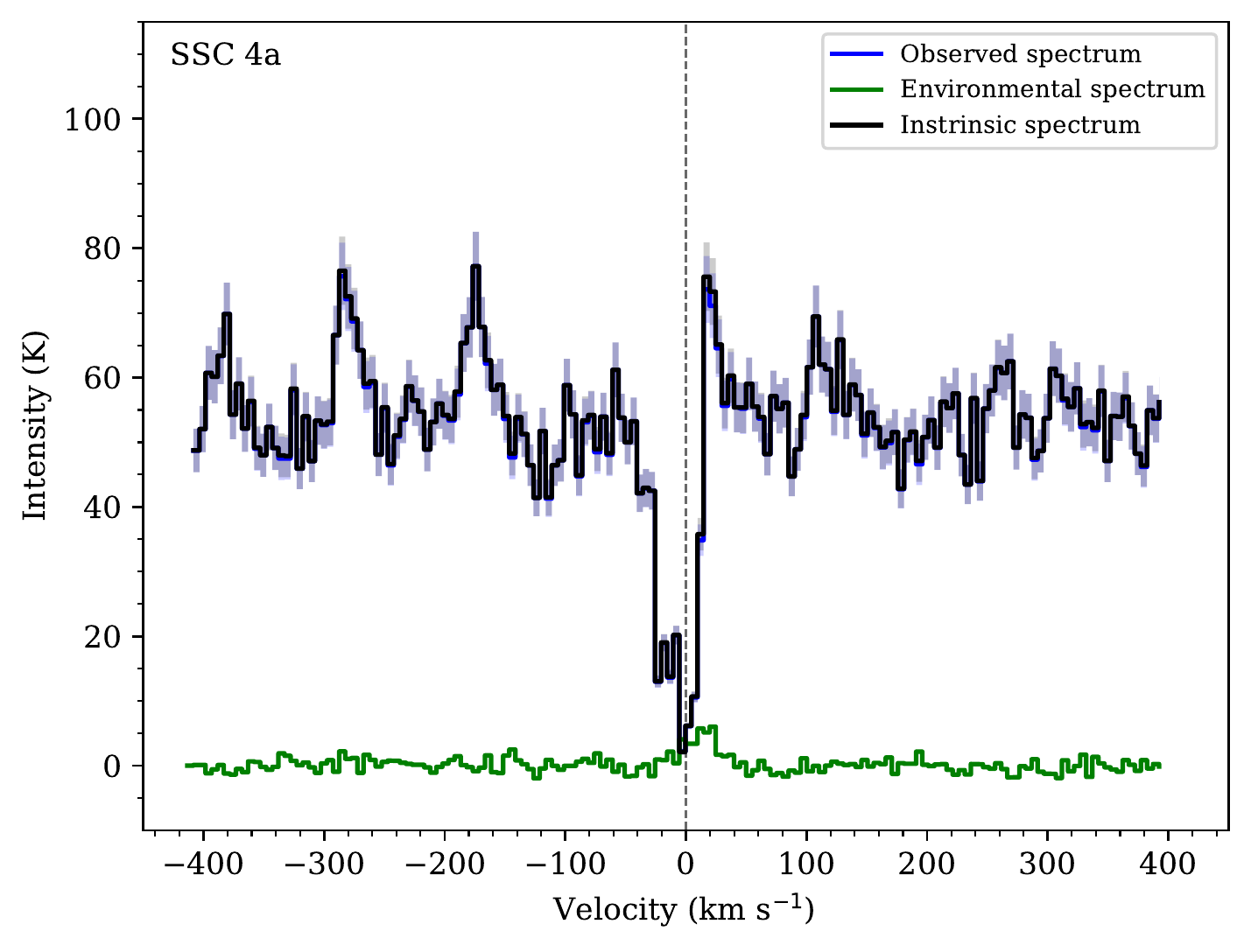}{0.925\columnwidth}{}}
\vspace{-10mm}
\gridline{\fig{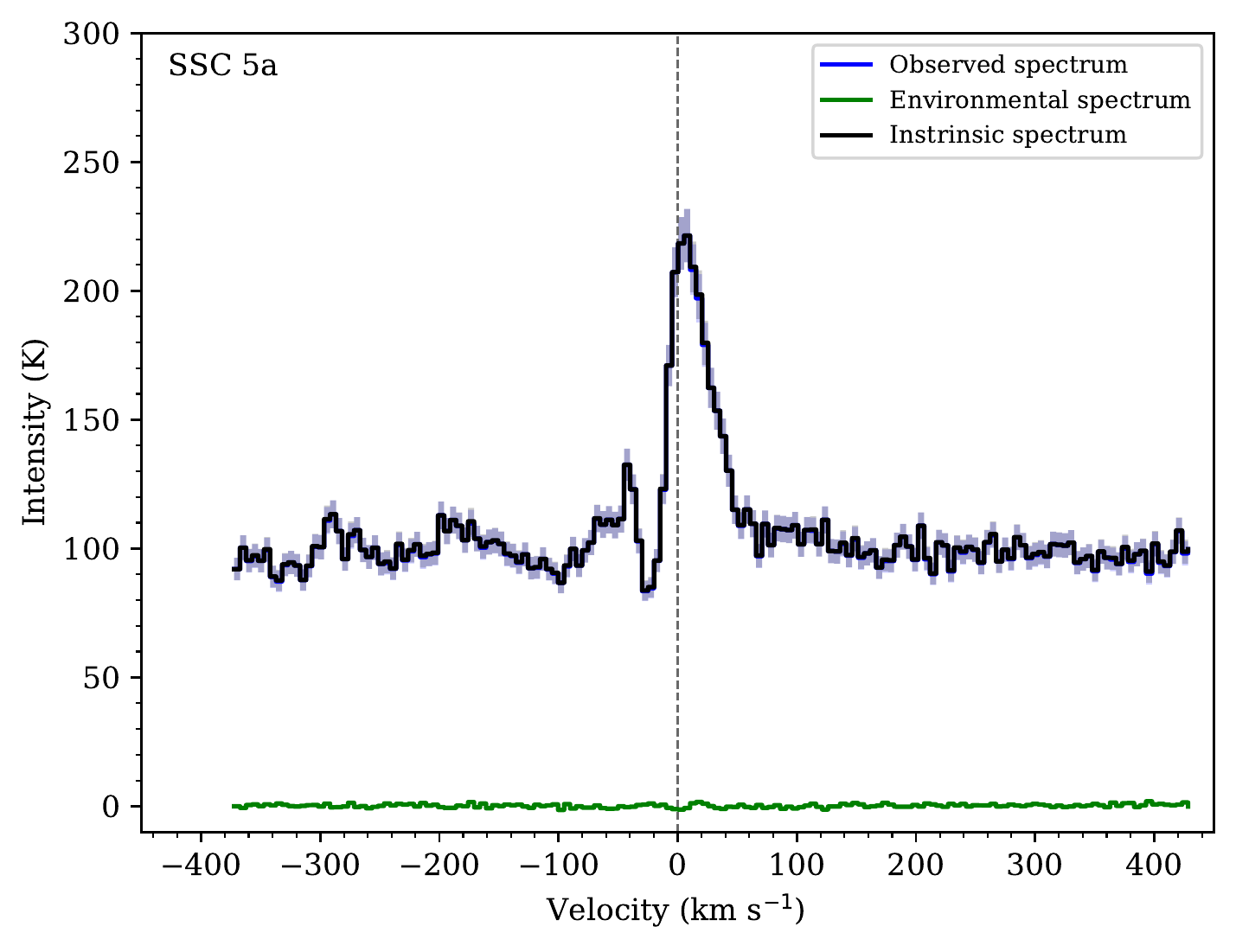}{0.925\columnwidth}{}}
\vspace{-10mm}
\gridline{\fig{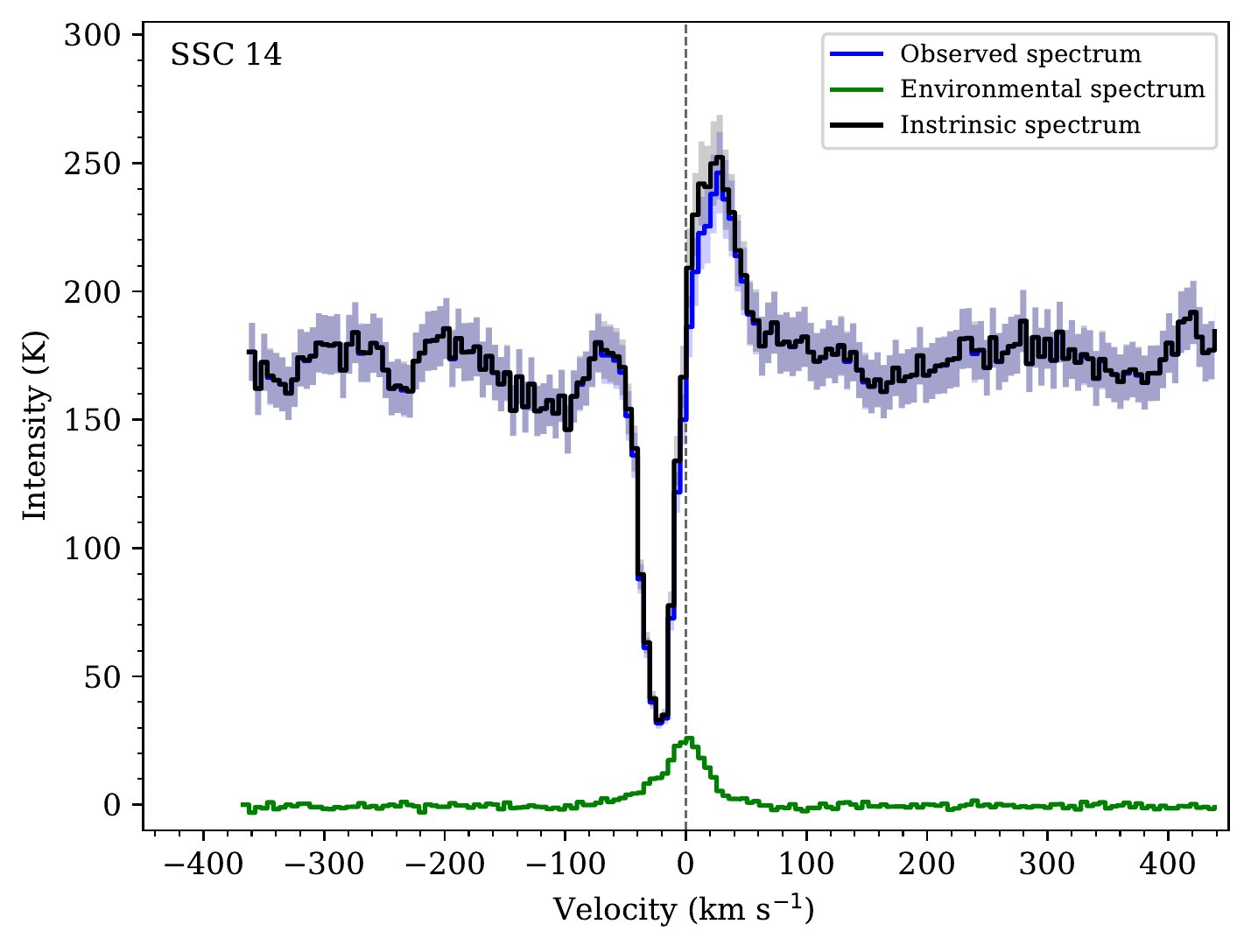}{0.925\columnwidth}{}}
\vspace{-5mm}
\caption{The \cs\ spectra for SSCs 4a (top), 5a (middle), and 14 (bottom) averaged over the 350 GHz continuum source area (blue; corresponding to the blue ellipses in Figure \ref{fig:chanmaps}), the foreground gas (green; corresponding to the green annuli in Figure \ref{fig:chanmaps}), and the corrected, intrinsic \cs\ spectrum accounting for the foreground gas (black). The blue shaded region around the observed spectrum reflects the standard deviation of the spectrum within the continuum source area. The gray shaded region around the intrinsic spectrum includes the propagated uncertainties from the assumed excitation temperature (${\rm T_{ex}=102\pm28}$ K). See Section \ref{ssec:foregroundgas} for details of the foreground gas correction.}
\end{figure}

Because the SSCs are embedded in the nucleus of NGC\,253, there is dense gas along the line of sight which may contribute to the observed spectra but is not associated with the SSCs themselves. This can be seen in the line channel maps around each SSC, shown in Figure \ref{fig:chanmaps} for \cs. This gas is most evident in SSCs 5a and 14, which both show changes in the gas morphology with velocity (relative to the cluster systemic velocity; see Section \ref{ssec:vsys}). The first panel of each set of channel maps shows the \cott\ velocity field in the same region as a tracer of the bulk gas motions not associated with the clusters themselves. Because we are interested in the large scale gas motions, we smooth the \cott\ velocity maps presented by \citet{krieger19} to 5$\times$ the beam major axis ($5\times0.17\arcsec = 0.85\arcsec= 14.4$ pc). The velocities shown in Figure \ref{fig:chanmaps} are relative to the cluster systemic velocities. All three clusters are located at different velocities compared to the bulk gas motions (i.e., \cott\ velocity maps are not centered on zero). In the case of SSC 14, the distinct morphological evolution with velocity matches qualitatively with the \cott, though there is a velocity offset. It is difficult to know what fraction of this is associated with the clusters themselves or with dense gas in the environment of the clusters but not necessarily bound to them. Here, we will assume that all of the emission in the green annuli in Figure \ref{fig:chanmaps} is not bound to the clusters and should be corrected to obtain the intrinsic spectrum of the clusters. The spectrum of the environmental gas (T$_{\rm env}$) is averaged in an annulus with an inner radius of 3$\times$ the half-flux radius and an outer radius of 5$\times$ the half-flux radius, as shown in Figure \ref{fig:chanmaps}. This effectively assumes the environmental dense gas forms a uniform screen in between the cluster and our line of sight. This is unlikely to be the case and is one of the uncertainties of this correction.

We calculate the optical depth ($\tau_\nu$) of the environmental gas following \citet{mangum15}. Combining their Equations 24 and 27, we have
\begin{equation}
    \label{eq:tauenv}
    \tau_{\nu} = -\ln\left[1-{\rm \frac{kT_{env}}{h\nu}}\left(\frac{1}{e^{\rm  h\nu/kT_{ex}}-1}-\frac{1}{e^{\rm  h\nu/kT_{bg}}-1}\right)^{-1}\right]
\end{equation}
 where the filling fraction is assumed to be unity, the excitation temperature (T$_{\rm ex}$) is $102\pm28$ K and the background temperature (T$_{\rm bg}$) is measured from the continuum level of the environmental spectrum (but is $\approx0$ K{\change; Figure \ref{fig:foregroundgas}}). The assumed value of T$_{\rm ex}$ is the average of the excitation temperatures found by \citet{meier15} over larger scales in the nucleus (74 K) and by \citet{krieger20} for regions more localized to the SSCs (130 K), since the true value of the excitation temperature of this dense gas near the clusters likely falls somewhere between these two measurements. The uncertainty reflects half of the difference of these two values. 
 
 We assume that half of the environmental emission comes from the foreground, so that the optical depth of the material along the line of sight in front of the cluster is $\tau_{\rm \nu,fg} = \tau_\nu/2$. We also assume that the environmental gas is colder than the gas in the cluster, so it will absorb emission from the cluster. We then derive the corrected (intrinsic) SSC spectra where 
\begin{equation}
\begin{split}
\label{eq:foregroundcorr}
{\rm T_{obs}} = {\rm T_{intrinsic}}e^{-\tau_{\rm \nu,fg}}+{\rm T_{env}}\left(1-e^{-\tau_{\rm \nu,fg}}\right)\\
\Rightarrow{\rm T_{intrinsic}} = e^{\tau_{\rm \nu,fg}}\left[{\rm T_{obs}}-{\rm T_{env}}\left(1-e^{-\tau_{\rm \nu,fg}}\right)\right],
\end{split}
\end{equation}
where the second equation is a rearrangement of the first. The observed, environmental, and intrinsic spectra of \cs\ for SSCs 4a, 5a, and 14 are shown in Figure \ref{fig:foregroundgas}. This process is repeated for \httcn\ and the full-band spectra in Figure \ref{fig:fullbandspec}.

As discussed earlier, it is difficult to determine what fraction of the extended emission seen in the channel maps (Figure \ref{fig:chanmaps}) is environmental or associated with the cluster. Moreover, our method for the foreground correction is quite simplistic, and there are uncertainties related to the fraction of gas that may be associated with the clusters, the geometry of the environmental material, and the assumed excitation temperature. As can be seen in Figure \ref{fig:foregroundgas}, however, the correction mostly affects the peak of the emission components. After the foreground correction, the emission peak for SSC 14 increases by 7\% (17 K), whereas the absorption {\change trough} only increases by 4.5\% (1.5 K). There is no appreciable change in the spectra for SSCs 4a and 5a. The absorption features are, therefore, largely unaffected by this correction, and the derived properties of the outflow (presented in Sections \ref{ssec:outflow} and \ref{ssec:outflowmodeling}) are robust. {\change The intrinsic (corrected) spectra are used throughout the analysis of this paper.}

\subsection{Cluster Systemic Velocities}
\label{ssec:vsys}
We constrain the systemic velocity of all of the detected clusters using the full-band spectra and the detected line list presented by \citet{krieger20}. Using the full-band spectrum for each cluster,  we simultaneously fit each line in the line list using a series of Gaussians of the form
\begin{equation}
    \label{eq:gaussians}
    I(\nu) = I_{\rm cont} + \sum_{\rm line} I_{\rm line}\exp{\left[\frac{-(\nu-\nu_{\rm line-rest})^2}{2\sigma_{\rm line}^2}\right]}
\end{equation}
where $I_{\rm cont}$ is the fitted continuum level over the whole spectrum, $I_{\rm line}$ is the fitted intensity of each line, $\nu_{\rm line-rest}$ is the fixed rest frequency of each line, $\sigma_{\rm line}$ is the fitted width of each line, and $\nu$ is the rest frequency of the observed spectrum which depends on the assumed systemic velocity. The primary lines used to determine the systemic velocity are marked in orange in Figure \ref{fig:fullbandspec}; these lines tend to have strong emission and simple line shapes. Other strong lines in the band (e.g., CO $3-2$, \hcn, \hcop) tend to have complicated line shapes that make accurately determining the systemic velocity difficult. Due to the presence of lines with complicated shapes and line-blending, the cross-correlation of each full-band spectrum and the multi-Gaussian fit is done by-eye. The systemic velocities of for all the clusters are listed in Table \ref{tab:contparams}. The uncertainty is estimated to be $\pm5$ \kms. For SSCs 4a, 5a, and 14, the systemic velocities are further refined and confirmed using other lines in the cleaned \cs\ and \httcn\ cubes (Figure \ref{fig:modelcomp}). Several of these lines --- the vibrational transitions, in particular --- have sharp peaks and are spatially localized to the clusters themselves. They, therefore, provide tighter constraints on the cluster systemic velocities. For SSCs 4a, 5a, and 14, we estimate that our cluster systemic velocities are accurate to $\pm1$ \kms.

\subsection{\change Other Cluster Properties}
\label{ssec:otherprop}
{\change In addition to cluster properties derived from these new 0.5 pc resolution data described in the previous subsections, we use stellar masses, total \htwo\ gas masses, and escape velocities of the clusters calculated by \citet{leroy18}. Here we briefly describe how these quantities were derived. These quantities are reproduced in Table \ref{tab:outflowparams} for the three clusters which are the focus of this work.}

\paragraph{\change Stellar Masses ($M_*$)} {\change The stellar masses are calculated based on the 36 GHz image from the VLA convolved to 0.11\arcsec\ (1.9 pc) resolution \citep{leroy18,gorski17,gorski19}. At 36 GHz, the emission is assumed to be entirely due to free-free (Bremsstrahlung) emission. From the 36 GHz luminosity, \citet{leroy18} derived $M_*$ for a zero-age main sequence (ZAMS) population with a standard initial mass function (IMF) (see their Section 4.3.1 and Table 2). As described below, we consider additional sources of uncertainty beyond those described by \citet{leroy18}. First, we assume that the primary clusters retain all of the stellar mass measured by \citet{leroy18}, which may lead to an overestimate of $M_*$ for those clusters that break apart. We estimate that this is at worst 50\% in those cases (although the higher resolution data presented here shows a number of satellite structures, the original one remains dominant in most). Secondly, \citet{leroy18} estimate that there is a $\pm$20\% systematic uncertainty due to assumptions about the Gaunt factor. Finally, E. A. C. Mills et al. (in prep.) measure $M_*$ of these clusters at 5 pc resolution using hydrogen radio recombination lines (RRLs). These agree with the $M_*$ estimated by \citet{leroy18} to within $\pm$0.3 dex on average. For SSCs 4a, 5a, and 14, the agreement is even better, and the RRL analysis finds $M_*$ which differ by $+0.1$, $-0.1$, and $-0.2$ dex from those derived from the 36 GHz emission{\changes, where $+$ ($-$) means the RRL measurements produce a larger (smaller) $M_*$.} Included in the calculations by E. A. C. Mills et al. (in prep.) is the effect of a synchrotron component. If synchrotron contaminates the 36 GHz flux (which is assumed to be entirely free-free in the $M_*$ calculation of \citealt{leroy18}), this would lead to an overestimate of $M_*$. There is negligible synchrotron emission in SSCs 4a, 5a, and 14, but this could affect the $M_*$ of other clusters (especially SSCs 1a, 10a, 11a, and 12a; E. A. C. Mills et al. in prep.). To determine the lower error bar on $M_*$, we take the largest of the above three sources of uncertainty for each cluster, where the uncertainty from the RRL measurements are included only if they yield a smaller $M_*$. To determine the upper error bars, we take the larger of the 20\% systematic uncertainty and RRL-measured $M_*$ where they yield a larger $M_*$. In addition to these quantifiable uncertainties on $M_*$, there are unquantified uncertainties relating to ionizing photons absorbed by dust and evolution beyond the ZAMS, both of which would result in the reported stellar masses being underestimates \citep{leroy18}.  While we do not attempt to calculate these unquantifiable uncertainties, we caution that they could be important. In this work, we use these stellar masses primarily to place the clusters in a mass-radius diagram and to evaluate the possible mechanisms powering the observed outflows.}

\paragraph{\change Total \htwo\ Gas Mass (${\rm M_{H_{2},tot}}$)} {\change 
The cluster gas masses are based on the 350 GHz image from ALMA at 0.11\arcsec\ (1.9 pc) resolution \citep{leroy18}. At 350 GHz, the majority of the emission is due to thermal dust emission, though there may be a small level of free-free contamination. \citet{leroy18} quantify this free-free contamination and correct the measured 350 GHz fluxes to determine a more accurate flux due to thermal dust emission (see their Section 4.1 and Table 1). Assuming a dust temperature of 130~K, dust-to-gas ratio of 1:100, and a dust mass absorption coefficient of 1.9~cm$^2$~g\per, \citet{leroy18} calculate the total \htwo\ mass of the clusters. Uncertainties are $\sim$0.4$-$0.5 dex, with these measurements likely biased low due to assumptions about the dust temperature, dust-to-gas ratio, and the dust opacity. See their Section 4.3.3 and Table 2 for more details. In this work, we consider these to be the total gas masses of the clusters, including gas still bound within the cluster and outflowing from it.}

\paragraph{\change Escape Velocity (${\rm V_{escape}}$)} {\change \citet{leroy18} calculate the escape velocity from the clusters using their measured stellar and gas masses and their measured sizes from Gaussian fits to the 0.11\arcsec\ (1.9 pc) resolution 350 GHz image. Uncertainties are dominated by those from $M_*$ and ${\rm M_{H_{2},tot}}$. See Section 4.3.6 and Table 2 of \citet{leroy18} for more details. In this work, we use these escape velocities to compare against the outflow velocities we measure to place constraints on whether the outflowing material will escape the cluster.}

\subsection{Outflow Candidates}
\label{ssec:outflowcandidates}

{\change We analyze the spectra towards all of the clusters (primary or otherwise) identified in the high resolution continuum data (Figure \ref{fig:cont}) in search of spectral outflow signatures.} We determine our final list of outflow candidates based on robust detections of blueshifted absorption and redshifted emission in the \cs\ and \httcn\ lines. We present SSCs 4a, 5a, and 14 as our conservative list of detected P-Cygni profiles indicative of outflows. The \cs\ and \httcn\ spectra towards SSCs 4a, 5a, and 14 are shown in Figure \ref{fig:modelcomp}, which are from the cleaned line cubes (Section \ref{ssec:lineimaging}) and have been corrected for foreground gas (Section \ref{ssec:foregroundgas}). 

While other SSCs show absorption features, these are not due to outflowing material as the absorption occurs at the line center. Absorption at the line center is likely due to self-absorption of the cold material surrounding the cluster against the embedded continuum source{\change(s)}.  Redshifted absorption would be indicative of material inflowing towards the cluster; a hint of inflow is perhaps seen towards SSC 8a (Figure \ref{fig:fullbandspec}, especially the \cs, \hcn\ and \hcop\ lines). Blueshifted absorption, on the other hand, means that the cold foreground material is outflowing from the cluster. For this analysis, we are focused on the blueshifted absorption indicative of outflows. 

We also verify that these absorption features are not induced by the spatial filtering by the interferometer. Due to the incomplete sampling of the Fourier plane and the lack of short spacings, emission on larger scales is filtered out and cannot be properly deconvolved, causing an increase in the RMS of the data at the velocities where bright extended emission would be present. This problem manifests itself as negative "bowls" around bright emission, but also as negative features that have velocity structure. While this mainly affects the continuum (which is not removed from these data), it can also induce false absorption features in the affected velocity range. This can be seen, for example, in the CO $3-2$ lines in Figure \ref{fig:fullbandspec}, where the absorption profiles are very complex. Our choice to focus on \cs\ and \httcn\ mitigates this problem, as these transitions require high densities to be excited and are hence fairly localized to the clusters themselves, with not much of an extended component. As shown in Figure \ref{fig:foregroundgas} the environmental \cs\ spectra around the SSCs are either flat or show the line in emission; the \httcn\ spectra show similar profiles. Therefore, we conclude that the \cs\ and \httcn\ absorption features are not a result of the spatial filtering by the interferometer, although it may contribute to uncertainties at the 10\% level. 

A careful measurement of the cluster systemic velocities (Section \ref{ssec:vsys}) is critical to determine whether the absorption is blueshifted with respect to the cluster or not. Critically, for a few clusters (e.g., SSCs 3a and 13a) there are pathological combinations of absorption at the line center coupled with strong vibrationally excited lines in emission at the redshifted edge of the absorption feature, which together could masquerade as P-Cygni profiles. We identify other line candidates present in the spectra around \cs\ and \httcn\ using the line list presented by \citet{krieger20} as well as Splatalogue\footnote{\label{ref:splat} \url{https://www.cv.nrao.edu/php/splat/advanced.php}}, which are listed in the shaded yellow regions in Figure \ref{fig:modelcomp}. It is outside the scope of this paper to verify precisely which other lines are present in the spectra, so we present them simply as possible candidates (or combinations of candidates) which confuse the \cs\ and \httcn\ spectra. This line blending can lead to false detections of P-Cygni profiles. For example, \httcn\ ${\rm v_2=1}$ is located 100 MHz ($\approx 88$ \kms) red-ward of \httcn, as can be seen in the right column of Figure \ref{fig:modelcomp}. Similarly, \hcn\ ${\rm v_2=1}$ is located only 45 MHz ($\approx 38$ \kms) red-ward of \hcn. Given the broad line profiles (in emission and absorption), these lines can easily blend together to produce a facsimile of a P-Cygni profile. A careful measurement of the cluster systemic velocities (using other lines with simpler profiles), cross-referencing with the line list presented by \citet{krieger20}, and further verification using Splatalogue\footref{ref:splat} reveals that these are not true P-Cygni profiles. It is also worth noting that there is an emission feature (likely OS$^{18}$O, an SO$_2$ isotopologue) seen in the \cs\ absorption trough of SSC 4a (upper left panel of Figure \ref{fig:modelcomp}).

That we detect outflows from SSCs 4a, 5a, and 14 is in large part due to our ability to spatially resolve the individual clusters, since the spectral signatures of the outflows are localized roughly to the central half-flux radius of the clusters. In retrospect, there are indications of these outflows in the \cs\ and \httcn\ line profiles presented by \citet[][see their Figure 3]{leroy18} in SSCs 4 and 14. The much weaker outflow in SSC 5a is not apparent in these lower resolution data where the clusters are only marginally resolved. There are also kinematic offsets between \hcn\ and H40$\alpha$ measurements towards these clusters which are consistent with the effects of these outflows at lower resolution (E. A. C. Mills et al., in prep.).

\subsection{Internal Cluster Kinematics}
\label{ssec:sscrotation}

Briefly here, we investigate whether SSCs 4a, 5a, and 14 exhibit signs of internal rotation. Observational detections of rotation in star and globular clusters in the Milky Way are mixed \citep[e.g.][]{kamann19,cordoni20}, though simulations predict that massive star clusters ($>1000$ \msun) should have appreciable rotation \citep{mapelli17}.

We investigate the internal kinematics by constructing position-velocity (PV) diagrams over a range of angles as well as peak-velocity and intensity-weighted velocity (moment 1) maps around each of the clusters. While we see no clear signs of internal rotation, the absorption hinders this analysis significantly as it affects the (effective) intensity weighting of the peak- and intensity-weighted velocity (moment 1) maps. It similarly confuses the interpretation of the PV diagrams. Therefore, we cannot claim that the SSCs are not rotating, only that, given the presence of absorption, we see no clear evidence for rotation. We also check clusters without absorption or outflow signatures, and also find no evidence of rotation.

\section{Results}
\label{sec:results}

Using ALMA data at 350 GHz at 1.9 pc resolution, \citet{leroy18} identified 14 marginally-resolved clumps of dust emission whose properties are consistent with forming SSCs. In our new 0.5 pc resolution data, many of these SSCs fragment substantially such that there is a bright, massive primary cluster surrounded by smaller satellite clusters (Figure \ref{fig:cont}). From the 0.5 pc resolution 350 GHz dust continuum image (Figure \ref{fig:cont}), we identify roughly three dozen clumps of dust emission. We {\change spectrally} identify candidate outflows towards three SSCs: SSCs 4a, 5a, and 14 (where the appended "a" denotes the primary cluster of the fragmented group). The locations of these clusters within the nucleus of NGC\,253 are shown in Figure \ref{fig:cont}, where the insets show the \cs\ line profiles towards these three clusters. In each of these SSCs, we see evidence for blueshifted absorption and redshifted emission in many lines (Figure \ref{fig:fullbandspec}), but we focus on the \cs\ and \httcn\ lines (Figure \ref{fig:modelcomp}) which provide the best balance between bright lines with sufficient SNR, absorption features which do not suffer from saturation effects, and which probe gas localized to the clusters. {\change This line shape---blueshifted absorption and redshifted emission---is commonly referred to as a P-Cygni profile and is indicative of outflows.}

We take two approaches to derive the physical properties of the outflows in each cluster. First, we fit the absorption component of the line profiles to measure the outflow velocity, column density, mass, mass outflow rate, momentum, etc (Section \ref{ssec:outflow}). Second, we model line profiles of the \cs\ and \httcn\ spectra towards each cluster with the goal of constraining the outflow opening angles and orientations to the line of sight (Section \ref{ssec:outflowmodeling}). While the primary goal of the line profile modeling is to constrain the outflow geometry, it also provides a measurement of the outflow velocity, column density, mass, mass outflow rate, etc. We compare common parameters of these methods in Section \ref{ssec:MethodComp} {\change and discuss our recommended values in Section \ref{ssec:recommendedvalues}.}

\subsection{Outflow Properties from Absorption Line Fits}
\label{ssec:outflow}

\begin{figure*}[p]
\label{fig:modelcomp}
\centering
\gridline{\fig{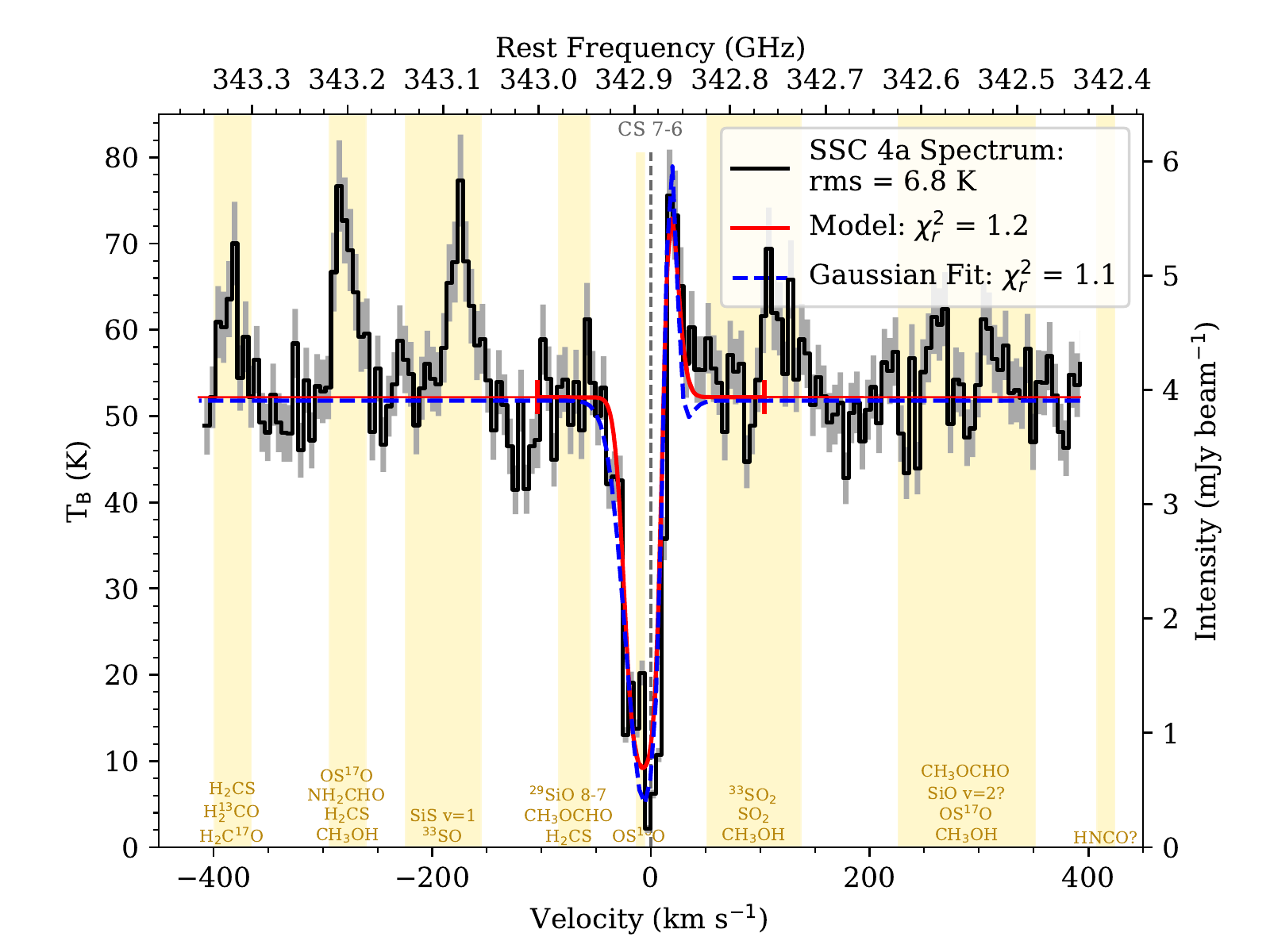}{0.5\textwidth}{}
\fig{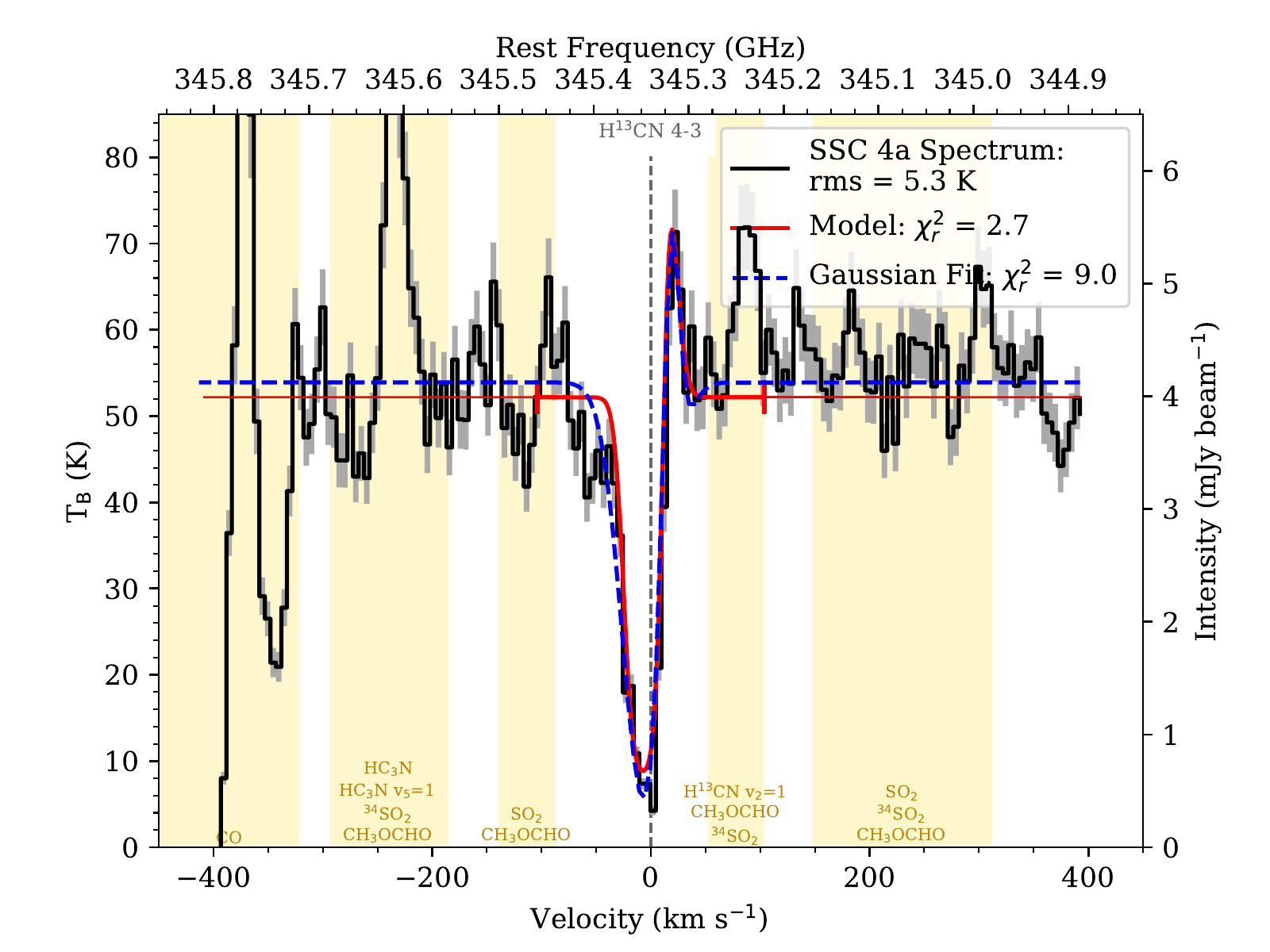}{0.5\textwidth}{}}
\vspace{-10mm}
\gridline{\fig{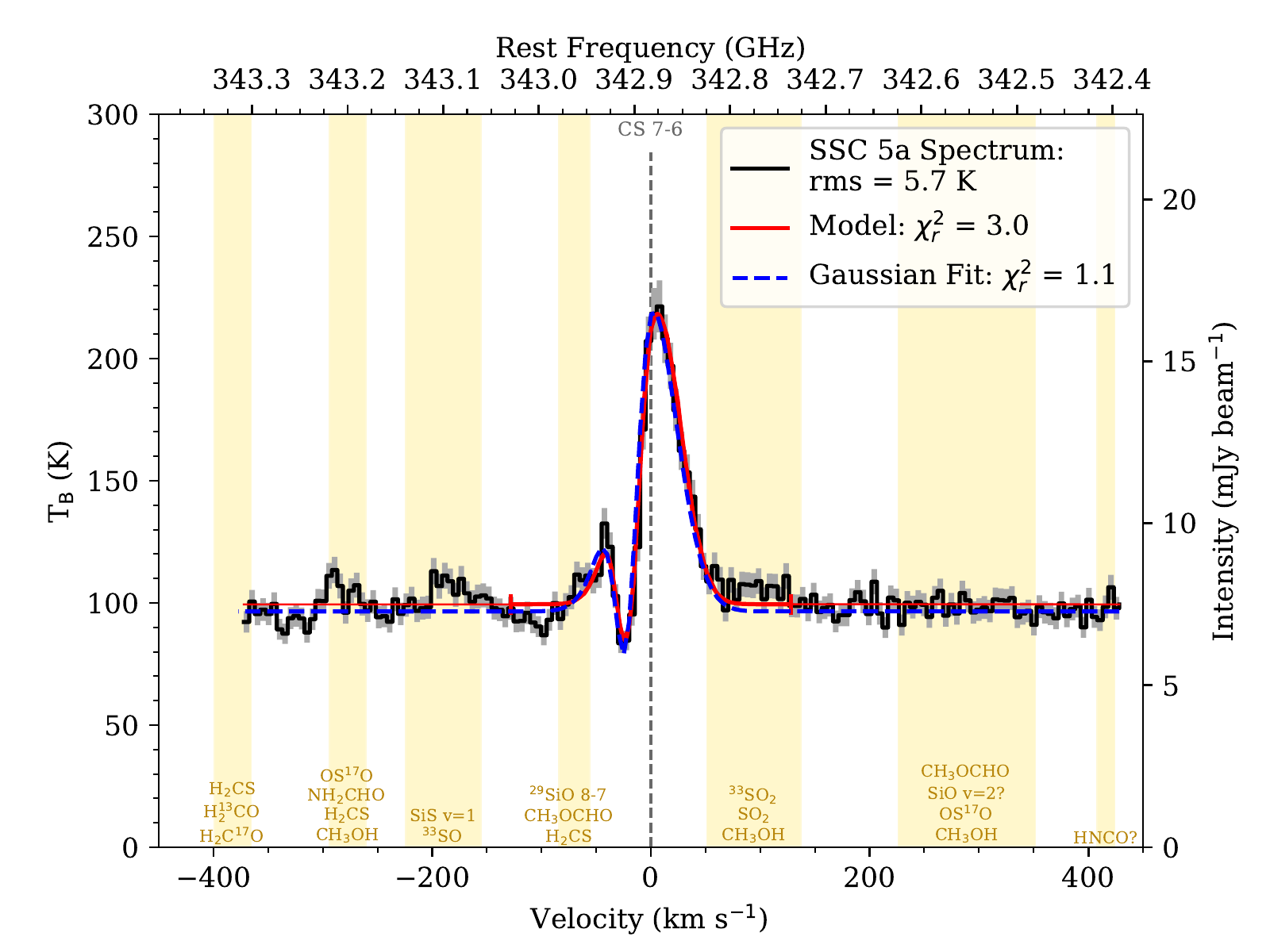}{0.5\textwidth}{}
\fig{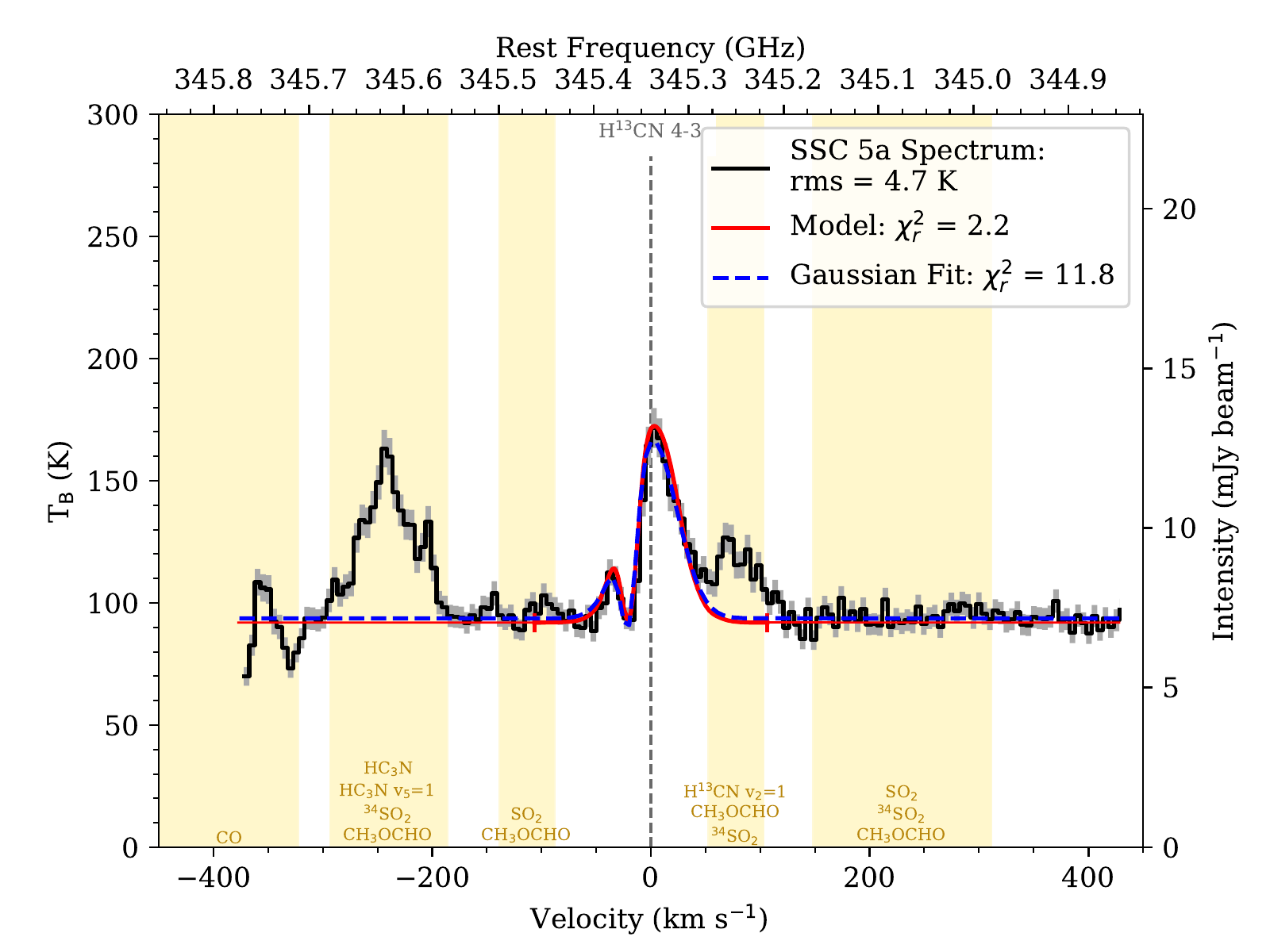}{0.5\textwidth}{}}\vspace{-10mm}
\gridline{\fig{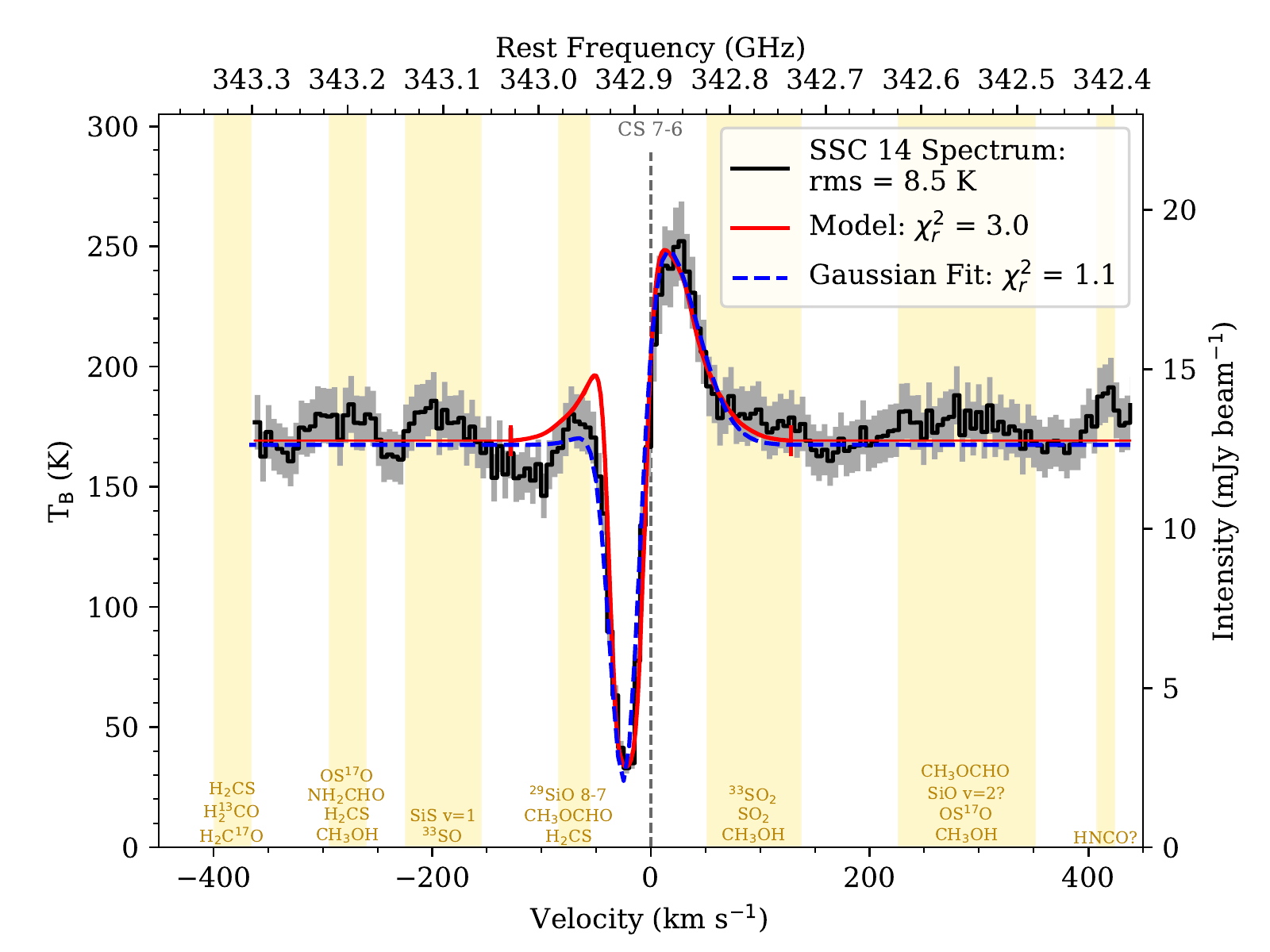}{0.5\textwidth}{}
\fig{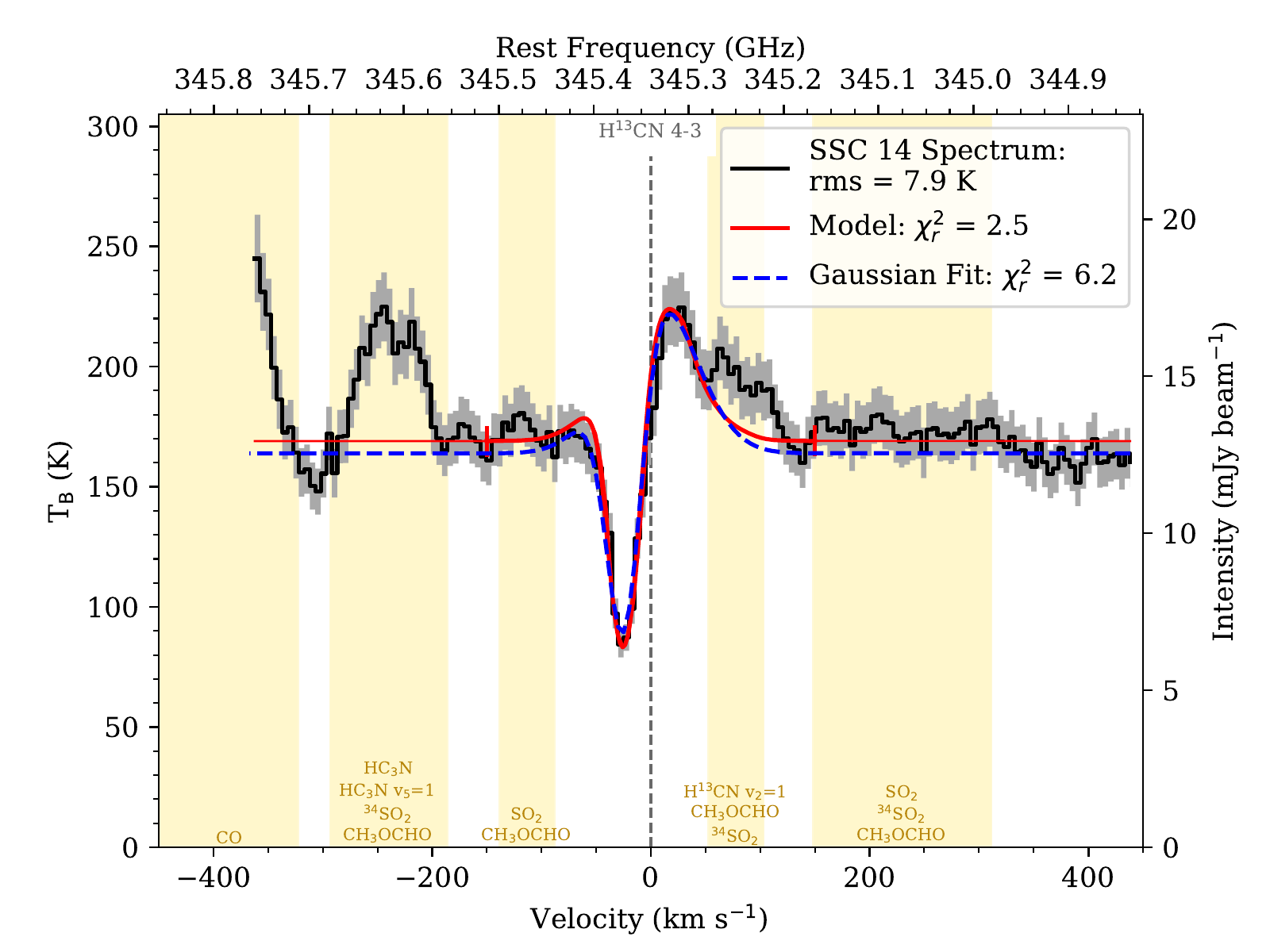}{0.5\textwidth}{}}
\vspace{-8mm}
\caption{The \cs\ (left) and \httcn\ (right) spectra for SSCs 4a (top), 5a (middle), and 14 (bottom). The yellow shaded regions show where other lines may contaminate the spectrum. Possible candidate lines are listed at the bottom of the yellow shaded regions. The dashed blue curves show the two-component Gaussian fit to the profile (Section \ref{ssec:outflow}). The solid red curves show the best-fitting spherical outflow models (Section \ref{ssec:outflowmodeling}). The vertical red line segments shows the range of velocities actually modeled, the red curve outside of these vertical line segments is an extrapolation.}
\end{figure*}

\begin{deluxetable*}{llcccccc}
\tablecaption{Cluster and Outflow Properties from Absorption Line Fits\label{tab:outflowparams}}
\tablehead{ & & \multicolumn{2}{c}{SSC 4a} & \multicolumn{2}{c}{SSC 5a} & \multicolumn{2}{c}{SSC 14} \\
\colhead{Quantity} & \colhead{Unit} & \colhead{CS $7-6$} & \colhead{H$^{13}$CN $4-3$} & \colhead{CS $7-6$} & \colhead{H$^{13}$CN $4-3$} & \colhead{CS $7-6$} & \colhead{H$^{13}$CN $4-3$}}
\startdata
$\log_{10}($M$_\mathrm{H_{2},tot})$ $^{a}$ & $\log_{10}$(M$_\odot$) & \multicolumn{2}{c}{5.1} & \multicolumn{2}{c}{5.3} & \multicolumn{2}{c}{5.7} \\
$\log_{10}($M$_\mathrm{*})$ $^{b}$ & $\log_{10}$(M$_\odot$) & \multicolumn{2}{c}{5.0} & \multicolumn{2}{c}{5.4} & \multicolumn{2}{c}{5.5} \\
V$_\mathrm{escape}$ $^{c}$ & km s$^{-1}$ & \multicolumn{2}{c}{22.0} & \multicolumn{2}{c}{33.2} & \multicolumn{2}{c}{45.5} \\
$\tau$ $^{d}$ &  & 2.3 & 2.3 & 0.2 & 0.0 & 1.8 & 0.6\\
V$_\mathrm{max-abs}$ $^{e}$ & km s$^{-1}$ & 6 & 6 & 22 & 20 & 24 & 24\\
V$_\mathrm{out,max}$ $^{f}$ & km s$^{-1}$ & 40 & 46 & 42 & 34 & 51 & 56\\
$\Delta\mathrm{V_{out,FWHM}}$ $^{g}$ & km s$^{-1}$ & 40 & 46 & 24 & 17 & 32 & 38\\
$\log_{10}$(t$_\mathrm{cross}$) $^{h}$ & $\log_{10}$(yr) & 5.0 & 5.0 & 4.5 & 4.6 & 4.4 & 4.4\\
$\log_{10}($N$_\mathrm{H_{2},out})$ $^{i}$ & $\log_{10}$(cm$^{-2}$) & 23.9 & 25.3 & 22.7 & 23.2 & 23.7 & 24.6\\
$\log_{10}($M$_\mathrm{H_{2},out})$ $^{j}$ & $\log_{10}$(M$_\odot$) & 4.7 & 6.1 & 3.8 & 4.3 & 4.6 & 5.5\\
$\log_{10}($M$_\mathrm{H_{2},out}$/M$_\mathrm{H_{2},tot}$) $^{k}$ &  & -0.4 & 1.0 & -1.5 & -1.0 & -1.1 & -0.2\\
$\log_{10}($M$_\mathrm{H_{2},out}$/M$_*$) $^{k}$ &  & -0.3 & 1.1 & -1.6 & -1.1 & -0.9 & -0.0\\
$\log_{10}(\dot{\rm M}_{\rm H_2,out})$ $^{l}$ & $\log_{10}$(M$_\odot$ yr$^{-1}$) & -0.3 & 1.1 & -0.7 & -0.3 & 0.2 & 1.1\\
$\log_{10}$(t$_\mathrm{remove-gas}$) $^{m}$ & $\log_{10}$(yr) & 5.4 & 4.0 & 6.0 & 5.6 & 5.5 & 4.6\\
$\log_{10}($p$_\mathrm{r}/$M$_*)$ $^{n}$ & $\log_{10}$(km s$^{-1}$) & 0.7 & 2.1 & 0.0 & 0.5 & 0.7 & 1.6\\
$\log_{10}$(E$_\mathrm{kin}$) $^{o}$ & $\log_{10}$(erg) & 49.2 & 50.6 & 49.6 & 49.9 & 50.4 & 51.3\\
\enddata
\tablecomments{Values (aside from the top three) are derived from fits to the absorption component of the line profiles. See Section \ref{ssec:outflow} and Appendix \ref{app:outflowproperties} for details on how these values were calculated. }
\tablenotemark{\footnotesize a}{The cluster gas mass measured from the dust continuum and assuming a gas-to-dust ratio of 100 \citep{leroy18}. Uncertainties are $\sim$0.4$-$0.5 dex, with these measurements likely biased low due to assumptions about the dust temperature, dust-to-gas ratio, and the dust opacity. {\change See Section \ref{ssec:otherprop} for a detailed discussion of how these values were calculated.}}

\tablenotemark{\footnotesize b}{The cluster zero age main sequence (ZAMS) stellar mass measured from the 36 GHz free-free emission \citep{leroy18}. {\change See Section \ref{ssec:otherprop} for a detailed discussion of the uncertainties on these stellar masses.}}

\tablenotemark{\footnotesize c}{The escape velocity from the cluster \citep{leroy18}. Uncertainties are dominated by those from the gas and stellar masses. {\change See Section \ref{ssec:otherprop} for a detailed discussion of how these values were calculated.}}

\tablenotemark{\footnotesize d}{The optical depth in the outflow calculated from the absorption to continuum ratio (Eq. \ref{eq:tau}).}

\tablenotemark{\footnotesize e}{The velocity where the absorption is maximized, corresponding to the velocity at which the bulk of the material traced by \cs\ and \httcn\ is outflowing.}

\tablenotemark{\footnotesize f}{The maximum outflow velocity, defined as 2$\sigma$ from the mean outflow velocity (Eq. \ref{eq:voutflow}). For a Gaussian line, this means that 95\% of the dense material has an outflow velocity slower than ${\rm V_{max,out}.}$}

\tablenotemark{\footnotesize g}{The FWHM line width from the Gaussian fit to the absorption feature.}

\tablenotemark{\footnotesize h}{The gas crossing time, or the time it would take a parcel of gas to travel from the center of the cluster to $r_{\rm SSC}$ at ${\rm V_{max-abs}}$ (Eq. \ref{eq:tcross}).}

\tablenotemark{\footnotesize i}{The \htwo\ column density in the outflow derived from $\tau$ (Eqs. \ref{eq:Nl}$-$\ref{eq:NH2}). This calculation assumes LTE with an excitation temperature of $130\pm56$ K and abundances of \cs\ and \httcn\ relative to \htwo\ from \citet{martin06}. The abundances may vary by an order-of-magnitude on these small scales, so all quantities which depend on the column density are limited to order-of-magnitude precision. The blueshifted outflow component only probes the portion of the outflow on the approching side, so the values are multiplied by 2 to account for the receding side of the outflow, assuming it is identical to the approaching side. Uncertainties are $\pm$1 dex.}

\tablenotemark{\footnotesize j}{The \htwo\ mass in the outflow derived from ${\rm N_{H_{2},out}}$ and the projected cluster size (Eq. \ref{eq:MH2}). This calculation and others which depend on it assume the outflow is spherical, which is supported by the results of our modeling (Section \ref{ssec:outflowmodeling}). Uncertainties are $\pm$1 dex.}

\tablenotemark{\footnotesize k}{The \htwo\ mass in the outflow compared to the total gas or stellar mass in the cluster. Uncertainties are $\pm$1 dex.}

\tablenotemark{\footnotesize l}{The mass outflow rate derived over one crossing time (Eq. \ref{eq:Mdot}). Uncertainties are $\pm$1 dex.}

\tablenotemark{\footnotesize m}{The gas removal time, or the time it would take for the current mass outflow rate to expel all of the cluster gas mass (${\rm M_{H_{2},tot}}$) from the clusters (Eq. \ref{eq:tremovegas}). Uncertainties are $\pm$1 dex.}

\tablenotemark{\footnotesize n}{The radial momentum carried by the outflow per unit stellar mass, which also assumes the outflow is spherical (Eq. \ref{eq:pr}). Uncertainties are $\pm$1 dex.}

\tablenotemark{\footnotesize o}{The kinetic energy of the outflow calculated from ${\rm M_{H_{2},out}}$ and ${\rm V_{max-abs}}$ (Eq. \ref{eq:Ekin}). Uncertainties are $\pm$1 dex.}

\end{deluxetable*}

 For a first estimate, we measure the outflow velocities, column densities, gas masses, and momentum of each outflow by fitting the profiles of \httcn\ and \cs\ with a two-component Gaussian (blue dashed curves in Figure \ref{fig:modelcomp}).  We exclude the portions of the spectra that may be contaminated by other lines, as marked by the yellow shaded regions in Figure \ref{fig:modelcomp}. The fit to the absorption feature yields the velocity at the maximum absorption (${\rm V_{max-abs}}$), or the velocity at which the bulk of the material is outflowing. We define the maximum outflow velocity as $2\sigma$ from the average velocity (${\rm{V_{max,out}\equiv V_{max-abs}+2\frac{\Delta V_{out,FWHM}}{2.355}}}$), which means that 95\% of the material traced by \cs\ and \httcn\ has an outflow velocity slower than ${\rm{V_{max,out}}}$ (for a truly Gaussian line). Given the mean outflow velocity and the cluster size, we calculate the crossing time, or the time for a parcel of gas to travel from the center of the cluster to the edge (${\rm t_{cross}\equiv r_{SSC}/V_{max-abs}}$), where ${\rm  r_{SSC}}$ is equivalent to the half-flux radius (Table \ref{tab:contparams}). From the fitted absorption to continuum intensity ratio, we derive the optical depth and hence the column density in absorption. We convert from the column density of the molecule to \htwo\ (${\rm N_{H_{2},out}}$) using abundances from \citet{martin06}. From the column density and projected size (Table \ref{tab:contparams}), we estimate the \htwo\ mass in the outflow (${\rm M_{H_{2},out}}$) assuming the outflow is spherical. Constraints on the opening angle of the outflows derived from modeling of the line profiles are discussed in Section \ref{ssec:outflowmodeling} and show that the outflows must be nearly spherical. Given the mass, crossing time, and mean outflow velocity, we calculate the mass outflow rate (${\rm \dot{M}_{H_{2},out}}$), the gas removal timescale (${\rm t_{remove-gas}}$), the radial momentum injected per unit stellar mass in the cluster (${\rm p_r/M_*}$), and the kinetic energy of the outflow (${\rm E_{kin}}$). Further details and equations used to derive these quantities are presented in Appendix \ref{app:outflowproperties}. The outflow parameters derived from the absorption profile fits are reported in Table \ref{tab:outflowparams}.  

The clusters have outflows with maximum velocities (${\rm{V_{max,out}}}$) of $\sim40-50$ \kms. For all three SSCs with outflows, ${\rm{V_{max,out}}}$ is larger than the escape velocity (${\rm{V_{escape}}}$; reproduced in Table \ref{tab:outflowparams} from \citealt{leroy18} {\change and see also Section \ref{ssec:otherprop}}). The bulk of the material traced by \cs\ and \httcn, however, has velocities less than ${\rm V_{escape}}$ (as traced by ${\rm{V_{max-abs}}}$). For SSCs 4a, 5a, and 14, 20\%, 7\%, and 7\% of the outflowing material has velocities larger than the escape velocity and will be able to escape from the cluster. {\change Gas which is outflowing with velocities below ${\rm{V_{escape}}}$ may be reaccreted by the cluster.} The gas crossing time (t$_{\rm cross}$) is $\sim$few$\times10^4$ years, which is a lower limit on the age of the outflow. {\change These timescales and the possibility of reaccretion will be discussed further in Section \ref{ssec:timescales}.}

The masses in the outflows are large. The molecular hydrogen column densities and masses derived from \httcn\ are larger than those derived from \cs\ in all cases. One possibility is that \cs\ is more optically thick than \httcn, which may be supported by the depth of the absorption in SSC 4a, but this cannot explain the discrepancy in SSC 5a. A more likely possibility is that the relative molecular abundances we assume are not correct for these small, extreme regions. We adopt molecular abundances for H$^{13}$CN and CS with respect to \htwo\ of [H$^{13}$CN]/[\htwo] $=(1.2\pm0.2)\times10^{-10}$ and [CS]/[\htwo] $=(5.0\pm0.6)\times10^{-9}$ from a study of the the center of NGC\,253 at 14--19\arcsec\ ($\sim240-320$ pc) resolution by \citet{martin06}, {\change where the brackets refer to the abundance of that species}. Adopting these abundances on the parsec scales of our clusters, however, is highly uncertain. Observations of envelopes around high-mass young stellar objects in the Milky Way, for example, reveal order-of-magnitude variations in the abundances of molecules, especially nitrogen and sulfur bearing species \citep[][and references therein]{vandishoeck98}. The impact of the uncertainty on the fractional molecular abundances is that they translate into order-of-magnitude accuracy for the H$_2$ column density measurements, as well as the subsequent calculations which depend on the column density, as reported in Table \ref{tab:outflowparams}. In the following subsection we discuss in more detail what we know about chemical abundances and their variation: our conclusion is that the masses derived from \httcn\ are likely more accurate.

\subsubsection{Fractional Abundance Variations}
\label{sssec:abun}
To {\change resolve} the discrepancies between our abundance-dependent measurements (${\rm N_{H_{2},out}}$, ${\rm M_{H_{2},out}}$, ${\rm \dot{M}_{H_{2},out}}$, and ${\rm t_{remove-gas}}$) from \cs\ and \httcn\ would require that either [CS]/[\htwo] is enhanced and/or [H$^{13}$CN]/[\htwo] is reduced in these clusters compared to the values measured by \citet{martin06}.  In the case of CS, modeling by \citet{charnley97} shows that [CS]/[\htwo] varies from $\sim10^{-10}-10^{-8}$ depending on age, O$_2$ abundance, and temperature, with CS being most abundant after $\sim10^4$ years, at low O$_2$ abundance, and at low temperatures (T$\sim$100 K) though the trend with temperature is not monotonic. Estimates of the kinetic temperatures of the (marginally resolved) clusters vary from $\sim200-300$ K \citep{rico-villas20}. At $\sim10^5$ years (the ZAMS ages of these clusters; \citealt{rico-villas20}) and a temperature of 300 K, \citet{charnley97} find [CS]/[\htwo] $\sim4\times10^{-9}$ depending on the assumed temperature and O$_2$ abundance, very close to our assumed ratio of $(5.0\pm0.6)\times10^{-9}$ measured by \citet{martin06}. From CS, SO, and SO$_2$ line ratios towards these clusters, \citet{krieger20} suggest that [CS]/[\htwo] may be enhanced by a factor of 2$-$3 from the values measured by \citet{martin06}. For SSCs 5a and 14, an enhancement of [CS]/[\htwo] by a factor of 2$-$3 brings our abundance-dependent measurements into much better agreement with those derived from \httcn, but this is not sufficient for SSC 4a. The gas in SSC 4a is likely more optically thick than in SSCs 5a and 14, as seen in the absorption to lower temperatures ($\sim 7$ K), which could help explain the lingering discrepancy in this cluster. 

In the case of [H$^{13}$CN]/[\htwo], \citet{colzi18} find that chemical evolution does not affect nitrogen fractionation, so the main driver of a different [H$^{13}$CN]/[\htwo] in these SSCs would be due to changes in the $^{12}$C/$^{13}$C ratio. If there is less $^{13}$C in these clusters compared to the environment, then [H$^{13}$CN]/[\htwo] may be lower. Towards these clusters, \citet{krieger20} find hints that $^{12}$C/$^{13}$C may be on the high side of the assumed ratio of $40\pm20$ \citep{martin10,martin19,henkel14,tang19}. If the the $^{12}$C abundance remains the same, this suggests that reductions in $^{13}$C and hence [H$^{13}$CN]/[\htwo] are possible. To bring our abundance-dependent measurements from \httcn\ into agreement with the lower values derived from \cs\ would imply $^{12}$C/$^{13}$C $\gtrsim$ 300, much larger than even the highest ratios measured in NGC\,253 \citep{martin10} while keeping $^{12}$C fixed. Changing [H$^{13}$CN]/[\htwo], therefore, likely plays a minor role in remedying the differences in our abundance-dependant quantities. 

If, therefore, the discrepancy between our abundance-dependent quantities measured from \cs\ and \httcn\ is due to a change in the abundances of those species, the effect is likely driven by CS which is enhanced relative to our assumed [CS]/[\htwo] with perhaps a  small contribution from reduced [H$^{13}$CN]/[\htwo]. As a result, the abundance-dependant quantities derived from \httcn\ are likely more accurate. In the values presented here and in the following section, we adopt the abundances measured by \citet{martin06} and maintain the conservative order-of-magnitude uncertainties on these quantities.

\subsection{Outflow Properties from Line Profile Modeling}
\label{ssec:outflowmodeling}

The short gas crossing times, large outflow masses, and outflow mass rates (Table \ref{tab:outflowparams}) suggest the outflow activity in these objects cannot be sustained for a long time. From this standpoint, it is reasonable that we detect outflows in $\lesssim$10\% of the SSCs, so it is possible that we are indeed catching a small fraction of the SSCs in this short-lived phase. The analysis in Section \ref{ssec:outflow}, however, implicitly assumes that the outflows are spherical. Another possibility, however, is that the outflows are biconical with a more-or-less narrow opening angle. In that case the outflows from the three clusters we identify are serendipitously pointed close enough to the line of sight to make them detectable.  If the observed outflows are not spherical, geometric correction factors will need to be applied to the measured quantities in Table \ref{tab:outflowparams}, and more outflows could exist that we do not detect because their geometry is unfavorable.

\begin{figure*}
\label{fig:modelcomponents}
\centering
\gridline{\fig{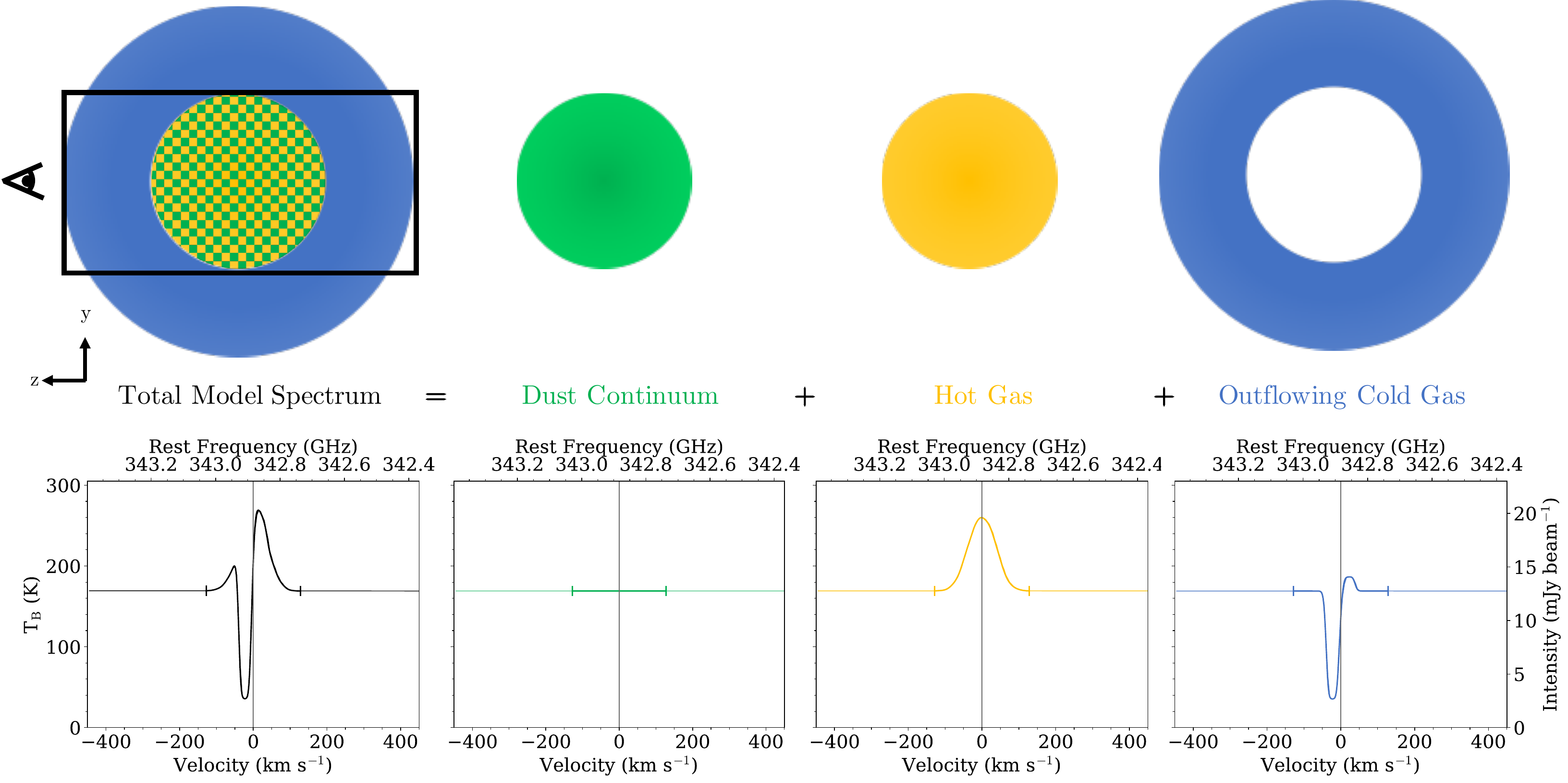}{\textwidth}{}}
\vspace{-5mm}
\caption{A 2D schematic showing the three physical components of the model at a slice through the box at $x=0$ and their spectral contributions. The black rectangle shows the slice of the field of view used to calculate the final spectrum. A representative spectrum for each component is shown below the corresponding 2D schematic. \cs\ towards SSC 14 is used here as an example. The thicker curves bounded by the vertical line segments show the velocity range modeled, whereas the thin curves outside are an extrapolation. The hot gas and cold outflowing gas spectra include the dust continuum component so that the absorption due to the outflowing components is visible.}
\end{figure*}

\begin{figure*}
\label{fig:modelgeom3d}
\centering
\includegraphics[width=\textwidth]{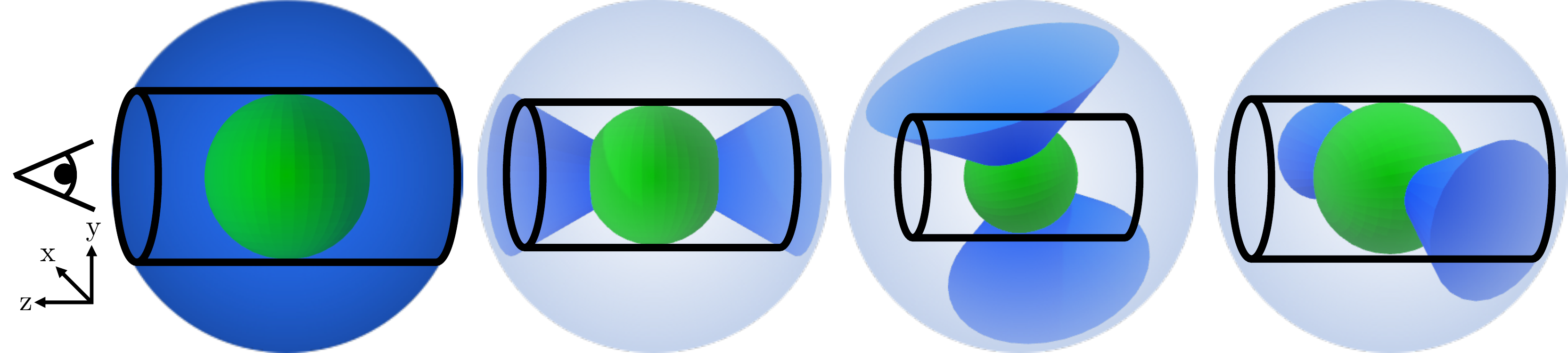}
\caption{3D schematics of (from left to right) a spherical ($\theta=180^\circ$) outflow, a biconical ($\theta=60^\circ$) outflow pointed along the line of sight, and biconical ($\theta=90^\circ$ and $\theta=45^\circ$) outflows that are inclined to the line of sight ($\Psi=95^\circ$ and $\Psi=60^\circ$). The green central sphere includes both the dust continuum and hot gas components and has $r=r_{\rm SSC}$. The dark blue outer shell or cones show the outflowing gas. The light blue component shows the ambient gas between the outflow cones. The black cylinder shows the line of sight and field of view over which the model spectrum is integrated and averaged, which has $r=r_{\rm SSC}$ (in the $xy$-plane) and spans $4\times r_{\rm SSC}$ along the $z$-axis.}
\end{figure*}

\begin{deluxetable*}{llcccccc}
\tablecaption{Line Profile Modeling Input Parameters and Calculated Values \label{tab:modelparams}}
\tablehead{ & & \multicolumn{2}{c}{SSC 4a} & \multicolumn{2}{c}{SSC 5a} & \multicolumn{2}{c}{SSC 14} \\
\colhead{Quantity} & \colhead{Unit} & \colhead{CS $7-6$} & \colhead{H$^{13}$CN $4-3$} & \colhead{CS $7-6$} & \colhead{H$^{13}$CN $4-3$} & \colhead{CS $7-6$} & \colhead{H$^{13}$CN $4-3$}}
\startdata
V${_\mathrm{max-abs}}$ $^{*,a}$ & km s$^{-1}$ & 7 & 7 & 25 & 22 & 25 & 28 \\
$\Delta\mathrm{V_{out,FWHM}}$ $^{*,b}$ & km s$^{-1}$ & 25 & 25 & 20 & 15 & 20 & 25 \\
V$_\mathrm{out,max}$ $^c$ & km s$^{-1}$ & 28 & 28 & 42 & 35 & 42 & 49 \\
$\log_{10}$(t$_\mathrm{cross}$) $^d$ & $\log_{10}$(yr) & 4.9 & 4.9 & 4.5 & 4.6 & 4.4 & 4.3 \\
$\Delta\mathrm{V_{hot,FWHM}}$ $^{*,b}$ & km s$^{-1}$ & 10 & 10 & 50 & 50 & 80 & 85 \\
T${_\mathrm{out}}$ $^{*,e}$ & K & 7 & 7 & 25 & 20 & 35 & 40 \\
T${_\mathrm{hot}}$ $^{*,f}$ & K & 900 & 900 & 800 & 800 & 800 & 800 \\
T${_\mathrm{cont}}$ $^{*,g}$ & K & 105 & 105 & 200 & 185 & 340 & 340 \\
$\log_{10}($n$_\mathrm{out})$ $^{*,h}$ & $\log_{10}$(cm$^{-3}$) & 7.8 & 7.5 & 5.5 & 5.9 & 6.2 & 6.8 \\
$\log_{10}($n$_\mathrm{hot})$ $^{*,h}$ & $\log_{10}$(cm$^{-3}$) & 8.6 & 9.9 & 8.8 & 9.8 & 8.8 & 10.0 \\
$\log_{10}(\tau{_\mathrm{cont,max}})$ $^{*,i}$ &  & -1.3 & -1.3 & -1.3 & -1.3 & -1.3 & -1.3 \\
$\log_{10}($N$_\mathrm{H_{2},out})$ $^{j}$ & $\log_{10}$(cm$^{-2}$) & 25.4 & 25.1 & 23.2 & 23.7 & 24.0 & 24.5 \\
$\log_{10}($M$_\mathrm{H_{2},out})$ $^{j}$ & $\log_{10}$(M$_\odot$) & 6.3 & 6.0 & 4.4 & 4.9 & 5.3 & 5.4 \\
$\log_{10}$($\dot{\mathrm{M}}_\mathrm{H_2,out}$) $^{j}$ & $\log_{10}$(M$_\odot$ yr$^{-1}$) & 1.4 & 1.1 & -0.1 & 0.3 & 0.9 & 1.1 \\
$\log_{10}$(t$_\mathrm{remove-gas}$) $^{j}$ & $\log_{10}$(yr) & 3.7 & 4.0 & 5.4 & 5.0 & 4.8 & 4.6 \\
\enddata
\tablecomments{The input and output or derived parameters for the best-fitting spherical outflow models. See Section \ref{ssec:outflowmodeling} and Appendix \ref{app:outflowmodeling} for details.}

\tablenotemark{\footnotesize *}{Best-fit input parameters to the model. The uncertainties listed below are reflective of the grid of tested parameters.}

\tablenotemark{\footnotesize a}{The outflow velocity, which is assumed constant. Uncertainties are $\pm$1 \kms.}

\tablenotemark{\footnotesize b}{The FWHM velocity dispersion of the given component. Uncertainties are $\pm$2 \kms.}

\tablenotemark{\footnotesize c}{The maximum outflow velocity defined as ${\rm V_{max-abs}+2\frac{\Delta V_{out,FWHM}}{2.355}}$. Propagated uncertainties are $\pm$2 \kms.}

\tablenotemark{\footnotesize d}{The gas crossing time defined as ${ \frac{r_{\rm SSC}}{\rm V_{max-abs}}}$. Propagated uncertainties are 4\% for SSCs 14 and 5 and  14\% for SSC 4a.}

\tablenotemark{\footnotesize e}{The gas temperature in the outflow at the line peak, which is constant spatially. Uncertainties are $\pm$2 K.}

\tablenotemark{\footnotesize f}{The gas temperature in the hot component at the line peak, which is constant spatially. Uncertainties are $\pm$25 K.}

\tablenotemark{\footnotesize g}{The continuum temperature, which is constant spatially and spectrally. Uncertainties are $\pm$5 K.}

\tablenotemark{\footnotesize h}{The log of the peak \htwo\ number density for the given component. Uncertainties are $\pm$0.25 dex.}

\tablenotemark{\footnotesize i}{The peak optical depth of the continuum component. Uncertainties are $\pm$0.025.}

\tablenotemark{\footnotesize j}{Uncertainties are $\pm$1 dex due to the molecular abundance ratios relative to \htwo.}

\end{deluxetable*}

In order to determine whether the observed outflows are spherical or biconical and how they are oriented with respect to the line of sight, we build a simple radiative transfer model to model the spectrum through the outflow\footnote{\change The code and best-fit input parameter files are available at \url{https://github.com/rclevy/ModelSSCOutflows}.}.
We consider spherical and biconical outflows, where the opening angle ($\theta$) and orientation to the line of sight ($\Psi$) of the biconical outflows can be varied. Opening angles are defined as the full-angle for one hemisphere; the maximum opening angle is $\theta=180^\circ$, corresponding to a sphere. Technical details and equations are presented in Appendix \ref{app:outflowmodeling}, but we summarize the basic scheme here. We construct a four-dimensional box ($x,y,z,\nu$), where radii are measured in spherical coordinates from the center of the box. We define the measured cluster radius such that $r_{\rm SSC} = r_{\rm half-flux}$ (Table \ref{tab:contparams}). The simulated box is scaled to $4\times r_{\rm SSC}$ in each spatial dimension ($x,y,z$) to fully encompass the line emission (e.g. Figure \ref{fig:chanmaps}) especially since the geometry along the line of sight ($z$-axis) is unknown. The velocity axis is scaled based on the input outflow velocity and velocity dispersion, so that the velocity resolution is optimized over the velocities relevant for the cluster and outflow; this is described more fully in Appendix \ref{app:outflowmodeling}. Three physical components are required to adequately model the \cs\ and \httcn\ spectra, as shown in Figures \ref{fig:modelcomponents} and \ref{fig:modelgeom3d}. These components and the input parameters for each are described below. The input parameters are denoted with $^*$ in Table \ref{tab:modelparams}, which lists input parameter values that yield the best model fits. Radial profiles of these components and the input parameters are summarized in Figure \ref{fig:paramprofiles} in Appendix \ref{app:outflowmodeling}.
\begin{enumerate}
\itemsep0em
\item Dust continuum component: Shown in green in Figures \ref{fig:modelcomponents} and \ref{fig:modelgeom3d}, this component is a sphere with $r=r_{\rm SSC}$ and a constant (in space and frequency) temperature (T$_{\rm cont}$). The optical depth in a single cell is a maximum ($\tau_{\rm cont,max}$) at the center, then decreases like a Gaussian with FWHM $=2\times r_{\rm SSC}$. In other words, the FWHM of the dust continuum optical depth profile matches the diameter of the 350 GHz continuum source. The temperature and optical depth are set to zero for $r>r_{\rm SSC}$. 
\item Hot gas: Shown in yellow in Figure \ref{fig:modelcomponents} (and encompassed within the green sphere in Figure \ref{fig:modelgeom3d}), this spherical component is required to reproduce the strong emission component of the P-Cygni profiles. This component is defined by an input hot gas temperature (T$_{\rm hot}$), a \htwo\ volume density (n$_{\rm hot}$) for the central ($r=0$) pixel, and a velocity dispersion ($\rm \Delta V_{hot,FWHM}$). T$_{\rm hot}$ is constant (spatially) for $r\leq r_{\rm SSC}$, and is set to zero outside. The density falls off $\propto r^{-2}$ from the center and is set to zero for $r>r_{\rm SSC}$. The line is centered on zero velocity along the frequency axis, and the Gaussian linewidth is given by ${\rm \Delta V_{hot,FWHM}}$. Using the equations in Appendix \ref{app:outflowmodeling}, this produces a spectrum at every pixel in the box.
\item Cold, outflowing gas: Shown in blue in Figures \ref{fig:modelcomponents} and \ref{fig:modelgeom3d}, this is the outflow component which produces the absorption features. This component is defined by an input gas temperature (T$_{\rm out}$), a \htwo\ volume density (n$_{\rm out}$) at the cluster boundary ($r=r_{\rm SSC}$), a constant outflow velocity (V$_{\rm out}$), a velocity dispersion ($\rm \Delta V_{out,FWHM}$), an opening angle ($\theta$), and an orientation to the line of sight ($\Psi$). The gas temperature is constant (spatially) within this component. The density is a maximum at $r=r_{\rm SSC}$ and deceases $\propto r^{-2}$ until the edges of the box; the density is set to zero inside the cluster ($r<r_{\rm SSC}$). In the spectral dimension, the line has a centroid velocity given by V$_{\rm out}$ and a FWHM linewidth of ${\rm \Delta V_{out,FWHM}}$. Using the equations in Appendix \ref{app:outflowmodeling}, this produces a spectrum at every pixel in the box. Since the outflow velocity is constant and the density $\propto r^{-2}$, the outflow conserves mass, energy, and momentum. To create a biconical outflow with the input opening angle ($\theta$), the velocity of the pixels outside the outflow cones is set to zero. This creates an ambient gas component (shown in light blue in Figure \ref{fig:modelgeom3d}), which has the same temperature, density, and velocity dispersion properties as the outflowing gas (but with V$_{\rm out}=0$). The box is then rotated to the input orientation from the line of sight ($\Psi$).
\end{enumerate}

Together, these three components are integrated from the back of the box forward (e.g. along the $-z$-axis in Figures \ref{fig:modelcomponents} and \ref{fig:modelgeom3d}). To obtain the final spectrum, only pixels within a cylinder along the line of sight with $r=r_{\rm SSC}$ are integrated (shown as the black cylinder in Figure \ref{fig:modelgeom3d}) to best compare with the measured spectra which are extracted only over an area corresponding to the continuum source half-flux radius. 
We adjust the input parameters component-by-component to find the model spectrum that best matches the observed \cs\ and \httcn\ spectra for SSCs 4a, 5a, and 14. There are degeneracies among input parameters, which are described below and in Appendix \ref{app:outflowmodeling}. These best-fit models are shown in red in Figure \ref{fig:modelcomp}, and the best-fit parameters for \cs\ and \httcn\ are listed in Table \ref{tab:modelparams}. For all spectra and sources, the spherical model provides the best fit, implying that the opening angles of the outflows need to be broad to explain the observed line profiles. A wide opening angle is in agreement with recent magneto-hydrodynamic (MHD) simulations, which show that cluster outflows are asymmetric and chaotic, but still wide-angle in general and regardless of the precise feedback mechanism {\change \citep[e.g.,][]{skinner15,kim18_jg,he19,geen21,lancaster20}}. From these best fit models, we also calculate the \htwo\ column density and mass in the outflows, which are also listed in Table \ref{tab:modelparams}.

We tested models with a fourth physical component representing a fast outflowing component. This was mainly motivated by SSC 14 and the mismatch between the spectrum and model at the blue-ward edge of the absorption feature for both \cs\ and \httcn\ (Figure \ref{fig:modelcomp}). In the model, this component is otherwise identical to the "slow" outflow component described above but with a larger outflow velocity and velocity dispersion and a different maximum \htwo\ volume. While including this component did marginally improve the fits --- especially for SSC 14 --- the improvement was not enough to justify the additional three parameters introduced into the model. 

Our model assumes a constant outflow velocity and $r^{-2}$ density profile to conserve mass, energy, and momentum in the outflow. In reality, a constant outflow velocity is unlikely to be precisely the case, due to turbulence within the outflow itself \citep[e.g.,][]{raskutti17} and because the outflow may decelerate as it encounters the surrounding medium. From the data, we investigate the location of the absorption trough around and across the sources. We see no strong evidence for systematic velocity shifts around or across SSCs 4a, 5a, or 14.

\begin{figure}
\label{fig:modeldegen}
\centering
\includegraphics[width=\columnwidth]{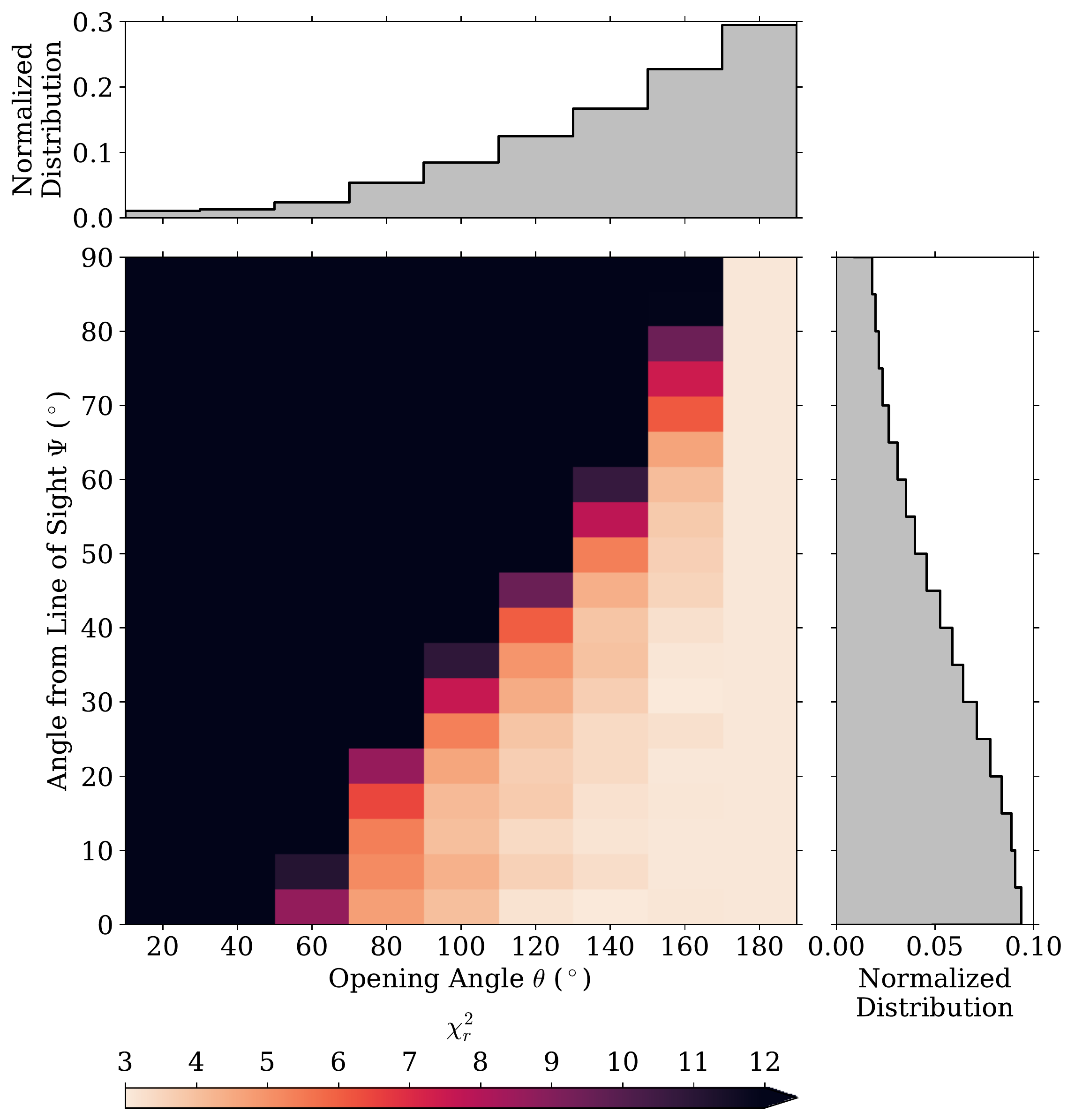}
\caption{A heatmap showing the reduced chi-squared ($\chi_r^2$) values for models with varying opening angles ($\theta$) and orientations from the line of sight ($\Psi$) for the same input parameters as listed in Table \ref{tab:modelparams} for \cs\ in SSC 14. The gray histograms show the marginalized distributions for $\theta$ (top) and $\Psi$ (right), normalized to unit area. This shows that the outflow opening angles must be wide and/or closely aligned with the line of sight.}
\end{figure}

There are ranges of opening angles ($\theta$) and orientations ($\Psi$) which are degenerate and will produce similar output spectra. To investigate how well we can constrain $\theta$ and $\Psi$, we run a grid of models with the same input parameters as in Table \ref{tab:outflowparams} with varying $\theta$ in steps of 20$^\circ$ and $\Psi$ in steps of 5$^\circ$. The results for \cs\ in SSC 14 are displayed in Figure \ref{fig:modeldegen}, which shows that wide outflows and/or those pointed close to the line of sight are strongly favored. Wide angle outflows from clusters are in agreement with results of numerical simulations as mentioned previously. Though the outflows in simulations are clumpy and highly non-uniform, they cover nearly 4$\pi$ steradians and hence approach the spherical limit of our simple modeling. Narrow and/or off-axis opening angles substantially increase the required input density in {\changes our} models and result in unphysical solutions for the outflowing mass. In any case, we cannot pin down the precise opening angle and/or line of sight orientation from this modeling, but we  place limits on them that suggest the dearth of clusters with observed outflows is not a selection bias due to geometrical effects. It is possible that we miss outflows if the dense gas is very optically thick and therefore obscures underlying outflow signatures \citep[e.g.,][]{aalto19}. We could also miss weak outflows below our detection limit of sensitivity and cluster mass.

\subsection{Comparing the Two Methods}
\label{ssec:MethodComp}
\begin{figure*}
\centering
\includegraphics[width=\textwidth]{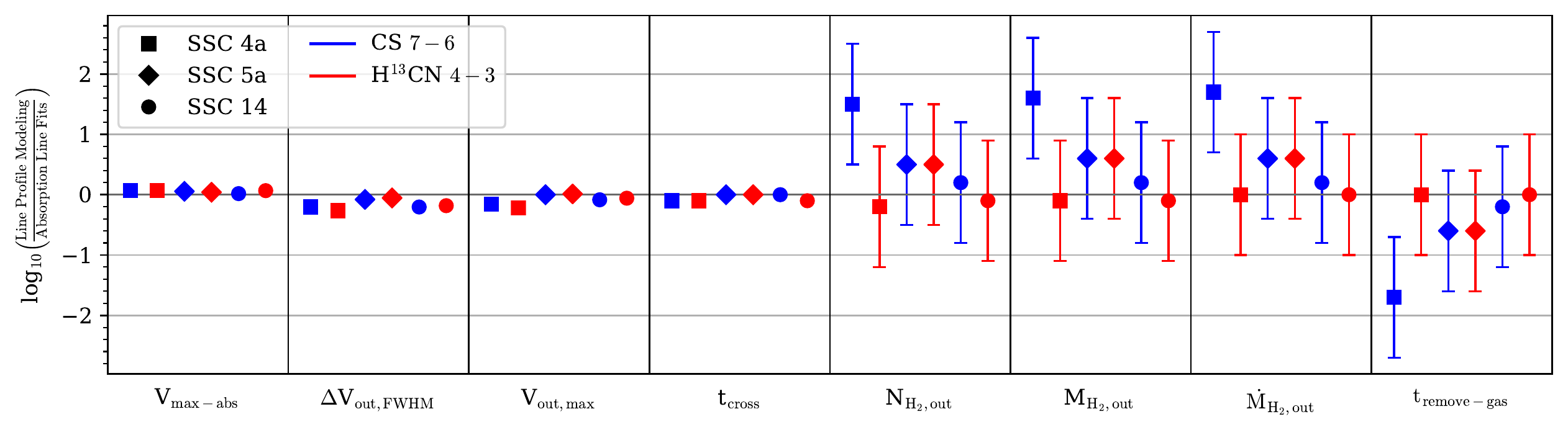}
\caption{A comparison of common quantities measured from the absorption line fits (Section \ref{ssec:outflow}) and the line profile modeling (Section \ref{ssec:outflowmodeling}). Each bin (separated by black vertical lines) shows one quantity listed in Tables \ref{tab:outflowparams} and \ref{tab:modelparams}. Within the bins, points are artificially offset along the horizontal axis for clarity. The different symbols show the three SSCs and the colors show quantities derived from the two spectral lines, as defined in the legend. The vertical axis shows the logarithm of the ratio of the quantities from each method, such that each major tick mark corresponds to an order of magnitude. The error bars for the first four quantities are negligible; the errorbars on the last four quantities reflect the order-of-magnitude uncertainty on the abundance ratio of the molecule with respect to \htwo.}
\label{fig:paramcomp}
\end{figure*}

The two methods of measuring the outflow properties described have calculated quantities in common. In Figure \ref{fig:paramcomp}, we compare the velocity where the absorption is maximum (V$_{\rm max-abs}$), the FWHM of the absorption feature ($\rm \Delta V_{out,FWHM}$), the maximum outflow velocity $({\rm V_{out,max}})$, the gas crossing time (${\rm t_{cross}}$), the \htwo\ column density in the outflow (${\rm N_{H_{2},out}}$), the \htwo\ mass in the outflow (${\rm M_{H_{2},out}}$), the mass outflow rate (${\rm \dot{M}_{H_{2},out}}$), and the time to remove all of the gas mass of the cluster at the current ${\rm \dot{M}_{H_{2},out}}$ (${\rm t_{remove-gas}}$). The equations used to calculate these parameters are given in Appendices \ref{app:outflowproperties} and \ref{app:outflowmodeling} for the absorption line fits and line profile modeling respectively. The vertical axis of Figure \ref{fig:paramcomp} shows the logarithm of the ratio of the quantities derived from each method; each major tick mark and horizontal gridline corresponds to an order of magnitude difference. In general, there is good agreement between the two methods for the quantities which depend on the velocity of the outflow {\changes (V$_{\rm max-abs}$, $\rm \Delta V_{out,FWHM}$, ${\rm V_{out,max}}$, and ${\rm t_{cross}}$)}. The uncertainties in Figure \ref{fig:paramcomp} reflect the order-of-magnitude adopted uncertainties around the abundance ratios, as discussed in Section \ref{ssec:outflow}.

Given the uncertainties the agreement between the absorption line fits and the line profile modeling is good, especially for \httcn. A notable exception is for \cs\ in SSC 4a, where the line profile modeling yields 1.5 dex larger \htwo\ columns and masses than derived from the absorption line fits. This is likely because the \cs\ absorption in SSC 4a is the most saturated (i.e., the closest to zero). For the absorption line fits, this renders a more uncertain optical depth (from the absorption to continuum ratio) as small changes in the absorption depth leads to large changes in the optical depth and hence the column density. For the line profile modeling, the depth of the absorption depends on the assumed gas temperature and \htwo\ volume density in the outflow. The bottom of the absorption trough sets an upper limit on the assumed gas temperature in the outflow (${\rm T_{out}}$); in the modeling, the temperature of the absorption {\change trough} cannot be lower than the assumed ${\rm T_{out}}$. In the case of SSC 4a, this temperature is low ($\sim7$ K), well below the temperature needed to excite the $J=7$ level of CS of 66 K \citep{lamda}\footnote{This information was retrieved from the Leiden Atomic and Molecular Database (LAMDA) on 2020-10-30.}. As a result, the \htwo\ density in the outflow must be increased, leading to a large outflowing mass. We assume a constant temperature in the outflowing gas, whereas a temperature gradient is likely needed to produce this absorption depth. Because the other sources and lines have shallower absorption depths, they do not suffer from this effect. Other strong lines---\hcn\ and \hcop---shown in Figure \ref{fig:fullbandspec} also suffer from this saturation, which is why they are not the focus of this analysis. We, therefore, suggest that the column density, mass, and mass outflow rate from the modeling are overestimated in the case of \cs\ in SSC 4a. This effect is not seen in the absorption line fitting because the absorption to continuum ratio is used to infer the optical depth, but an excitation temperature of 130~K is assumed to derive the level populations (Appendix \ref{app:outflowproperties}). In the line profile modeling, the input gas temperature (7~K in the case of \cs\ in SSC 4a) is used to derive the level populations instead (Appendix \ref{app:outflowmodeling}). Discrepancies between the two methods are also seen in both lines towards SSC 5a, though they are not as extreme as for SSC 4a. The discrepancies in SSC 5a cannot, however, be explained by saturation. The same abundance ratios are used for both methods and these enter into the calculations in the same way, so this discrepancy cannot be remedied by changing [CS]/[\htwo].

\subsection{Recommended Values}
\label{ssec:recommendedvalues}
Given the number of assumptions of the line profile modeling and the parameter covariances, we suggest that the outflow properties from the absorption line fits presented in Table \ref{tab:outflowparams} are more reliable and should be adopted. The assumption of spherical outflows in computing those numbers is substantiated by the modeling, which strongly favors wide outflows (Figure \ref{fig:modeldegen}). It is encouraging that the line profile modeling generally finds similar values for the outflowing mass, but there are cases where the model likely overestimates the outflowing mass (e.g., SSC 4a). Both sets of measurements are limited in the same way by the uncertainty on the molecular abundances with respect the \htwo. As discussed in Section \ref{sssec:abun}, discrepancies between quantities derived from \cs\ and \httcn\ may be driven primarily by changes in [CS]/[\htwo] relative to the assumed abundance, so that quantities derived from \httcn\ may be more reliable. Studies, such as the ALMA Comprehensive High-Resolution Extragalactic Molecular Inventory (ALCHEMI)\footnote{\url{https://alchemi.nrao.edu}}, that measure abundances at higher spatial resolution than currently available in these extreme environments may improve the accuracy of our column density, mass, mass outflow rate, and gas removal timescale estimates.

\section{Discussion}
\label{sec:discussion}

The existence of outflows in SSCs 4a, 5a, and 14 suggests these clusters may be in a different evolutionary stage compared to the other SSCs in the starburst. In the following section, we investigate the relationship between various timescales relevant to the clusters (\ref{ssec:timescales}) and possible outflow mechanisms (\ref{ssec:outflowpower}). We also investigate SSC 5a in more detail, as it is the only cluster visible in the NIR and shows evidence for a shell of dense gas surrounding it (\ref{ssec:nirclusters}).

\subsection{Timescales, Ages, and Evolutionary Stages} 
\label{ssec:timescales}

\begin{deluxetable*}{ccccccc}
\tablecaption{A comparison of relevant cluster timescales and ages\label{tab:timescales}}
\tablehead{\colhead{SSC Number} & \colhead{$\log_{10}({\rm t_{ZAMS-age}})$\tablenotemark{\tiny a}} & \colhead{$\log_{10}({\rm t_{ff}})$\tablenotemark{\tiny b}} & \colhead{${\rm t_{ZAMS-age}/t_{ff}}$} & \colhead{$\log_{10}({\rm t_{cross}})$\tablenotemark{\tiny c}} & \colhead{$\log_{10}({\rm t_{remove-gas}})$\tablenotemark{\tiny d}} & \colhead{$\log_{10}({\rm t_{dep}})$\tablenotemark{\tiny e}}} 
\startdata
4a & 4.95 & 4.9 & 1.1 & 5.0 & 4.3$\pm$0.7 & 6.6 \\
5a & $\gtrsim 5$ & 4.7 & $\gtrsim$2.0 & 4.6 & 5.5$\pm$0.4 & 6.4 \\
14 & 4.88 & 4.5 & 2.4 & 4.4 & 4.9$\pm$0.4 & 6.8 \\
\enddata
\tablecomments{For details, see the discussion in Section \ref{ssec:timescales}. }

\tablenotetext{\tiny a}{The age of the cluster since the zero-age main sequence (${\rm t_{ZAMS-age}}$) calculated by the ratio of the luminosity in proto-stars to that in ZAMS stars from \citet{rico-villas20}.}

\tablenotetext{\tiny b}{The free-fall time (${\rm t_{ff}}$) calculated by \citet{leroy18}. For SSC 4a, the value for SSC 4 is used because SSC 4a is the dominant component of the SSC 4 complex. The uncertainty introduced by this assumption is likely minor compared to other uncertainties which are $\sim0.4-0.5$ dex \citep[c.f. Table 2 of][]{leroy18}.}

\tablenotetext{\tiny c}{The time for gas to travel from the center of the cluster to the radius of the continuum source ($r_{\rm SSC}$) at the typical outflow velocity (${\rm V_{max-abs}}$). This timescale (${\rm t_{cross}}$) is a proxy for the age of the outflow. For each SSC, this is the average of the values in Table \ref{tab:outflowparams} and \ref{tab:modelparams}.}

\tablenotetext{\tiny d}{The time for the entire gas mass of the cluster \citep[from][]{leroy18} to be depleted at the current ${\rm \dot{M}_{H_{2},outflow}}$. For each SSC, this is the average of the values in Table \ref{tab:outflowparams} and \ref{tab:modelparams}. Uncertainties reported in the table are the standard deviations of the mean values from Table \ref{tab:outflowparams} and the values in Table \ref{tab:modelparams}. Propagated systematic uncertainties (due to the uncertainty in the molecular abundances) are 0.4 dex.}

\tablenotetext{\tiny e}{The gas depletion timescale, defined as ${\rm t_{dep}\equiv\frac{M_{H_{2},tot}}{SFR_{36\ GHz}}}$.  For SSC 4a, the gas mass for SSC 4 is used because SSC 4a is the dominant component of the SSC 4 complex.}

\end{deluxetable*}

There are several timescales and ages calculated from this and previous analyses for these clusters, which we summarize here and list in Table \ref{tab:timescales}:

\paragraph{ZAMS Age (\tzamsage)} \citet{rico-villas20} uses the ratio of the luminosity in protostars to that of ionizing zero-age main sequence (ZAMS) stars to estimate the ages of the clusters (\tzamsage), finding that SSC 4a is $\sim10^{4.95}$ yrs old, SSC 5a is $\gtrsim10^5$ yrs old, and SSC 14 is $\sim10^{4.88}$ yrs old (Table \ref{tab:timescales}). These measurements are based on the same 36 GHz emission used by \citet{leroy18} to calculate the stellar masses. The stellar masses and ages associated with the ionizing photon rates derived from the ZAMS assumption may be underestimated if the cluster stellar population has evolved beyond the ZAMS stage, or if some fraction of the ionizing photons are absorbed by dust. These three SSCs have negligable synchrontron components of their SEDs, so synchrotron contamination of the 36 GHz emission is a small effect. These ages are, therefore, likely lower limits on the true "age" of the cluster.

\paragraph{Cluster Formation Timescales}  Given the \tzamsage\ and free-fall times (\tff) calculated by \citet{leroy18}, we can try to place the clusters in a relative evolutionary sequence. An important caveat is that \citet{leroy18} estimated \tff\ based on the marginally resolved data. With these spatially resolved data, the radii of the clusters decreased (e.g., Figure \ref{fig:massradius}), meaning that these values of \tff\ may be overestimated. SSCs 4a, 5a, and 14 have \tzamsage/\tff\ = 1.1, $\gtrsim$2.0, and 2.4 respectively (Table \ref{tab:timescales}). Modeling by \citet{skinner15} suggests that gas is actively collapsing to form stars on timescales $\sim1-2$\tff, the typical timescale for cluster formation is $\sim 5$\tff, and the gas is completely dispersed by $\sim8$\tff. These clusters should be nearing the end of the period of active gas collapse. SSC 5a may possibly be transitioning to the initial stages of gas dispersal, especially because it is the only cluster visible in the NIR (Section \ref{ssec:nirclusters}), though the gas dispersal has not yet finished because we still see evidence for an outflow. It is important to note, however, that these evolutionary stages are not clear-cut divisions, as gas accretion can continue while the cluster is forming and while outflows are present, and that the \tzamsage\ are likely lower limits on the cluster ages. {\change Moreover, the possibility that expelled gas is reaccreted onto the clusters may mean that this cluster formation sequence is more cyclic.}

\paragraph{Crossing Time (\tcross)} The crossing times (\tcross) we report in Tables \ref{tab:outflowparams} and \ref{tab:modelparams} are short: $10^{4.5-5.2}$ years. Similarly short crossing times are also seen in a SSC candidate in the Large Magellanic Cloud \citep[$\sim10^{4.8}$ yr; ][]{nayak19} and in simulations {\change \citep[e.g.,][]{lancaster20}}. At least one crossing-time has passed since the outflows turned on, as P-Cygni line profiles are detected out to $\gtrsim r_{\rm SSC}$. If the outflows are present beyond $r_{\rm SSC}$, they are increasingly difficult to detect in absorption away from the continuum source. Therefore, this timescale places a lower limit on the age of the outflow. 

\paragraph{Gas Removal Time (\tremovegas)} The gas removal times (\tremovegas) are longer in general than the crossing or free-fall times, though the uncertainties on \tremovegas\ are large. The average \tremovegas$\sim10^{5.0\pm0.6}$ years, where the uncertainty is the standard deviation of the mean \tremovegas\ for each SSC and line. The gas removal times are $\gtrsim$\tzamsage, except for SSC 4a. Assuming a constant mass outflow rate, this would imply that there is still gas in the clusters to be removed, though a constant mass outflow rate is unlikely \citep[e.g.][]{kim18_jg}. {\change This timescale also assumes that none of the expelled gas is reaccreted later on. Given that the bulk of the gas has outflow velocities below the escape velocity (Section \ref{ssec:outflow}), reaccretion of material is a likely scenario.}

\paragraph{Gas Depletion Timescale (\tdep)} This timescale is the duration of future star formation, assuming a constant SFR for each cluster and no mass loss: \tdep${\rm \equiv M_{H_{2},tot}/SFR}$ where ${\rm M_{H_{2},tot}}$ is from \citet{leroy18}. The SFRs we use are also from \citet{leroy18} and are based on the measured 36 GHz fluxes which trace the free-free emission from each cluster \citep{gorski17,gorski19}. This estimate of \tdep\ based on the SFR assumes continuous star formation (over $\sim10$~Myr; \citealt{murphy11}), whereas we would expect the actual star formation in these clusters to be bursty. These are the longest timescales for each cluster listed in Table \ref{tab:timescales}. {\changes Compared to the clusters' \tzamsage, this may suggest that the clusters are early in their star formation process and that there is plenty of fuel to form new stars and for the clusters to continue grow.} This assumes, however, that all of the molecular gas remains in the cluster. The current gas removal times of the outflows (\tremovegas) are much shorter than \tdep, indicating that these outflows will substantially affect the cluster's star formation efficiency (SFE). {\change The possibility that gas is reaccreted by the cluster will affect the available gas reservoir for future star formation.}

\bigskip

That we detect outflows only in three sources, or $\sim$8\% of the three dozen SSCs in the center of NGC\,253 (Figure \ref{fig:cont}), gives credence to the idea that this outflowing phase must be short-lived. It is unlikely that we miss many sources with outflows due to their orientation and geometry because the modeling presented in Section \ref{ssec:outflowmodeling} as well as simulations \citep[e.g.,][]{geen21} suggest that the outflows are wide. We could, however, be missing outflows if the outer layers of dense gas are very optically thick, which could obscure the outflows \citep[e.g.,][]{aalto19} or if there are weak outflows below our sensitivity or cluster mass detection limits. Given that it is expected that the SSCs begin disrupting their natal clouds after $\sim10^{5-6}$ years \citep{johnson15} and taking ${\rm t_{cross}}=10^{4.5}$ years as the lower limit on the age of the outflow, we would expect to find outflows in at least $3-30$\% of SSCs, which agrees well with our detection rate of 8\%. This percentile range is a lower limit because ${\rm t_{cross}}$ is the minimum possible age of the outflow and there could be additional outflows below our detection limit, though they would be weak. This also assumes that the clusters formed at the same time, which is also unlikely.

In general the chemistry-based age sequences presented by \citet{krieger20} lead to different relative cluster ages than the dynamical progression presented here, which are also different from the ZAMS age sequence of \citet{rico-villas20}. It is important to keep in mind, however, that the oldest clusters are not necessarily the most evolved, and vice versa. Using HCN/HC$_3$N as a relative age tracer, \citet{krieger20} suggest that SSCs 4 and 14 are in the younger half of the SSCs studied while SSC 5 is among the oldest. An age sequence using the chemistry of sulfur bearing molecules suggests instead that SSCs 5 and 14 are younger whereas SSC 4 is older \citep{krieger20}. This is in disagreement with the age progression suggested by \citet{rico-villas20}, who suggest an inside out formation with SSCs 4$-$12 being the oldest and SSCs 1$-$3, 13, and 14 being the youngest. The detections of outflows towards SSCs 4a, 5a, and 14 would suggest that they are the most evolved clusters in the young burst, in the simplest model where the clusters completely and finally clear their gas at the end of their formation periods. As described in the following section, SSC 5a may be among the most evolved clusters, as it is the only one of these clusters visible in the NIR. \citet{krieger20} also find the lowest dense gas ratios in SSC 5a, suggesting that it has expelled and/or heated and dissociated much of its natal molecular gas. Given the gas-rich environment surrounding these clusters, however, it is possible that other clusters are older and more evolved, but have reaccreted gas from the surrounding medium or that was not completely expelled. Given that the mean velocities of the outflows in SSCs 4a, 5a, and 14 are less than the escape velocities, this is perhaps a likely scenario.  

\subsection{Outflow Mechanics}
\label{ssec:outflowpower}

\begin{figure*}
    \centering
    \includegraphics[width=\textwidth]{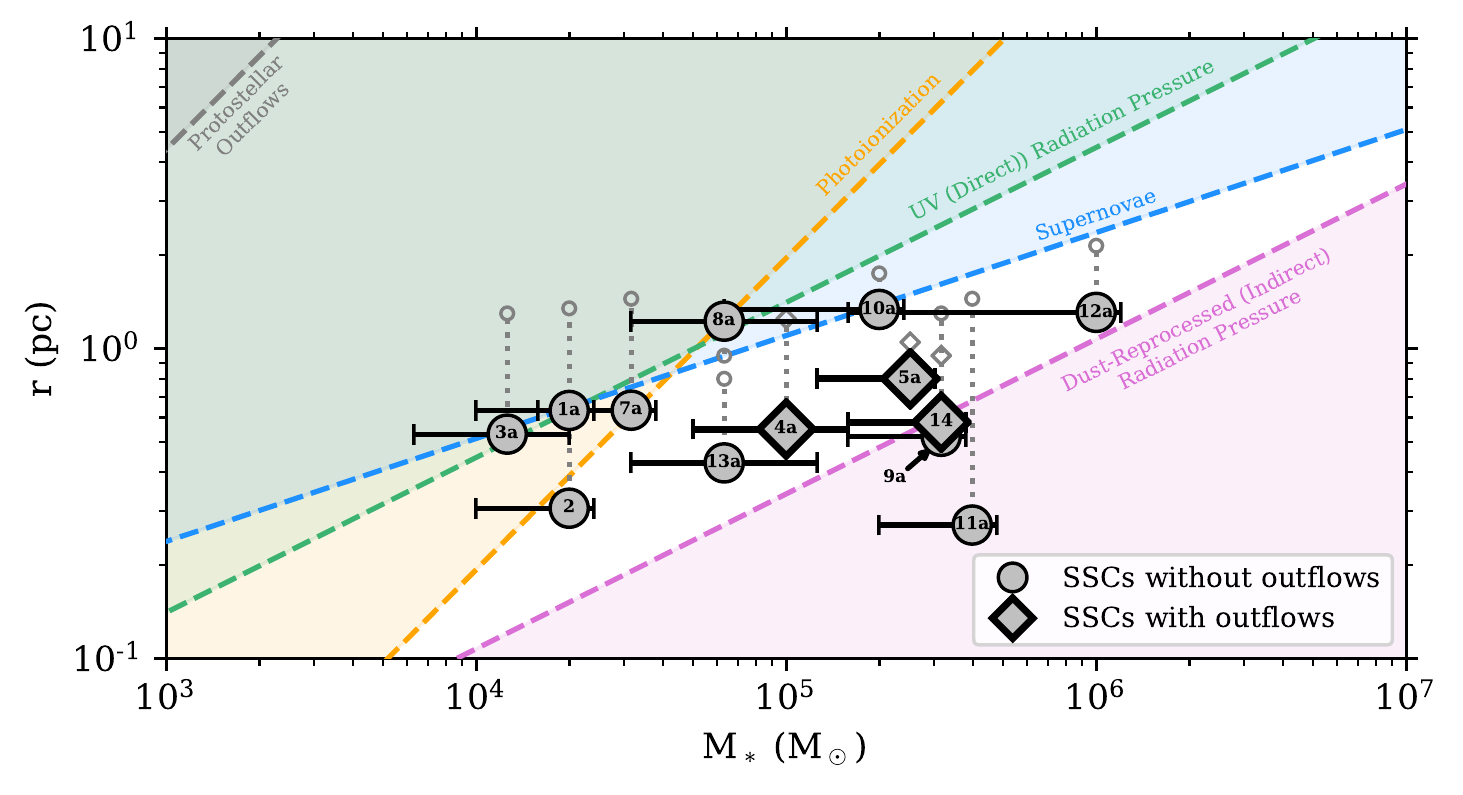}
    \caption{The cluster mass-radius diagram, adapted from Figure 12 of \citet{krumholz19}. The colored shaded regions bounded by dashed lines show the regions of this parameter space where the corresponding feedback mechanism is efficient. There is a locus where none of these feedback mechanisms are expected to be efficient, resulting in high star formation efficiencies. The circles show the primary SSCs in NGC\,253 without evidence for outflows, and the diamonds show those with outflows. The cluster stellar masses ($M_*$) are from \citet{leroy18}. The error bars include differences with measurements from RRLs (E. A. C. Mills et al., in prep.), systematics due to assumptions about the Gaunt factor, and conservative estimates of the effects of the clusters resolving into multiple smaller clusters. Not shown are unquantified uncertainties related to absorption of ionizing photons by dust and evolution beyond the ZAMS, both of which would result in higher values of $M_*$ than reported. See Section {\change \ref{ssec:otherprop}} for details on the calculation of the error bars. The radii are our measured 350 GHz half-flux radii (Table \ref{tab:contparams}), which are smaller than the radii previously measured by \citet{leroy18} ({\change gray open symbols in the same style as in the legend}) due to the increase in the spatial resolution of these observations. The change in radius is shown by the vertical gray dotted lines {\change connecting the symbols}. Notably, the SSCs with outflows lie within or near the locus where feedback is expected to be inefficient. }
    \label{fig:massradius}
\end{figure*}

There are a handful of feedback mechanisms relevant for setting the SFE of star clusters. These include proto-stellar outflows, supernovae, photoionization, UV (direct) radiation pressure, dust-reprocessed (indirect) radiation pressure, and stellar winds. Each of these processes is efficient in driving outflows for different cluster masses, radii, and ages. One way to visualize this is through a mass-radius diagram \citep[e.g.,][]{ fall10,krumholz19}, as shown in Figure \ref{fig:massradius}. There is a locus where none of the feedback mechanisms considered by \citet{krumholz19} are efficient, and so clusters with those masses and radii should grow with high SFEs. An important caveat of this figure is that there are other parameters relevant to whether a feedback mechanism is efficient, such as the momentum carried by each mechanism and the timescales over which it operates, which are not shown in this representation.

We show in Figure \ref{fig:massradius} the primary SSCs in NGC\,253, where SSCs without outflows are shown as circles and those with outflows are shown as diamonds. The radii are our measured half-flux radii listed in Table \ref{tab:contparams}. Because of the factor of 4 improvement in the linear spatial resolution with respect to the observations reported by \citet{leroy18}, the radii of these clusters are smaller than reported by \citet{leroy18}. As a result, the clusters are systematically shifted down in Figure \ref{fig:massradius} as shown by the vertical dotted gray lines. {\change The stellar masses are taken from \citet{leroy18}, as described in Section \ref{ssec:otherprop}. We note, in particular, that the reported stellar masses may be underestimated (beyond the error bars) due to uncertainties about dust absorption and evolution beyond the ZAMS.}

Many of the SSCs in NGC\,253 fall inside the white area of inefficient feedback in Figure \ref{fig:massradius}, which suggests that none of those mechanisms may be efficient for these clusters. The detection of outflows in SSCs 4a, 5a, and 14, however,  means there is direct evidence of strong feedback. The \htwo\ masses in the outflows are significant compared to the total mass in the cluster itself (Table \ref{tab:outflowparams}). From the previous section (and quantified in Table \ref{tab:timescales}), the gas removal times (\tremovegas) based on the current mass outflow rates are much shorter than the timescale for future star formation (\tdep). Comparison of these timescales implies that the outflows will remove the molecular gas faster than it would otherwise be used up by star formation. Note that these timescales assume either a constant mass outflow rate or a constant star formation rate, respectively, neither of which is likely to be the case in reality. Moreover, \tremovegas\ assumes there is no new infall replenishing the reservoir, and \tdep\ assumes no molecular gas is removed. Nevertheless, the outflows will have a non-negligible effect on the reservoir of gas available to the cluster to form stars and hence on the cluster's SFE.

What mechanism, then, is powering these outflows, given that SSCs 4a and 5a lie in the locus where no mechanism is expected to be efficient and SSC 14 is near a boundary? We explore four plausible mechanisms shown in Figure \ref{fig:massradius} below \citep[excluding proto-stellar outflows which are only important for much lower mass clusters; e.g.,][]{guszejnov21}. We also consider winds from high mass stars which may be important for clusters of these masses and ages {\change \citep[e.g.,][]{gilbert07,agertz13,geen15,lancaster20}}. In addition to their location in the mass-radius diagram (a re-framing of their surface densities), we also compare the momentum expected to be carried by each of these processes and the timescales over which they operate to the values estimated for these clusters. It is also possible that a synergistic combination of mechanisms is at work {\change \citep[e.g.,][]{rahner17,rahner19}}. These potential scenarios are discussed in the reminder of this section.

\subsubsection{Supernovae}
\label{sssec:sne}
These clusters are young \citep[$\sim10^5$ years;][]{rico-villas20}, and so it is not expected that many, if any, supernovae have exploded since it typically takes $\sim$3 Myr before the first supernova explosion {\changes \citep[e.g.,][]{zapartas17}}. Moreover, the expected cloud lifetimes in a dense starburst like NGC\,253 are expected to be shorter than 3 Myr \citep[e.g.,][]{murray10}, meaning that clouds would be disrupted before supernova feedback would be important. The gas removal times estimated for these outflows are $\ll 3$ Myr in all cases (Table \ref{tab:timescales}), supporting the idea that the clusters are too young for supernovae. Even with the large uncertainties, the radial momentum we measure in their outflows (Table \ref{tab:outflowparams}) is an order of magnitude or more lower than expected for supernova-driven outflows \citep[$1000-3000$ \kms; e.g.][]{kim15_cg,kim17_cg}. Finally, E. A. C. Mills et al. (in prep.) construct millimeter spectral energy distributions of these SSCs at 5 pc resolution and find that all three of these sources have negligible synchrotron components, further ruling out supernovae as the mechanism driving the cluster outflows. 

\subsubsection{Photoionization}
\label{sssec:pion}

Photoionization can remove a substantial amount of gas from a cluster, depending primarily on the density of the cluster \citep[e.g.,][]{kim18_jg,he19,geen21,geen20,dinnbier20}. \citet{he19} study the effects of photoionization on massive star clusters, finding that photoionization is most efficient at suppressing star formation in lower density clusters. Similarly, \citet{dinnbier20} find that photoionization (and winds) are inefficient at clearing the natal gas from clusters with masses $>5\times10^3$ \msun, unless they form with a high SFE. Though simulating different size and mass scales, \citet{dinnbier20} and \citet{geen20,geen21} find that feedback by photoionization is most important at the outer layers of the cloud or \HII\ region. \citet{kim18_jg} find that the momentum carried by a photoionization-driven outflow decreases with increasing cluster surface density. {\change Assuming that the dense gas we measure follows the trends for the neutral gas simulated by \citet{kim18_jg} and that the trends continue to these higher surface densities, the radial momentum per unit mass from photoionization should be very small (${\rm p_r/M_*\lesssim1}$ \kms) for  the molecular gas surface densities of SSCs 4a, 5a, and 14 of $\Sigma_{\rm H_{2}} \sim 10^{4.2-4.9}$ \msun\ pc\pers\ \citep{leroy18}.} Using the stellar masses and radii, we can calculate the estimated ionized gas masses and momentum \citep[using Equations 3, 4, and 11 of][]{leroy18} and assuming ${\rm V_{ion}=15}$~\kms\ (as below), finding the ${\rm p_{ion}/M_*}\approx 0.13,\ 0.15,\ {\rm and\ } 0.08$~\kms\ for SSCs 4a, 5a, and 14. We note that the ionizing photon rates estimated by \citet{leroy18} for these sources are consistent with a new analysis by E. A. C. Mills et al. (in prep.). These estimated momenta based on the ionizing photons are much smaller than the measured values of ${\rm p_r/M_*} \approx\ 5-126,\ 1-3,\ {\rm and\ } 5-40$~\kms\ (Table \ref{tab:outflowparams}). 


\citet{krumholz19} parameterize the effect of photoionizaion in terms of the ionized gas sound speed (${\rm c_{s,ion}}$) and the cluster escape velocity (${\rm V_{esc}}$), such that photoionization will be efficient when ${\rm c_{s,ion}} > {\rm V_{esc}}$. It is thought that the photoionization will be efficient for clusters with  ${\rm V_{esc}}\approx{\rm c_{s,ion}}\leq10$ \kms. Figure \ref{fig:massradius} and \citet{krumholz19} assume ${\rm V_{esc}}\approx{\rm c_{s,ion}}=15$ \kms. The escape velocities of SSCs 4a, 5a, and 14 are approximately 23, 34, and 50 \kms, respectively. This suggests photoionization could be important for SSC 4a, but the escape velocities of SSCs 5a and 14 are likely too large for this mechanism to be efficient. This is in agreement with the qualitative results of simulations discussed above.

Therefore, feedback from photoionization likely plays a minor role in driving these outflows as these clusters are dense and hence their escape velocities are too large for photoionization to efficiently drive outflows.

\subsubsection{UV (Direct) Radiation Pressure}
\label{sssec:drp}

There are many studies that investigate the role of UV (direct) radiation pressure in driving outflows and expelling gas from a molecular cloud or \HII\ region \citep[e.g.,][]{kim16_jg,kim17_jg,kim18_jg,kim19_jg,raskutti17,crocker18b,crocker18a,barnes20,dinnbier20}. In general, many of these studies find that UV radiation pressure can play an important --- if not dominant --- role in expelling gas from a cluster, especially for low surface density clouds \citep[e.g.,][]{skinner15,raskutti17}. Using a numerical radiation hydrodynamic approach, \citet{raskutti17} study the effects of radiation pressure from  non-ionizing UV photons on massive star forming clouds, finding that over a cloud's lifetime $\gtrsim 50$\% of the UV photons escape through low-opacity channels induced by turbulence and hence do not contribute to radiation pressure-driven outflows. An important caveat is that the clouds simulated by \citet{raskutti17} are larger in size and so have lower densities than the clusters in NGC\,253. \citet{raskutti17} find that the mean outflow velocity of their radiation-pressure driven outflows is $\sim1.5-2.5\ {\rm V_{esc}}$, independent of cloud surface density. For the clusters in NGC\,253, this translates to expected mean outflow velocities of $\sim33-114$ \kms, faster than the observed mean velocities of $6-28$ \kms. The radial momentum per unit stellar mass (${\rm p_r/M_*}$) of SSCs 4a, 5a, and 14 are an order of magnitude or more less than measured by \citet{raskutti17} for clouds of roughly the same initial mass, though the clusters in NGC\,253 are denser than the simulated clouds. In contrast, the clusters in NGC\,253 have slightly larger radial momenta per unit outflowing mass (${\rm p_r/M_{out}}$). This means that the clusters in NGC\,253 have less outflowing mass relative to their stellar masses than the simulated clusters, though this could also be due to the difference in cloud densities. 

Simulations of feedback from photoionization and UV radiation pressure by \citet{kim18_jg} find \tdep\ $\sim2.5$ Myr for their densest clouds ($\Sigma_{\rm neutral} = 10^3$ \msun\ pc\pers). Our clusters, which have $\Sigma_{\rm H_{2}} \sim 10^{4.4-5.3}$ \msun\ pc\pers\ \citep{leroy18}, have \tdep $= 2.5-6.3$ Myr, whereas extrapolation of the \citet{kim18_jg} calculations would result in lower values of \tdep\ if the trends continue to these higher densities. \citet{kim18_jg} also find relatively constant outflow velocity of $\sim$ 8 \kms, approximately independent of surface density in the neutral phase. This is similar to the mean outflow velocities of the dense gas traced by \cs\ or \httcn\ for SSC 4a ($\approx$7~\kms), but lower than those for SSCs 5a and 14 ($\approx$ 22 and 25 \kms, respectively). In terms of momentum, according to Equation 18 in \citet{kim18_jg}, an outflow driven by photoionization and UV radiation pressure would have a momentum per unit stellar mass of $\sim1$ \kms\ at the surface densities observed in our clusters. This is about an order of magnitude lower than the ${\rm p_r/M_*}$ measured for dense gas traced by \cs\ or \httcn\ in these clusters (though our uncertainties are large). {\changes An important caveat to this is that the clusters in NGC\,253 are denser than those simulated by \citet{kim18_jg}. Additionally, these simulations probe the neutral outflowing phase whereas our observations probe the outflowing dense molecular gas.}

The efficiency of UV radiation pressure can be parameterized in terms of the surface mass density of the cluster ($\Sigma$) compared to the the outward force of radiation pressure. \citet{skinner15} and \citet{krumholz19} define a critical surface density, ${\rm \Sigma_{DR}=\frac{\Psi}{4\pi Gc}}$ where $\Psi$ is the light-to-mass ratio, below which UV radiation pressure becomes important. A population of ZAMS stars that fully samples a \citet{chabrier03} IMF has $\Psi\approx1100$ L$_\odot$ \msun$^{-1}$ \citep[e.g.][]{fall10,kim16_jg,crocker18b}, resulting in $\Sigma_{\rm DR}\approx 340$ \msun\ pc$^{-2}$. Models and simulations show that UV radiation is only effective for surface densities ${\rm \Sigma\lesssim10\,\Sigma_{DR}}$ because turbulence will introduce low-column sight-lines that allow the radiation to escape \citep[e.g.][]{thompson16,grudic18,krumholz19}. The value of $\Sigma_{\rm DR}$ depends principally and linearly on the assumed light-to-mass ratio and hence on the IMF. SSCs 4a, 5a, and 14 are all well below this boundary shown in green in Figure \ref{fig:massradius}, meaning that a significantly top-heavy IMF resulting in a much higher value of $\Psi$ is required for these outflows to be powered solely by UV radiation pressure. A top-heavy IMF has been suggested in other massive or super star clusters \citep[e.g.][]{turner17,schneider18}. \citet{turner17} find a top-heavy IMF in a SSC in NGC\,5253 with ${\rm \Psi\approx2000\ L_\odot\ M_\odot^{-1}}$. However, an unphysically large light-to-mass ratio ${\rm \sim10,000 \ L_\odot\ M_\odot^{-1}}$ is needed for UV radiation to explain the feedback in SSCs 4a, 5a, and 14.

It is, therefore, unlikely that UV (direct) radiation pressure is the dominant mechanism responsible for driving the outflows in SSCs 4a, 5a, and 14.

\subsubsection{Dust-Reprocessed (Indirect) Radiation Pressure}
\label{sssec:irp}

Given the large dust columns in these SSCs, dust-reprocessed (indirect) radiation pressure is a promising mechanism to power the outflows.  In a study of massive star clusters in the Milky Way's Central Molecular Zone, \citet{barnes20} find that indirect radiation pressure is important at early times (< 1 Myr). Similarly, \citet{olivier21}  find that dust-reprocessed radiation pressure is the dominant feedback mechanism in ultra compact \HII\ regions in the Milky Way. Moreover, given the possible underestimate of the stellar masses due to absorption of ionizing photons by dust or evolution beyond the ZAMS (which are not reflected in the error bars in Figure \ref{fig:massradius}), indirect radiation pressure is a plausible mechanism to drive the observed outflows.

Whether the dust-reprocessed radiation pressure can drive an outflow fundamentally depends on the balance between the outward force of the radiation pressure and the inward force of gravity (i.e. the Eddington ratio, $f_{\rm Edd}$). This depends on the dust opacity and the input luminosity from the cluster, which can be parameterized by the light-to-mass ratio of the assumed IMF. The dust opacity ($\kappa_d$) quantifies a dust grain's ability to absorb infrared radiation. As explored by \citet{semenov03}, this quantity varies with temperature, dust composition, grain shape, grain size distribution, and the opacity model. For example, grains in dense molecular clouds exhibit larger $\kappa_d$ than those in the diffuse ISM, which is thought to be due to the growth of mantles \citep[e.g.,][]{ossenkopf94}. Given the high density and intense radiation environment in these clusters, the dust opacity may be quite different than found in Galactic regions, though we have no precise observational constraints on how much $\kappa_d$ can vary in these environments. Often times, a dust-to-gas ratio (DGR) is assumed to convert the dust opacity to the opacity in the gas, and the standard Solar Neighborhood value is DGR = 0.01. Finally, the light-to-mass ratio ($\Psi$) depends on the assumed IMF, where a top-heavy IMF will have a larger value of $\Psi$. Top-heavy IMFs have been claimed in massive clusters with $\Psi\sim2000~{\rm L_\odot~M_\odot^{-1}}$ \citep[e.g.,][]{turner17,schneider18} compared to either a \citet{kroupa01} or \citet{chabrier03} IMF which have $\Psi = 883~{\rm L_\odot~M_\odot^{-1}}$ and $\Psi=1100~{\rm L_\odot~M_\odot^{-1}}$ respectively. {\change Models also find that the IMFs of globular cluster progenitors --- thought to be SSCs --- are more top-heavy with increasing density and decreasing metallicity \citep{marks12}. For clusters with the stellar masses and densities of those in NGC\,253 \citep{leroy18}, we would expect a typical high-mass IMF slope.} Towards constraining the nature of the IMF in these clusters, E. A. C. Mills et al. (in prep.) measure the fraction of ionized helium towards these SSCs using H and He radio recombination lines at 5 pc resolution. This ratio is expected to be somewhat dependent on the IMF, as a top-heavy IMF will result in more massive stars capable of ionizing He. E. A. C. Mills et al. (in prep.) measure mass-weighted He$^+$ fractions which are consistent with He$^+$ fractions found in \HII\ regions in the center of the Milky Way \citep[e.g.,][]{mezger76,depree96,lang97} which is not thought to have a globally top-heavy IMF \citep[e.g.,][]{lockmann10} though there are individual clusters which do seem to favor a top-heavy IMF \citep[e.g.,][]{lu13}. While the measured He$^+$ fractions do not completely rule out the possibility of a top-heavy IMF (it is unclear precisely how much the He$^+$ fraction changes with the IMF), it is unlikely that the IMF in these clusters is extremely top-heavy. 

Below we describe two approaches taken by simulations in determining the efficiency of dust-reprocessed radiation pressure in driving outflows, and discuss how adjustments to the {\changes fiducial} assumptions on $\kappa_d$, DGR, and $\Psi$ may help explain the outflows observed in SSCs 4a, 5a, and 14. 

Numerical simulations of dust-reprocessed radiation pressure by \citet{skinner15} assume that the dust opacity ($\kappa_d$) is constant with temperature (and hence distance from the UV source). \citet{skinner15} find:
\begin{equation}
    \label{eq:fedd}
    f_{\rm Edd} = 0.68\left(\frac{\rm DGR}{0.01}\right)\left(\frac{\kappa_d}{0.1~{\rm cm^2~g^{-1}}}\right)\left(\frac{\Psi}{883~{\rm L_\odot}~M_\odot^{-1}}\right)
\end{equation}
where we have explicitly included the dependence on the dust-to-gas ratio (DGR). For the fiducial values of DGR and $\Psi$, \citet{skinner15} find $f_{\rm Edd} > 1$ only for $\kappa_d > 0.15$~cm$^2$~g\per, which is unphysically large for Solar Neighborhood-like dust properties (typically 0.03~cm$^2$~g\per; \citealt{semenov03}). If, however, the dust is different in these environments than in the Solar Neighborhood, $\kappa_d$ could be larger, though we have no observational constraints to evaluate this. Assuming Solar Neighborhood-like dust and a \citet{kroupa01} IMF, we can achieve $f_{\rm Edd} > 1$ if DGR~$>0.05$. Although the center of NGC\,253 is known to have a super-solar metallicity ($Z=2.2Z_\odot$; \citealt{davis13}) and the clusters may be even more dust-rich \citep{turner15,consiglio16}, a DGR~$>0.05$ seems quite high. Turning instead to the possibility of a top-heavy IMF, Equation \ref{eq:fedd} yields $f_{\rm Edd} > 1$ for $\Psi > 4300~{\rm L_\odot~M_\odot^{-1}}$ (for Solar Neighborhood-like values of $\kappa_d$ and DGR,), which is much more top-heavy than has been found in other SSCs \citep[e.g.,][]{turner17}. Given that E. A. C. Mills et al. (in prep.) do not find strong evidence for an increased He$^+$ fraction in these clusters, a very top-heavy IMF is unlikely. Finally, \citet{skinner15} note that for clustered sources of UV photons --- almost certainly the case in the SSCs in NGC\,253 --- there can be appreciable cancellation of the radiation pressure terms from each source, so that the net momentum to drive a cluster-scale outflow will be lower, independent of increases in the DGR, $\kappa_d$, or $\Psi$. Therefore, while there may be some combination of increased $\kappa_d$, DGR, and $\Psi$ that enables dust-reprocessed radiation pressure to efficiently drive outflows in these clusters, it is unclear how much internal cancellation will affect the net momentum. 

\citet{crocker18b} take a slightly different approach to study the efficiency of dust-reprocessed radiation pressure. {\changes Modeling of the Rosseland mean opacity by \citet{semenov03} found $\kappa_d\propto T^2$ for $T<100$~K. \citet{crocker18b} convert this temperature dependence into a radial dependence from the central source of UV photons, introducing a gradient in $\kappa_d$.} This has the effect of boosting the effective dust opacity for clusters that are very dense, allowing for outflows to be driven more easily compared to \citet{skinner15}. Secondly, this allows \citet{crocker18b} to express their $f_{\rm Edd}$ in terms of a critical surface density:
\begin{equation}
    \label{eq:SigmaIR}
    \begin{split}
    \Sigma_{\rm *,IR} = \left(1.3\times10^5 {\rm\ M_\odot\ pc^{-2}}\right)\times\\ \left(\frac{\kappa_d}{0.03 {\rm\ cm^{2}\ g^{-1}}}\right)^{-1}\left(\frac{\Psi}{1100 {\rm\ L_\odot\ {M_\odot}^{-1}}}\right)^{-1}\left(\frac{\rm DGR}{0.01}\right)^{-1}
    \end{split}
\end{equation}
where we have again explicitly shown the dependence on the DGR. The dashed pink boundary in Figure \ref{fig:massradius} assumes the fiducial values of $\kappa_d$, $\Psi$, and DGR. While SSC 14 lies on this boundary, SSCs 4a and 5a fall above it, suggesting that dust-reprocessed radiation-pressure is not sufficient for driving outflows in these clusters. 

Note, however, that there are considerable uncertainties in this picture. The cluster stellar masses may be underestimated if appreciable ionizing photons are absorbed by dust or if the stellar population is evolved beyond the ZAMS. Therefore, they may be closer to the region where dust-reprocessed radiation pressure is efficient than shown in Figure \ref{fig:massradius}. Moreover, we do not fully understand the properties of dust in these conditions. As discussed above, the DGR in their surrounding gas may be $\gtrsim0.022$ given the observed super-solar metallicity in the center of NGC\,253 \citep{davis13} and the high density conditions in the starburst molecular gas, which may favor dust formation. Assuming DGR = 0.022 in Equation \ref{eq:SigmaIR},  moves the $\Sigma_{\rm *,IR}$ boundary up to encompass SSCs 4a and 5a. This difference compared to the results of \citet{skinner15} comes from the assumed temperature dependence in $\kappa_d$, which provides a boosted dust opacity for very compact sources. This itself is very uncertain, since the dust models are designed for proto-planetary disks, and the growth of $\kappa_d$ with $T$ saturates at $T\sim100$~K \citep[see][]{semenov03}. Considering changes to the IMF using the fiducial value of $\kappa_d$ and the DGR would require $\Psi\gtrsim3000 {\rm\ L_\odot\ M_\odot^{-1}}$ to explain SSCs 4a and 5a, or 1.5$\times$ more top-heavy than the IMF in NGC\,5253 \citep{turner17}. Even with the boost in $\kappa_d$, there would still be cancellations in the radiation pressure due to clustered UV sources, although these cancellations may not be as severe as in the constant $\kappa_d$ case. 

Therefore, it is possible that dust-reprocessed (indirect) radiation pressure could drive the outflows observed in SSCs 4a, 5a, and 14, though this hinges critically on the behavior of the dust opacity ($\kappa_d$) for which there are no observational constraints in these extreme environments. A likely elevated dust-to-gas ratio (DGR) in these sources helps, but cancellations from clustered UV sources hinders the efficiency with which dust-reprocessed radiation pressure can drive outflows. Therefore, whether dust-reprocessed radiation plays a dominant role in powering the outflows observed from SSCs 4a, 5a, and 14 is an open question.

\subsubsection{Winds from High Mass Stars}
\label{sssec:winds}

Given the stellar masses of these clusters ($M_*=10^{5.0-5.5}$ \msun), we would expect $\gtrsim1000-3000$ O stars in each cluster, assuming a \citet{kroupa01} IMF. It has been suggested that outflows from young, massive SSCs in the Antennae are driven by a combination of O and Wolf Rayet (WR) stellar winds \citep{gilbert07}, although other possible mechanisms are not evaluated. The combined power of the winds from these massive stars could, therefore, play an important role in powering the outflows from the clusters in NGC\,253. 

It has been suggested that winds from WR stars significantly impact how a cluster clears its natal gas, as they impart $\sim10\times$ the energy of O-star winds over a shorter period of time \citep[e.g.,][]{sokal16}. The mass-loss rates of WR stars are metallicity dependent, with higher mass loss rates at higher metallicity for both carbon- and nitrogen-rich WR stars \citep{vink05}. Because WR stars are evolved O-stars, it is thought that they should not contribute much towards the cluster feedback until after $\sim3-4$ Myr, when other processes such as supernovae are becoming important and when much of the natal gas has already been dispersed. In an observational study of massive embedded clusters, however, \citet{sokal16} find that WR winds are important even at earlier stages, though the clusters they study all have ages $>$1 Myr. Moreover, they find that clusters without WR stars tend to stay embedded longer than those with WR stars, indicating that WR winds may accelerate the gas-clearing stage.

It is unclear, however, whether the clusters in NGC\,253 harbor WR stars yet, given their very young ages (\tzamsage\ $\approx0.1$ Myr). There is evidence of a WR population towards SSC 5 --- perhaps the most evolved cluster --- but higher spatial and spectral resolution observations are needed to confirm this \citep{kornei09,davidge16}. There is a known WR X-ray binary in NGC\,253, but it is outside of the nuclear region studied here by $\sim250$ pc in projection \citep{maccarone14}. Given the young ages of these clusters, it is unlikely that there are many, if any, WR stars present in these clusters, especially SSCs 4a and 14. Once a portion of the O star population evolves into WR stars, however, their winds could potentially strongly contribute to driving outflows.

Simulations of stellar wind feedback on the clearing of a cluster's natal gas have mixed conclusions. Some simulations show that stellar winds are important at early times in a cluster's life, especially for clearing the natal dense gas before supernovae start occurring at around 3 Myr \citep[e.g.,][]{agertz13,geen15,geen20,geen21}. {\changes Given that the clusters in NGC\,253 are substantially younger than this (\tzamsage $\approx 0.1$ Myr), stellar winds may play a prominent role in driving the outflows we observe from these SSCs.} Other simulations, however, find that stellar winds are most effective after 3 Myr \citep{calura15}. There are also simulations that find that stellar winds (and photoionization) alone cannot expel the natal gas for massive ($>5\times10^3$ \msun) clusters at any time point unless they form with a SFE~${\rm \equiv M_*/(M_{gas}+M_*)} > 1/3$ \citep{dinnbier20}. 

In the absence of gas cooling, stellar winds can impart momentum into the surrounding material typically ${\rm p_r/M_*}\approx50-65$~\kms\ \citep{weaver77}, assuming a wind luminosity of $10^{34}$ erg s\per\ \citep[Starburst99;][]{leitherer99}, the ages of these clusters to be $\approx10^5$ yrs \citep{rico-villas20}, and $n_{\rm H}=10^5$ cm$^{-3}$. These estimates are on the high side of our observed range of ${\rm p_r/M_*} \approx\ 5-126,\ 1-3,\ {\rm and\ } 5-40$~\kms\ for SSCs 4a, 5a, and 14 respectively (Table \ref{tab:outflowparams}), and are fairly insensitive to the assumed average gas density (an order of magnitude in $n_{\rm H}$ results in a factor of $\approx 1.5$ change in the expected momentum).  

If the gas can cool, however, the momentum imparted will be substantially lower, and this is especially relevant in the large ambient densities found near these SSCs {\change\citep[e.g.,][]{silich04,palous14,wunsch17,lochhaas17,el-badry19,gray19,lancaster20}}. An outflow stalled by cooling has been claimed in a SSC in NGC\,5253 \citep{cohen18}. Nonetheless, recent simulations by {\change \citet{lancaster20}} find that although cooling is important for the gas densities in the central starburst of NGC\,253, {\change the momentum imparted with significant cooling still provides} a modest enhancement over a momentum-conserving wind. The computed enhancement factor ($\alpha_p$) is sufficient to power outflows even with efficient cooling (with $\alpha_p\approx1-4$ {\change assuming a normal IMF and solar metallicity}). For the typical ages of these SSCs ($\sim0.1$~Myr), {\change \citet{lancaster20}} predict a momentum injection of ${\rm p_r/M_*\approx0.8}$~\kms, lower than momenta measured for SSCs 4a, 5a, and 14 of ${\rm p_r/M_*} \approx\ 5-126,\ 1-3,\ {\rm and\ } 5-40$~\kms\ (Table \ref{tab:outflowparams}). Given the predicted shell velocity from {\change \citet{lancaster20}}, we can estimate the value of $\alpha_p$ implied by the observed outflows:
\begin{equation}
\label{eq:lan71b}
    \alpha_p = \left(\frac{\rm V_{max-abs}}{2.0~{\rm km~s^{-1}}}\right)^2\left(\frac{\rm M_{H_{2},out}}{\rm M_*}\right)\left(\frac{\rm r_{SSC}}{1~{\rm pc}}\right)^{-1}.
\end{equation}
For the measured outflow properties of SSCs 4a, 5a, and 14, this suggests $\alpha_p\approx 9-227,\ 4-10,\ {\rm and\ } 34-272$ respectively (using the parameters in Tables \ref{tab:contparams} and \ref{tab:outflowparams}) {\change compared to $\alpha_p\approx1-4$ from the simulations. For all SSCs, our values of ${\rm M_*}$ may be underestimated (Section \ref{ssec:otherprop}), reducing the values of $\alpha_p$ inferred from Equation \ref{eq:lan71b}. A top-heavy IMF could provide up to a factor of 4 enhancement in the strength of the wind \citep{lancaster20}. As described in the Section \ref{sssec:irp}, we do not expect the IMF in these clusters to be particularly top-heavy \citep[e.g.,][E. A. C. Mills et al. in prep.]{marks12}, although this possibility cannot be ruled out entirely. The wind strength can also be enhanced if the metallicity is super-solar, as is likely the case for these clusters \citep{davis13,turner15,consiglio16}. 

Given the uncertainties in our measured masses, the IMF, and the metallicity, it is possible that the outflow in 5a is powered by stellar winds. Given the concerns about saturation in SSC 4a, the outflowing mass may be overestimated and so our inferred $\alpha_p$ from Equation \ref{eq:lan71b} may also be overestimated, meaning that the outflow from SSC 4a may also be driven (at least in part) by stellar winds. The outflow from SSC 14, on the other hand, is unlikely to be solely by stellar winds.}

Therefore, O star winds are unlikely to be driving the outflow observed towards SSC 14 in NGC\,253, although it is possible they play a role in driving the current outflows in SSCs 4a and 5a, especially if the current stellar masses of these clusters are underestimated {\change and if the metallicities are super-solar}. The O star populations in these clusters are likely too young to host many WR stars, except perhaps in the case of SSC 5a, so the effect of WR star winds is likely negligible at this stage in the SSCs evolution.

\subsubsection{Summary: What Mechanisms Power the Outflows?}
\label{sssec:outflowsynthesis}

To summarize the above exploration of possible feedback mechanisms, we find that the outflows from SSCs 4a, 5a, and 14 are difficult to explain, though they are likely powered by a combination of dust-reprocessed radiation pressure and stellar winds. All three clusters are too young for supernovae to have exploded and too dense for photoionization or UV (direct) radiation pressure to be efficient. Whether dust-reprocessed radiation pressure is efficient depends on the properties of the dust opacity ($\kappa_d$), for which there are virtually no constraints in extreme environments like these SSCs, and likely requires some combination of a higher dust opacity, an increased dust-to-gas-ratio, and a top-heavy IMF, all of which are currently poorly constrained in these clusters. Moreover, clustered UV sources within the SSCs can have the effect of cancelling out the radiation pressure terms from other sources \citep{skinner15}, reducing the net momentum to drive a cluster-scale outflow. In the case of stellar winds, cooling is expected to be important for these clusters. Although recent simulations find that even in the presence of strong cooling O star stellar winds may be sufficient to power outflows {\change \citep{lancaster20}}, we find that the expected momentum is insufficient to explain the observed properties of the outflows in SSC 14 and possibly in SSC 4a. For SSC 14, the outflows are likely dominated by dust-reprocessed radiation pressure, whereas the outflow in SSC 5a may be dominated by stellar winds. For SSC 4a, the deep absorption renders the outflowing mass estimate especially uncertain and likely overestimated, so we can only say that the outflow in that cluster is likely a combination of dust-reprocessed radiation pressure and stellar winds. Therefore, the precise mechanism(s) powering these outflows remains uncertain.

\subsection{SSC 5a and \hst\ NIR Clusters}
\label{ssec:nirclusters}
\begin{figure*}
    \centering
    \gridline{\fig{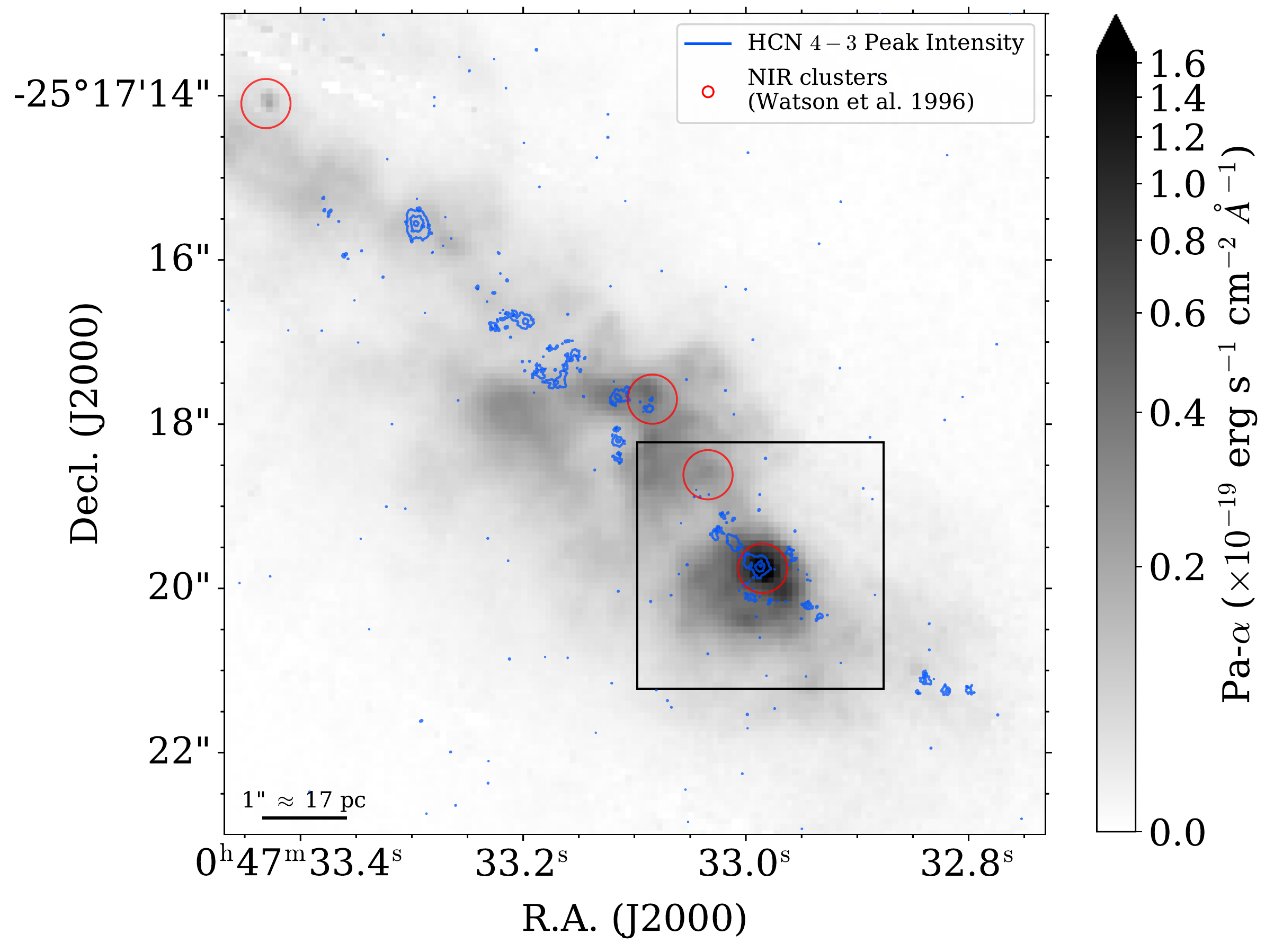}{0.5\textwidth}{}
    \fig{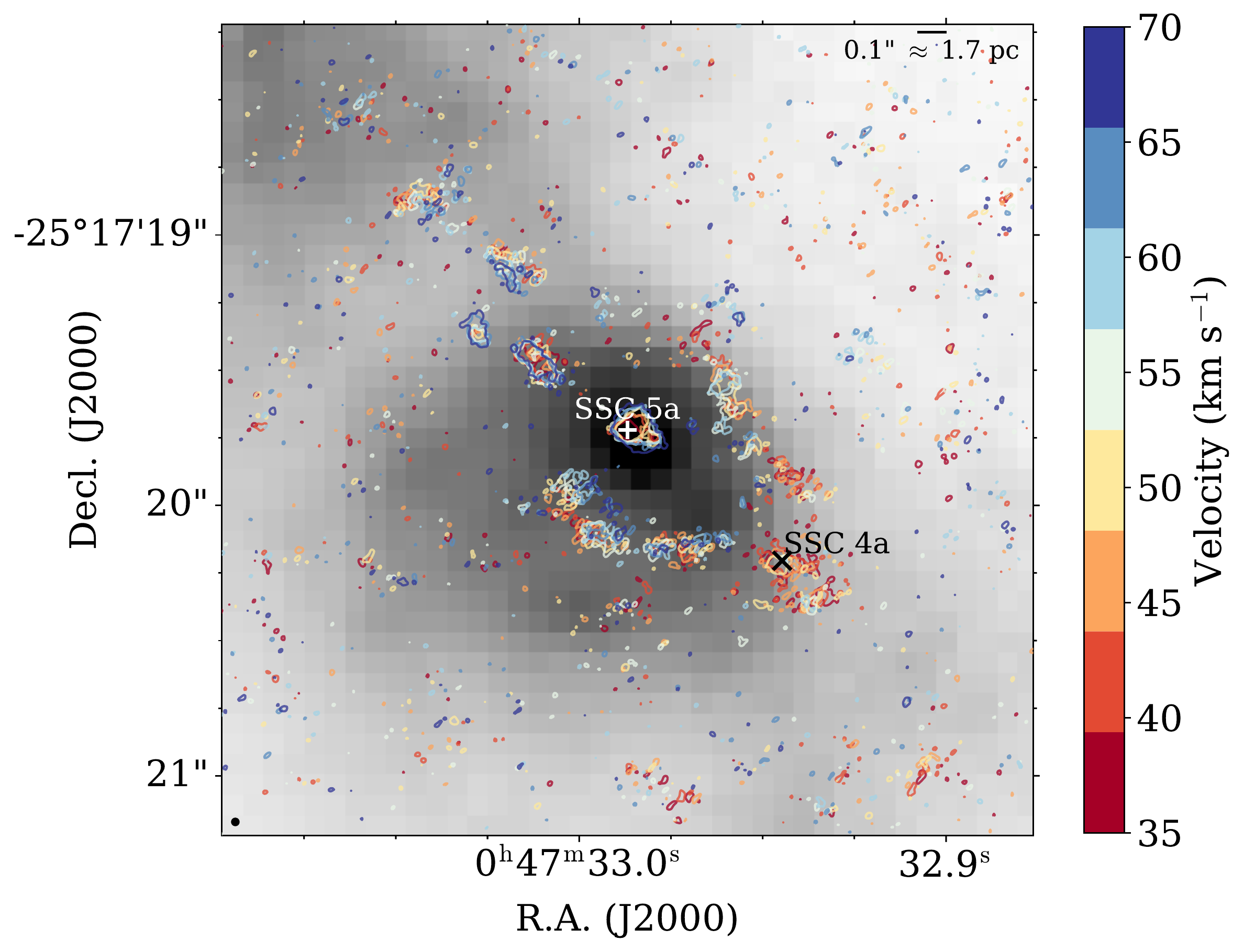}{0.5\textwidth}{}}
    \caption{(Left) The \hst\ Pa-$\alpha$ image. The blue contours show the ALMA \hcn\ peak intensity at 2, 5, and 10$\times$ the rms of the peak intensity image. The red circles show the NIR clusters identified by \citet{watson96}, where the size of the circle reflects their 0.3$\arcsec$ positional uncertainty. Only one of the primary clusters (SSC 5a) corresponds to the previously identified NIR clusters. (Right) The \hst\ Pa-$\alpha$ image centered on SSC 5a (white plus sign) over the 3$\arcsec\times$3$\arcsec$ black square from the left panel. There is an asymmetry in the Pa-$\alpha$ emission, which \citet{kornei09} posit may be due to an outflow. The contours show the \hcn\ emission in selected the velocity channels relative to the systemic velocity of SSC 5a (Table \ref{tab:outflowparams}) showing a shell-like structure with a velocity gradient across the shell. The contours are drawn at 3$\times$ the rms of the \hcn\ cube. The location of SSC 4a is also marked for context (black cross). The asymmetry in the Pa-$\alpha$ emission does not align with the \hcn\ shell. This shell of \hcn\ may be the signature of an earlier (stronger) outflow phase from SSC 5a.}
    \label{fig:shell}
\end{figure*}

{\change Several clusters and one SSC have been previously identified in the center of NGC\,253 based on \hst\ near-infrared (NIR) images \citep{watson96,kornei09}, with the NIR-detected SSC corresponding to SSC 5 \citep{leroy18}. To be able to see the NIR emission, the cluster must have dispersed much of its natal gas and dust or it aligns with a serendipitous hole in the extinction screen. If the presence of NIR emission at this cluster location is from gas clearing, this implies that the cluster must be older than the other, highly extincted clusters.  As shown in Figure \ref{fig:modelcomp}, we see evidence for a weak outflow in the dense gas tracers from this cluster, which may be the tail-end of this gas dispersal process from the timescale arguments in Section \ref{ssec:timescales}. The measured mass outflow rate and momentum injection are also lower than for SSCs 4a and 14. It is important to note, however, that SSC 5a still has a high overall gas fraction (${\rm M_{H_{2},tot}/M_*}$) of 0.8, though it is not as high as SSCs 4a and 14 (1.3 and 1.6, respectively).

Further support for this picture comes from the discovery of a shell near SSC 5a in \hcn\ as shown in Figure \ref{fig:shell}. This feature is too faint to be seen in \cs\ or \httcn\ at this resolution. The shell is visible from $\sim35-70$ \kms\ relative to the cluster systemic velocity (Table \ref{tab:contparams}) and has a projected radius of $\sim6$ pc, though it is not perfectly centered on SSC 5a. Using these projected velocities, this implies an age of $\sim8-16\times10^4$ years. At its largest extent, the shell reaches the location of SSC 4a (in projection) which also has a dense gas outflow. This shell-like structure is not seen around any other SSCs.

\citet{kornei09} note that the Pa-$\alpha$ emission around SSC 5a is asymmetric (Figure \ref{fig:shell}) and could be indicative of an outflow. We investigate how this asymmetric geometry corresponds to the observed shell in \hcn. There are systematic offsets between the location of SSC 5a in the ALMA data, the NIR clusters identified in the \hst\ NICMOS data by \citet{watson96}, and the F187N and F190N \hst\ NICMOS data used by \citet{kornei09}. \citet{leroy18} corrected the positions of the \citet{watson96} NIR clusters by $\Delta\alpha,\ \Delta\delta = +0.32\arcsec,\ -0.5\arcsec$, so that the NIR cluster corresponding to SSC 5a has $\alpha,\ \delta = 00^{\rm h}47^{\rm m}32.985^{\rm s},\ -25^\circ17^{\rm m}19.76^{\rm s}$ (J2000). The positions of the F187N and F190N mosaics\footnote{These calibrated and reduced \hst\ NICMOS (NIC2) images were downloaded from the Mikulski Archive for Space Telescopes (MAST) database from M. Rieke's August 1998 program with exposure times of $\approx$384 s. The mosaic tiles were stitched together using the {\fontfamily{cmtt}\selectfont reproject} package in Astropy.} were corrected by matching the locations of the NIR clusters identified by \citet{watson96}. A linear shift of $\Delta\alpha,\ \Delta\delta = +1.48\arcsec,\ -0.85\arcsec $ brings the \hst\ images into agreement with the NIR clusters and the ALMA datasets. Figure \ref{fig:shell} shows the Pa-$\alpha$ (F187N-F190N) image around SSC 5a with the \hcn\ contours overlaid. The expanding \hcn\ shell is not particularly aligned with the asymmetry seen in the Pa-$\alpha$ emission, though it does seem to align with the north-western edge.

It is therefore possible that this shell is a previous (stronger) version of the outflow detected spectrally in this work. The shell age of $\sim10^5$ years is in good agreement with the minimum ZAMS age and the end of the period of active gas accretion (Section \ref{ssec:timescales}). It is unlikely that the outflow is due to a supernova explosion both because synchrotron emission is a negligible component of SSC 5's spectral energy distribution (E. A. C. Mills et al., in prep) and because the current momentum injection is well below what is expected from supernovae \citep[e.g.][]{kim15_cg,kim17_cg}.}

\section{Summary}
\label{sec:summary}
The central starburst in NGC\,253 harbors over a dozen massive ($M_*\sim10^5$~\msun, a likely lower limit inferred from radio recombination lines and radio continuum) and extremely young star clusters which are still very rich in gas and likely in the process of formation (\citealt{leroy18}, E. A. C. Mills et al. in prep.). Using high resolution ($\theta\approx0.028\arcsec$, equivalent to $0.48$~pc) data from ALMA we study the 350 GHz (0.85~mm) spectra of these objects. We summarize our main results below, indicating the relevant figures and/or tables. 
\begin{enumerate}
\itemsep0em
    \item We observe P-Cygni line profiles --- indicative of outflowing gas --- in three super star clusters in the center of NGC\,253, sources 4a, 5a, and 14 (c.f., Table \ref{tab:contparams}). These line profiles can be seen in the full-band spectra in many lines  (Figure \ref{fig:fullbandspec}), and particularly cleanly in the \cs\ and \httcn\ lines which are the focus of this analysis (Figures \ref{fig:cont} and \ref{fig:modelcomp}). Among these clusters, 5a is notable for being the only one of the massive clusters that is observable in the near IR (Figure \ref{fig:shell}), suggesting it has cleared much of its surrounding gas. 
    
    \item By fitting the absorption line profiles (Figure \ref{fig:modelcomp}), we measure outflow velocities, column densities, masses, and mass outflow rates (Table \ref{tab:outflowparams}). The outflow crossing times --- a lower limit on the outflow age --- are short ($\sim\rm few\times10^4$ yr), suggesting we are witnessing a short-lived phase. The outflowing mass in these objects is a non-negligible fraction of the total gas or stellar mass.
    
    \item To place limits on the opening angles and line of sight orientations of the outflows we construct a simple radiative transfer model aiming at reproducing the observed P-Cygni profiles in \cs\ and \httcn\ (Figures \ref{fig:modelcomponents} and  \ref{fig:modelgeom3d}). By varying the input temperatures, densities, velocities, and velocity dispersion of each component as well as the opening angle and orientation of the outflow component, we find that very wide opening angle models best reproduce the observed P-Cygni profiles (Figure \ref{fig:modelcomp}). While we cannot precisely determine the opening angle for each cluster, the outflows must be almost spherical, although somewhat smaller opening angles are acceptable if the outflows are pointed almost perfectly along the line of sight (Figure \ref{fig:modeldegen}). This modeling also provides measurements of the outflow velocity, column density, masses, and mass outflow rates (Table \ref{tab:modelparams}). In general, these two sets of measurements agree, given the large uncertainties (Figure \ref{fig:paramcomp}). 
    
    \item We compare measurements of the ZAMS age \citep{rico-villas20} to the gas free-fall time \citep{leroy18}, outflow crossing time, gas removal time implied by the mass outflow rate, and the gas depletion time (Table \ref{tab:timescales}). These estimates are consistent with the SSCs still being in a period of active gas collapse, though SSC 5a could be past this phase. The gas removal timescale (assuming a constant mass outflow rate) is about an order of magnitude smaller than the gas depletion time due to star formation, showing that the outflows will have a substantial effect on the star formation efficiency of these SSCs. {\change Reaccretion of gas that was expelled by the outflows is perhaps a likely scenario, which complicates the interpretation of the clusters' evolutionary sequences.}
    
    \item Given our measured velocities, masses, radii, surface densities, and momentum per unit stellar mass, we investigate the mechanisms responsible for driving the observed outflows. Possibilities are supernovae, photoionization, UV (direct) radiation pressure, dust-reprocessed (indirect) radiation pressure, and O star stellar winds. While none of these mechanisms completely explains the observations, the two explanations that are potentially in play are dust-reprocessed radiation pressure and stellar winds. It is possible that the outflows are powered by a combination of both mechanisms, with the feedback in SSC 14 dominated by dust-reprocessed radiation pressure and the feedback in SSC 5a dominated by stellar winds (Figure \ref{fig:massradius}, Section \ref{ssec:outflowpower}). 
    
    \item We report the discovery of an expanding shell (seen in \hcn) around SSC 5a with $r\sim 6$ pc (Figure \ref{fig:shell}). As mentioned above, SSC 5a is the only cluster visible in the near IR, which coupled with it having only a lower limits on the ZAMS age suggests that SSC 5a is the most evolved cluster. SSC 5a also has the weakest P-Cygni profile among the three detected and the smallest current mass outflow rate. Given its velocity and size, the shell is $\sim10^5$ yrs old, in good agreement with the minimum ZAMS age of the cluster and estimates of the end of the period of active gas collapse (Table \ref{tab:timescales}). It is thus likely that the expanding shell is a remnant from the earlier stages of the gas clearing phase when the outflow was stronger. It is unlikely that this shell was created by a past supernova explosion as synchrotron emission is a negligible component of this cluster's spectral energy distribution on 5 pc scales (E. A. C. Mills et al. in prep.). 
\end{enumerate}

While the SSCs in the heart of NGC\,253 constitute a very young population of clusters, there is evidence for differing evolutionary stages among them. A major step towards better characterizing these clusters is better measurements of their stellar masses. Current stellar mass estimates use the 36 GHz continuum emission --- assuming that it is all due to free-free emission --- to calculate the ionizing photon rate and hence the stellar mass \citep{leroy18}. Due to the enormous extinctions towards these clusters, it is not feasible to use traditional optical and near IR recombination lines as tracers of the ionized gas. High resolution hydrogen radio recombination lines (RRLs) offer direct probes of the ionizing photon rate and hence the stellar mass, and are uninhibited by dust extinction (\citealt{emig20} in NGC\,4945, E. A. C. Mills et al. in prep. in NGC\,253). In the near future, {\em James Webb Space Telescope} observations may allow us to independently establish {\change stellar masses, radiation fields, and ages} by accessing mid IR spectral line indicators. In the next decade, the combination of sensitivity and exquisite resolution of the Next Generation Very Large Array (ngVLA) may make studies at this high resolution possible for galaxies out to the Virgo cluster and beyond. 

\acknowledgments
{R.C.L. thanks Eve Ostriker for very helpful discussions and feedback which greatly improved this paper. R.C.L. also thanks Todd Thompson, Stuart Vogel, {\change Lachlan Lancaster, and Pavel Kroupa} for useful conversations and advice. The authors thank Nicolas Bolatto for his help in making the 3D visualizations for Figures \ref{fig:modelgeom3d} and \ref{fig:simsetup}. {\change Publication support was provided by the NRAO.} R.C.L. acknowledges support for this work provided by the NSF through Student Observing Support Program (SOSP) award 7-011 from the NRAO. K.L.E. acknowledges financial support from the Netherlands Organization for Scientific Research through TOP grant 614.001.351. {\change E.R. acknowledges the support of the Natural Sciences and Engineering Research Council of Canada (NSERC), funding reference number RGPIN-2017-03987.} This paper makes use of the following ALMA data: ADS/JAO.ALMA\#2017.1.00433.S. ALMA is a partnership of ESO (representing its member states), NSF (USA) and NINS (Japan), together with NRC (Canada), MOST and ASIAA (Taiwan), and KASI (Republic of Korea), in cooperation with the Republic of Chile. The Joint ALMA Observatory is operated by ESO, AUI/NRAO and NAOJ. The National Radio Astronomy Observatory is a facility of the National Science Foundation operated under cooperative agreement by Associated Universities, Inc. Based on observations made with the NASA/ESA Hubble Space Telescope, obtained from the data archive at the Space Telescope Science Institute. STScI is operated by the Association of Universities for Research in Astronomy, Inc. under NASA contract NAS 5-26555.}

\facilities{ALMA, HST} 
\software{Astropy \citep{astropy}, CASA \citep{casa}, \emcee\ \citep{emcee}, MatPlotLib \citep{matplotlib}, NumPy \citep{numpy}, pandas \citep{pandas}, photutils \citep{photutils}, SciPy \citep{scipy}, seaborn \citep{seaborn}}

\bibliographystyle{aasjournal}

\appendix
\restartappendixnumbering
\section{Measuring Outflow Properties from Fitting the Absorption Features}
\label{app:outflowproperties}
For each outflow candidate SSC, we measure important outflow properties based on the \httcn\ and \cs\ spectra (Section \ref{ssec:outflow}, Table \ref{tab:outflowparams}). Below, we explain how each of these properties is calculated. First, we fit the foreground-removed {\changes (Section \ref{ssec:foregroundgas})} \httcn\ and \cs\ spectra with a two-component Gaussian of the form
\begin{equation}
\begin{split}
    \label{eq:gauss2}
    {\rm I(V) = }\\{\rm I_{\rm max-emis}e^{-(V-V_{\rm max-emis})^2/2\sigma_{\rm emis}^2}+I_{\rm max-abs}e^{-(V-V_{\rm max-abs})^2/2\sigma_{\rm abs}^2} + I_{\rm cont}}
    \end{split}
\end{equation}
with terms to fit the emission, absorption, and continuum components respectively. These fits are shown in Figure \ref{fig:modelcomp} as the blue dashed curves. We define the outflow crossing time, which the time it takes a gas parcel to travel from the center of the cluster to $r_{\rm SSC}$ (half the major axis FWHM sizes listed in Table \ref{tab:contparams}) moving at the typical outflow velocity (${\rm V_{max-abs}}$) as
\begin{equation}
    \label{eq:tcross}
    {\rm t_{cross}} = \frac{r_{\rm SSC}}{\rm V_{max-abs}}.
\end{equation}
This is a lower limit to the outflow age, assuming a constant outflow velocity, since the outflows are observed out to at least $r_{\rm SSC}$ and could be present at larger distances from the cluster. The maximum outflow velocity is defined as
\begin{equation}
    \label{eq:voutflow}
    {\rm V_{\rm out,max} \equiv V_{\rm max-abs}+2\sigma_{abs}}.
\end{equation}
We determine the optical depth of the center of the absorption feature where
\begin{equation}
    \label{eq:tau}
    {\rm \tau_{\rm max-abs} = -\ln\Big(\frac{I_{\rm max-abs}}{{}I_{\rm cont}}\Big)}.
\end{equation}
From there, we calculate the column density in the \textit{lower} state of each molecule following \citet{mangum15} 
\begin{equation}
    \label{eq:Nl}
    {\rm N_\ell = 16\pi\sqrt{2\ln2}\Big(\frac{\nu_{u\ell}}{c}\Big)^2\sigma_{\rm abs}\frac{\tau_{\rm max-abs}}{A_{u\ell}}\Big[e^{h\nu_{u\ell}/kT_{ex}}+1\Big]^{-1}}
\end{equation}
(combining their equations  29, A1, and A7), where ${\rm \nu_{u\ell}}$ is the frequency of the transition, ${\rm c}$ is the speed of light, ${\rm A_{u\ell}}$ is the Einstein A coefficient for the transition, and ${\rm T_{ex}}$ is the excitation temperature. By replacing ${\rm T_\ell}$ with ${\rm T_{ex}}$, we are assuming LTE, which we will assume throughout. We assume ${\rm T_{ex}} = 130\pm56$ K, which is the excitation temperatures found in these clusters \citep{krieger20}. The uncertainty is the difference between this value and the excitation temperature measured at lower resolution \citep[74 K;][]{meier15}. Our assumption on ${\rm T_{ex}}$ is a major source of uncertainty in these calculations. We then calculate the column density in the upper state
\begin{equation}
    \label{eq:Nu}
    {\rm N_u = N_\ell\frac{g_u}{g_\ell}e^{-h\nu_{u\ell}/kT_{ex}}}
\end{equation}
\citep[e.g. equation 6 of ][assuming the upper and lower states have the same density distribution along the line of sight]{mangum15}. The total column density of each molecule (e.g. of all energy levels) is
\begin{equation}
    \label{eq:Ntot}
    {\rm N_{\rm mol} = \frac{N_u}{g_u}\mathscr{Z}e^{h\nu_{u\ell}/kT_{ex}}}
\end{equation}
\citep[e.g. equation 31][]{mangum15}, where $\mathscr{Z}$ is the partition function calculated assuming LTE
\begin{equation}
    \label{eq:Z}
    \mathscr{Z} = \sum_{i}g_ie^{-E_i/{\rm kT_{ex}}}
\end{equation}
where $g_i$ and $E_i$ are the degeneracy and excitation energy of each level $i$. We calculate $\mathscr{Z}$ assuming LTE up to $i=19$ for CS and $i=16$ for H$^{13}$CN\footnote{\label{ref:levelpop}Level population data are from \url{https://www.astro.umd.edu/rareas/lma/lgm/properties/cs.pdf} for CS and \url{https://www.astro.umd.edu/rareas/lma/lgm/properties/h13cn.pdf} for H$^{13}$CN. \label{fnref}}. To convert from the column density of each molecule to the column density of \htwo\ in the outflow (${\rm N_{H_{2},out}}$), we need to know the abundance ratio of each molecule with respect to \htwo:
\begin{equation}
    \label{eq:NH2}
    {\rm N_{{\rm H}_{2},out} = 2\frac{[{\rm H}_2]}{[{\rm mol}]}N_{\rm mol}}
\end{equation}
where the factor of 2 accounts for the redshifted outflowing material, assuming it is the same as the blueshifted component. The abundance ratios vary with environment, so we use those calculated in the center of NGC\,253 by \citet{martin06}, where [CS]/[\htwo] $=5.0\times10^{-9}$ and [H$^{13}$CN]/[\htwo] $=1.2\times10^{-10}$. The assumed abundance ratio is another large source of uncertainty in our measurements and can vary substantially by environment \citep[e.g.][and references therein]{vandishoeck98}. We take an order-of-magnitude uncertainty on the abundance fractions, which limits subsequent calculations to order-of-magnitude precision as well (see the discussion in Section \ref{sssec:abun}).  

To calculate the \htwo\ mass along the line of sight, 
\begin{equation}
    \label{eq:MH2}
    {\rm M_{{\rm H}_{2},out} = 4\Sigma_{{\rm H}_{2},out}A = 4m_{\rm H_{2}}N_{{\rm H}_{2},out}A}
\end{equation}
where ${\rm m_{H_{2}}}$ is the mass of a hydrogen molecule, $A$ is the area of the source measured from the continuum (Table \ref{tab:outflowparams}), and the factor of 4 converts the projected area to a sphere. 

From the mass in the outflow and the crossing time, we calculate the mass outflow rate (${\rm \dot{M}_{H_{2},out}}$) where
\begin{equation}
    \label{eq:Mdot}
    {\rm \dot{M}_{H_{2},out} = \frac{M_{H_{2},out}}{t_{cross}}}.
\end{equation}
From the total gas mass of the cluster \citep{leroy18} and the mass outflow rate, we calculate the gas-removal time, or the time it would take to expel all of the gas in the cluster at the current mass outflow rate:
\begin{equation}
    \label{eq:tremovegas}
    {\rm t_{remove-gas} = \frac{M_{H_{2},tot}}{\dot{M}_{H_{2},out}}}.
\end{equation}
The timescale assumes that the mass outflow rate is constant with time, which is likely not the case \citep[e.g.][]{kim18_jg}. The momentum injected into the environment normalized by the stellar mass (${\rm M_*}$) by the outflow is 
\begin{equation}
    \label{eq:pr}
   {\rm \frac{p_r}{M_*} = \frac{\sqrt{3}V_{\rm max-abs}M_{{\rm H}_{2},out}}{M_*}}
\end{equation}
assuming spherical symmetry \citep{leroy18}. We assume the SSC stellar masses reported by \citet{leroy18}{\changes, and the uncertainties in these ${\rm M_*}$ measurements are discussed in Section \ref{ssec:otherprop}.} The kinetic energy in the outflow is
\begin{equation}
    \label{eq:Ekin}
    {\rm E_k = \frac{1}{2}M_{H_{2},out}V_{max-abs}^2}.
\end{equation}

Values calculated using \cs\ and \httcn\ for each outflow candidate are listed in Table \ref{tab:outflowparams} along with the propagated uncertainties.

\section{Outflow Modeling Details}
\label{app:outflowmodeling}

We perform simple radiative transfer modeling of the source with varying outflow geometries and input physical parameters to constrain the outflow opening angles and orientations as described in Section \ref{ssec:outflowmodeling}.

To set up the model, we define a three-dimensional grid that is 65 ($=2^{\change 6}$+1) pixels in each dimension; we refer to this grid as the simulated box. The physical scale of the box is such that the length of each size is $4\times r_{\rm SSC}$, or twice the diameter of the SSC to be modeled. Every pixel in the box is given a fourth dimension, corresponding to the spectral axis (in terms of frequency or velocity, which we use interchangeably here). The spectral axis has 129 ($=2^7+1$) channels. The velocity range of the spectral axis and hence the spectral resolution of the model is defined adaptively for each model to maximize the number of channels over the emission and absorption features. The spectral axis is centered on zero velocity (the rest frequency of the line to be modeled) and spans ${\rm \pm4\sqrt{V_{max-abs}^2+\Delta V_{out,FWHM}^2}}$, where ${\rm V_{max-abs}}$ and ${\rm \Delta V_{out,FWHM}}$ are the outflow velocity and FWHM outflow velocity dispersion input into the model. This is done to optimize the velocity resolution over the velocities relevant for the cluster and outflow as opposed to picking a fixed velocity range. This is shown most clearly in Figure \ref{fig:modelcomponents}, where the portions of the spectra within the vertical {\change line segments} show the velocities actually modeled and the thin lines are extrapolations. 

\begin{figure}
\label{fig:simsetup}
\centering
\includegraphics[width=0.5\textwidth]{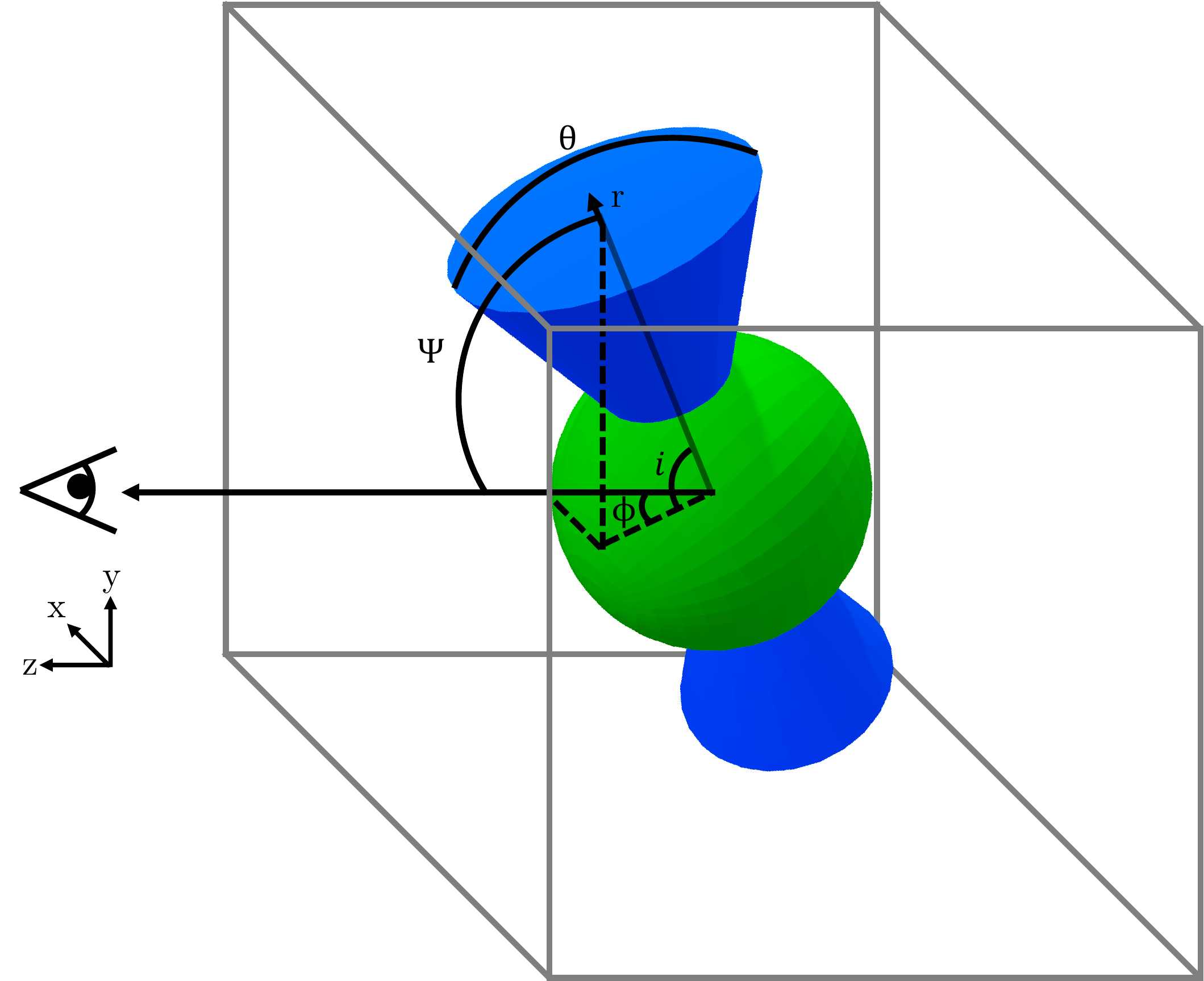}
\caption{An example outflow, in the style of Figure \ref{fig:modelgeom3d}; for clarity, the ambient gas is not shown. The simulated box is shown in gray. The orientation angle definitions are marked, where $\theta$ is the outflow opening angle and $\Psi$ is the orientation angle to the line of sight measured from the center of the outflow cone with components $i$ and $\phi$ to the $x$- and $y$-axis respectively.}
\end{figure}

\begin{figure}
    \centering
    \includegraphics[height=0.75\textheight]{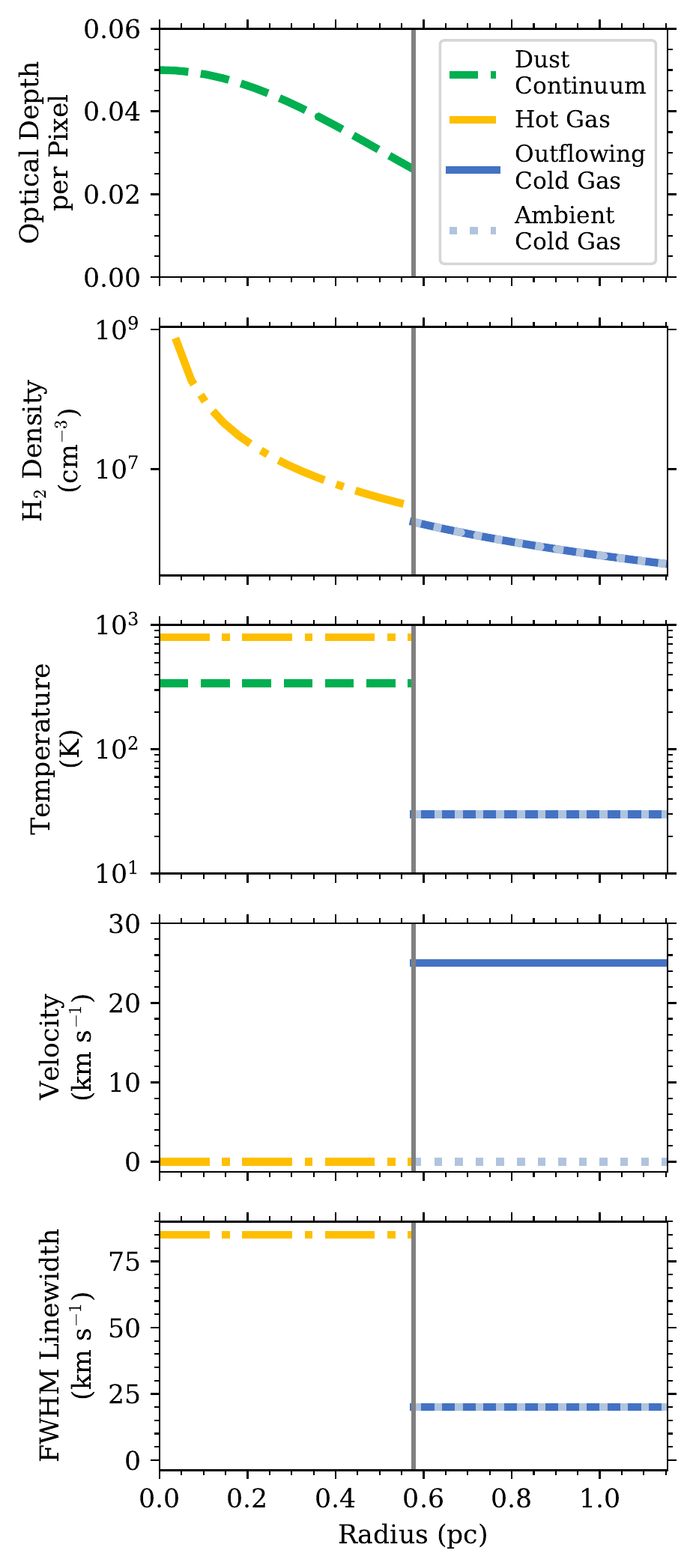}
    \caption{Radial profiles of the input parameters to the modeling for the dust continuum (green dashed), hot gas (gold dash-dotted), outflowing cold gas (blue solid), and the ambient cold gas (light blue dotted) as described in Section \ref{ssec:outflowmodeling}. For the dust continuum, the optical depth per pixel is specified, whereas for the other components, the \htwo\ density per pixel is specified. For the ambient cold gas, input parameters are identical to the outflowing cold gas except for the velocity. Cold gas inside the outflow cones has the properties of the outflowing cold gas, whereas cold gas outside the outflow cones has the properties of the ambient cold gas. The vertical gray line marks $r_{\rm SSC}$ which is the boundary between the cluster (e.g. continuum and hot gas components) and the cold gas components (outflowing or ambient). These input parameters are scaled to the best-fit spherical model for the \cs\ line in SSC 14 (Table \ref{tab:modelparams}).}
    \label{fig:paramprofiles}
\end{figure}

Once the four-dimensional box is defined, the three physical components representing the cluster and outflow are constructed. It is helpful to define coordinates related to the simulated box in cartesian coordinates, whereas coordinates pertaining to the cluster and outflow are in spherical coordinates, as shown in Figure \ref{fig:simsetup}. As described in Section \ref{ssec:outflowmodeling}, the three components of the system are:
\begin{enumerate}
\itemsep0em
\item Dust continuum component: Shown in green in Figures \ref{fig:modelcomponents}, \ref{fig:modelgeom3d}, \ref{fig:simsetup}, and \ref{fig:paramprofiles} this component is a sphere with $r=r_{\rm SSC}$ and a constant (in space and velocity) temperature (T$_{\rm cont}$). The optical depth is a maximum ($\tau_{\rm cont,max}$) at the center, then decreases like a Gaussian with FWHM $=2\times r_{\rm SSC}$, so that the continuum source is semi-transparent. The temperature and optical depth are set to zero for $r>r_{\rm SSC}$.

\item Hot gas: Shown in yellow in Figures \ref{fig:modelcomponents} and \ref{fig:paramprofiles} (and encompassed within the green sphere in Figures \ref{fig:modelgeom3d} and \ref{fig:simsetup}), this spherical component is required to reproduce the strong emission component of the P-Cygni profiles. This component has a hot gas temperature (T$_{\rm hot}$), a central \htwo\ volume density (n$_{\rm hot}$), and a velocity dispersion ($\rm \Delta V_{hot,FWHM}$). T$_{\rm hot}$ is constant (spatially) for $r\leq r_{\rm SSC}$, and is set to zero outside. The density falls off $\propto r^{-2}$ from the center, and is set to zero for $r>r_{\rm SSC}$. The line is centered on zero velocity along the spectral axis, and the Gaussian linewidth is given by ${\rm \Delta V_{hot,FWHM}}/2.355$.

\item Cold, outflowing gas: Shown in dark blue in Figures \ref{fig:modelcomponents}, \ref{fig:modelgeom3d}, \ref{fig:simsetup}, and \ref{fig:paramprofiles} this is the outflow component which produces the absorption features. This component is defined by a gas temperature (T$_{\rm out}$), a maximum \htwo\ volume density (n$_{\rm out}$), a constant outflow velocity (V$_{\rm out}$), a velocity dispersion ($\rm \Delta V_{out,FWHM}$), and opening angle ($\theta$), and an orientation to the line of sight ($\Psi$). The gas temperature is constant (spatially) within this component and is set to zero for $r<r_{\rm SSC}$. The density is a maximum at $r=r_{\rm SSC}$ and deceases $\propto r^{-2}$ until the edges of the box; the density is set to zero for $r<r_{\rm SSC}$. The line is centered on along the frequency axis at V$_{\rm outflow}$, and the Gaussian linewidth is given by $\Delta V_{\rm out,FWHM}$. Since the outflow velocity is constant and the density $\propto r^{-2}$, the outflow conserves energy and momentum. The outflow cones are masked to the given opening angle ($\theta$) and rotated to the given orientation from the line of sight ($\Psi$) (Figure \ref{fig:simsetup}). For outflows with $\theta<180^\circ$, the outflowing gas component outside the outflow cones is replaced by an ambient gas component (shown in light blue in Figures \ref{fig:modelgeom3d} and \ref{fig:paramprofiles}), which has the same properties as the outflowing gas but with V$_{\rm out}=0$. We refer to these together simply as the "cold" component.
\end{enumerate}

The optical depths of the hot and cold (outflowing and ambient) gas are needed at every pixel and as a function of frequency ($\tau_{\rm \nu,hot}$ and $\tau_{\rm \nu,outflow}$ respectively) for the radiative transfer. For simplicity in the following equations, we will drop the "hot" and "cold" subscripts since the calculations are the same for both components. First, we calculate the intrinsic line shape assuming Doppler broadening is dominant
\begin{equation}
    \label{eq:intrinlineshape}
    \phi_\nu = \frac{1}{\sqrt{\pi}\nu_{u\ell}}\frac{c}{b}e^{-V^2/b^2}
\end{equation}
where $b\equiv\sqrt{2}\sigma_V=\frac{\Delta V_{\rm FWHM}}{2\sqrt{\ln{2}}}$, $\nu_{u\ell}$ is the rest frequency of the transition being modeled, and $V$ is the velocity along the spectral axis (with $V=0$ corresponding to $\nu=\nu_{u\ell}$) and where $\int\phi_\nu d\nu = 1$ \citep{draine11}. The absorption cross section is
\begin{equation}
    \label{eq:abscrosssec}
    \sigma_{\ell u}(\nu) = \frac{g_u}{g_\ell}\frac{c^2}{8\pi\nu_{u\ell}^2}A_{u\ell}\phi_\nu
\end{equation}
where $g_u$ and $g_\ell$ are the upper and lower level degeneracies and $A_{u\ell}$ is the Einstein A value of the transition\footref{fnref} \citep{draine11}. Given the \htwo\ number density at every pixel (${\rm n_{H_{2}}}(x,y,z)$), the number density in the lower state of the modeled transition is
\begin{equation}
    \label{eq:nl}
    n_\ell(x,y,z) = \frac{\rm [mol]}{\rm [H_2]}\frac{g_\ell}{\mathscr{Z}}e^{-T_\ell/T}{\rm n_{H_{2}}}(x,y,z)
\end{equation}
where $\frac{\rm [mol]}{\rm [H_2]}$ is the fractional abundance of the molecule being modeled with respect to \htwo, $\mathscr{Z}$ is the partition function (Equation \ref{eq:Z}), $T_\ell\equiv E_\ell/k$ is the excitation energy of the lower energy state, and $T$ is the input temperature of the gas. The absorption coefficient is then
\begin{equation}
    \label{eq:asbcoeff}
    \kappa_\nu(x,y,z) = n_\ell(x,y,z)\sigma_{\ell u}(\nu)\left(1-e^{-k\nu/kT}\right)
\end{equation}
\citep{draine11}. Finally, the optical depth of each pixel and as a function of frequency is
\begin{equation}
    \label{eq:dtaunu}
    \Delta\tau_\nu(x,y,z) = \kappa_\nu(x,y,z)\Delta z
\end{equation}
where $\Delta z$ is the size of each pixel in the $z$ direction (though in this simulation the pixels are equal size in all spatial dimensions).

For each component---now also including the continuum component--- the intensity (expressed in Rayleigh-Jeans brightness temperature units) at every pixel and as a function of frequency is 
\begin{equation}
    \label{eq:dTnu}
    \Delta T_{\nu}(x,y,z) = T\left[1-e^{-\Delta\tau_\nu(x,y,z)}\right].
\end{equation}
We perform the radiative transfer along the line of sight from the back of the box to the front (e.g. along the $+z$-axis):
\begin{equation}
    \label{eq:radtrans}
    T_{\nu}(x,y,z_i) = T_\nu(x,y,z_{i+1})e^{-\sum_{\rm comp}\Delta\tau_\nu(x,y,z_i)}+\sum_{\rm comp}\Delta T_\nu(x,y,z_i)
\end{equation}
where $z_i$ denotes an individual step along the $z$-axis and $\sum_{\rm comp}$ means a sum over the dust continuum, hot gas, and cold (outflowing and ambient) gas components.

The observed spectra are averaged over the FWHM continuum source size. To best compare with the observed spectra, we mask the simulated box of $T_\nu(x,y,z)$ to a cylinder along the $z$-axis with $r=r_{\rm SSC}$ as shown in Figure \ref{fig:modelgeom3d}, setting pixels outside this region equal to zero. The final modeled spectrum is
\begin{equation}
    \label{eq:specout}
    T_\nu = \frac{1}{N_{\rm pix}}\sum_{x,y}T_\nu(x,y,z=z_{\rm max})
\end{equation}
where $z=z_{\rm max}$ denotes the slice (in the $xy$ plane) at the front of the box and $N_{\rm pix}$ is the number of non-masked pixels in that slice. The best-fitting modeled spectra are shown in red in Figure \ref{fig:modelcomp}.

There are degeneracies between input parameters. The observed continuum level is a combination of the intrinsic continuum temperature (T$_{\rm cont}$) and the continuum optical depth ($\tau_{\rm cont}$); a higher $\tau_{\rm cont}$ with a lower T$_{\rm cont}$ can produce the same fit as a lower $\tau_{\rm cont}$ with a higher T$_{\rm cont}$. The temperature and density of the hot gas component (T$_{\rm hot}$ and n$_{\rm hot}$) are linked in a similar way. More, a lower $\tau_{\rm cont}$ means less of the hot gas and redshifted outflow are attenuated, and so lower T$_{\rm hot}$ and n$_{\rm hot}$ values are required. These degeneracies, especially with regards to the density, lead to a very uncertain total mass for the modeled cluster. The outflow parameters, however, are more robust. The gas temperature in the outflow (T$_{\rm out}$) is linked to the density in the outflow (n$_{\rm out}$), but are not as degenerate as for the hot gas component {\changes for the following reason. For absorption, the minimum modeled brightness temperature cannot be less than the given T$_{\rm out}$, even for an arbitrarily high density. That is, in order to match the models with the observations, the observed temperature at maximum absorption sets the maximum possible T$_{\rm out}$ in the model.} For these reasons, though we report all input parameters for the best-fitting models, we only report derived parameters for the outflow in Table \ref{tab:modelparams}.

\end{document}